\documentclass[aps,pra]{revtex4}
\usepackage{amsmath}
\usepackage{graphicx}
\usepackage{epsfig}

\begin{document}

\title{Fundamental noise in matter interferometers}

\author{Adilet Imambekov, Vladimir Gritsev and Eugene Demler}
\affiliation{Department of Physics, Harvard University, Cambridge
MA 02138}

\date{\today}

\begin{abstract}
These lecture notes discuss two effects which contribute to the
reduction of the interference fringe contrast in matter
interferometers. The first effect is the shot noise arising from a
finite number of atoms used in experiments. Focusing on a single
shot measurement we provide explicit calculations of the full
distribution functions of the fringe contrast for the interference
of either the coherent or the number states of atoms.  Another
mechanism of the suppression of the amplitude of interference
fringes discussed in these lecture notes is the quantum and thermal
fluctuations of the order parameter in low dimensional condensates.
We summarize recent theoretical and experimental studies
demonstrating that suppression of the interference fringe contrast
and its shot to shot variations can be used to study correlation
functions within individual condensates.  We also discuss full
distribution functions of the fringe amplitudes for one and two
dimensional condensates and review their connection to high order
correlation functions. We point out intriguing mathematical
connections between the distribution functions of interference
fringe amplitudes and several other problems in field theory,
systems of correlated electrons, and statistical physics.
\end{abstract}
\maketitle
\tableofcontents

\section{Introduction}
\label{introsection}

\subsection{Interference experiments with cold atoms}

From the earliest days of quantum mechanics its probabilistic nature
was the cause of many surprises and controversies \cite{EPR}.
Perhaps the most unusual manifestation of the quantum uncertainty is
a quantum noise: measurements performed on identical quantum
mechanical systems can produce results which are different from one
experimental run to another. At the level of a single particle
quantum mechanics, the quantum noise is no longer a research topic
but is discussed in undergraduate physics textbooks
\cite{COHEN-TANNOUDJI}.  However, the situation is different when we
talk about quantum mechanics of many-body systems. One can ask
seemingly simple questions to which there is no obvious answer: does
it still make sense to talk about quantum noise when discussing
measurements on many-body quantum states? How does the quantum noise
manifest itself?  Can one use this noise to extract nontrivial
information about the system?

The idea of the quantum noise analysis of many-particle systems is
common to many areas of condensed matter physics \cite{levitov,
meso, beenakker, fqhe_noise} and quantum optics
\cite{HBT,scully,Walls,Glauber63,KlaudSud,MaWo}.  In the field of
ultracold atoms it has been successfully employed in a variety of
recent experiments
\cite{greiner_noise,bloch_noise,Spielman_noise,bloch_noise_fermions,Esslinger_g2,Aspect,Aspect2}
with many more theoretical proposals awaiting their turn
\cite{pra_noise,Cherng,Galitski,Bruder,Lamacraft,Kuklov,Lewenstein_noise}.
These lecture notes address a very specific experimental probe of
the cold atomic ensembles -- interference experiments. Our
discussion focuses on a variety of interesting and important
phenomena which originate from the fundamental quantum and/or
thermal noise of cold atoms condensates and can be studied in
interference experiments. Although focused on the specific type of
experiments, the general methodology discussed in these lecture
notes can be extended to a variety of other measurements on systems
of cold atoms.

Interference experiments constitute an important part of the
modern toolbox for studying ultracold atoms. Original experiments
used large three dimensional Bose-Einstein condensates (BEC) to
demonstrate macroscopic coherence \cite{andrews}. More recently
interference experiments have been done with one and two
dimensional condensates \cite{Thywissen, Schumm, Hofferberth,stock,zoran_KT, flatland}  and demonstrated the
important role of fluctuations in low dimensional systems. Matter
interferometers using cold atoms
\cite{Kasevich,Kasevich2,Chu,Hansch_int,Clade,Sr,Schumm, Hofferberth, Dumke,hinds,hansel,Andersson,Wang,Kreutzmann,phasedyn,Shin2,MIT_squeezing,Jo2,Chip_review}
have been considered for applications in accelerometry,
gravitometry, search for quantum gravity, and many other areas
(for a  review see Ref. \cite{Interferometrybook}). Interference
experiments have been used to measure the condensate formation
\cite{Bloch_formation, Esslinger_formation} as well as the
critical properties of the BEC transition \cite{Esslinger_formation2}.
What is common to most interference experiments is the focus on
the {\it phase} of interference patterns. Suppression of the
{\it fringe contrast} is considered to be a spurious effect caused by
noise and fluctuations. On the contrary, these lecture notes focus
on understanding physical phenomena that underlie the imperfect
visibility of interference fringes. As we discuss below,
suppression of the fringe visibility comes from fundamental
physical phenomena, such as the noise  intrinsic to
performing a classical measurement on a quantum mechanical wave
function (shot noise) or classical and quantum fluctuations of the
order parameter.

In these lecture notes we discuss how one can use analysis of the
contrast of interference fringes to learn about fluctuations of the
order parameter \cite{pnas,Gritsev,Gritsev_proc,fcsshort,zoran_KT,flatland}.
We will demonstrate that important information is contained not only in the
average contrast but also in its shot to shot variations. For example,
when we discuss fluctuating condensates, we will show that high
moments of the interference fringe amplitudes contain information about
high order correlation functions and thus provide valuable information
about the system.

\begin{figure}
\includegraphics[width=13.0cm]{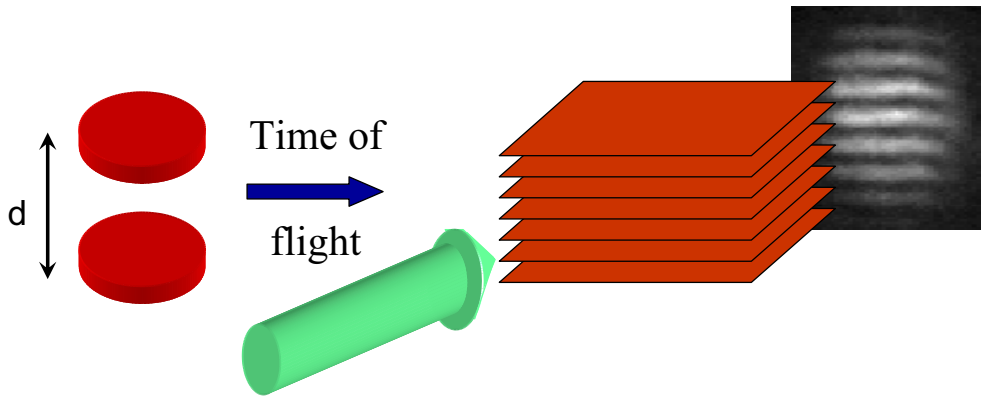}
\caption{\label{fartistic}  Schematic view of the interference
experiments with ultracold atoms. Two condensates, originally separated by the
distance $d,$  are released from the traps and expand until the clouds overlap.
 The imaging beam measures the density of the atoms after the expansion. Quantum interference leads to the periodically modulated density projected on the screen. Projected density image is taken from the actual experimental data of Hadzibabic et al. \cite{zoran_KT}.
}
\end{figure}

\begin{figure}
\psfig{file=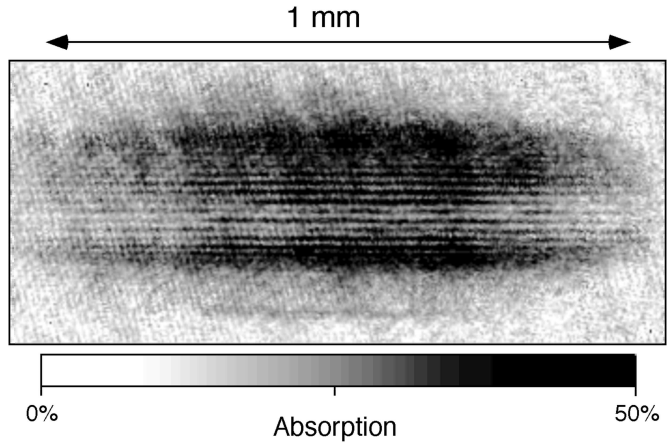} \caption{\label{ketterlefig} Interference pattern
from the first set of experiments with sodium atoms. Taken from Ref. \cite{andrews}. }
\end{figure}

The basic scheme of interference experiments is shown in Fig.
\ref{fartistic}. Originally two condensates are located at distance
$\bf d$ away from each other. At some point they are allowed to
expand ballistically \cite{ballist} until  sizes of the clouds
become much larger than the original separation between the clouds
$d$. After the expansion the density is measured by shining a laser
beam through the cloud. Interference leads to the appearance of the
density modulation at a wave vector ${\bf Q} = m {\bf d}/ \hbar t$
(see Fig.~\ref{fartistic} and discussion in section
\ref{3dinterference}). When the two condensates are coherent, the
position of interference fringes is determined by the relative phase
between the two clouds. Surprisingly the interference pattern will
be observed even for two independent condensates which do not have a
well defined relative phase (see e.g. Fig. \ref{ketterlefig}). To
beginning readers it may seem confusing that we can observe
interference in the absence of coherence between the two clouds. Or
even more confusing, we discuss interference patterns when both
clouds are number states and  phases of individual condensates are
not well defined. Several theoretical frameworks have been
introduced to understand the origin of interference patterns in the
absence of phase coherence \cite{Glauber, Javanainen, Castin,
Zoller,Pethick}. In this paper we explain the origin of interference
fringes using the language of correlation functions
 and point out connections to Hanbury Brown and
Twiss (HBT) experiments \cite{HBT} in optics. This section
provides a simple heuristic picture and a more formal discussion
is left to section \ref{3dinterference}.

What one measures in experiments is the density profile after the
expansion $\rho({\bf r})$. Interference pattern appears as the
density modulation $\rho({\bf r}) = \rho_{\bf Q} e^{- i{\bf Q}{\bf
r}} + {\rm c.c. + const}$. The absolute value of $\rho_{\bf Q}$
determines the amplitude of interference fringes and its phase
defines the position of the fringe maxima and minima. Schematically
we can write $\rho_{\bf Q} \sim e^{i \phi_1 - i \phi_2},$ where
$\phi_{1,2}$ are phases of the two condensates before the expansion
(see discussion in section \ref{3dinterference}). In the absence of
coherence $\langle e^{i \phi_1 - i \phi_2} \rangle =0, $ which
implies that $ \langle \rho_{\bf Q} \rangle =0 $. Vanishing of the
average, however, does not mean that interference fringes are absent
in each individual shot.  In the present case it only shows that the
phase of interference fringes is random in each shot. We remind the
readers that taking an expectation value in quantum mechanics
implies averaging over many measurements. On the other hand we can
focus on the amplitude of interference fringes and accept the fact
that we can not predict their phase. Then we need to consider the
quantity which does not vanish after averaging over the
unpredictable phase difference. One such quantity is given by the
density-density correlation function $ \rho({\bf r})\rho({\bf r'}) =
|\rho_{\bf Q}|^2 ( e^{i{\bf Q}({\bf r}-{\bf r}')} + {\rm c.c.}) +
{\rm other \,\, terms}.
$
The right hand side of the last equation  does not vanish when
we average over random phases $\phi_{1,2},$ and we find finite
expectation value
\begin{eqnarray}
\langle \rho({\bf r})\rho({\bf r'}) \rangle =2 \langle |\rho_Q|^2
\rangle \cos ({{\bf Q}({\bf r}-{\bf r}')})+ \mbox{const}. \label{intro_eq1}
\end{eqnarray}
What this correlation function tells us is that in a single shot
we can not predict whether at a given point $r$ we will have a
minimum or a maximum of the density modulation. However what we
can say is that if there is a maximum at point $r$, it will be
followed by another maximum a distance $2\pi/Q$ away. While this
simple argument explains the origin of interference patterns from
independent condensates, it leaves many questions unanswered. For
example, it is not obvious how accurately one can represent two
independent condensates using states with a well defined
but unknown phase difference. Also it is not clear how to
generalize this analysis to elongated condensates, when we need to
go beyond the single mode approximation and include phase
fluctuations within individual systems. These lecture notes will
present a uniform approach for addressing these and many other
questions.

When the focus of interference experiments is on measuring the phase,
one usually averages interference patterns obtained in several
shots. The result is easy to interpret: an average of many
experimental runs is precisely what we define as a quantum mechanical
average. However in experiments with independent condensates, summing
interference patterns is not appropriate. The phase of interference
patterns is random from shot to shot and adding individual images
washes out interference fringes completely (for a nice experimental
demonstration see Ref. \cite{zoran}).  Hence, in this case one needs to
focus on interference patterns obtained in individual shots. In the
absence of averaging, a single shot measurement contains noise. Thus
to characterize such experiments we need to provide both the average
value and the shot to shot fluctuations of the fringe contrast.  The
most comprehensive description of the fluctuating variable comes from
providing its {\it full distribution function}. Theoretical calculations
of the distribution functions of the fringe contrast will be the
central part of these lecture notes.

\begin{figure}
\includegraphics[width=5cm]{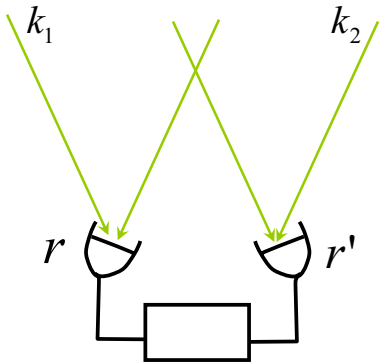}
\caption{\label{hbtfig}  Schematic view of Hanbury Brown and Twiss
noise correlation experiment as an example of intensity
interferometry. Detectors at positions $r,r'$ measure the {\it
intensity} of light coming from two distant incoherent sources. The
"correlator" (denoted by the box) measures the coincidence events and
thus the intensity-intensity correlation function.}
\end{figure}

It is useful to point out the analogy between the approach
discussed in this paper and the famous Hanbury Brown and Twiss
experiments in optics. The original motivation for HBT experiments
came from astronomy: the goal was to measure the angle between two
incoherent stellar sources such as two different points on the
surface of the star. Since the two sources are incoherent, this can
not be done using a single detector: first order interference is
absent and the measured signal is simply the sum of the two
intensities \cite{scully}. The insight of HBT was to use two
detectors and measure the correlation function of the two
intensities as a function of the relative distance between the two
detectors. One finds that this correlation function is given by
\begin{eqnarray}
\langle I({\bf r}) I({\bf r'} )\rangle \sim \cos\left( ({\bf k}_1 -{\bf k}_2)
 ({\bf r} -{\bf r}') \right) + \mbox{const},
 \label{intro_eq2}
\end{eqnarray}
where ${\bf k}_{1,2}$ are wave vectors of photons arriving from the
two points on the surface of the star, ${\bf r}$, ${\bf r}'$ are
positions of the detectors, and $I({\bf r})$, $I({\bf r}')$ are the
intensities measured in the two detectors (see Fig. \ref{hbtfig}).
Hence, the main idea of HBT experiments is that information is
contained not only in the average signal $\langle I({\bf r})
\rangle,$ but also in the noise. Such noise can be characterized by
looking at higher order correlations.  In astronomy, HBT experiments
were used to measure several important properties of distant stars,
including their angular sizes and the surface temperature
\cite{Baym}.

There is  an obvious analogy between equations (\ref{intro_eq1})
and (\ref{intro_eq2}), but there is also an important difference.
HBT stellar interferometers operate in real time, and averaging
over time is built into the measuring procedure.  In these
experiments, fluctuations of $I({\bf r}) I({\bf r'}) - \langle
I({\bf r}) I({\bf r'}) \rangle$, which would correspond to higher
order correlations in $I({\bf r})$, are not easy to measure. On
the other hand interference experiments with cold atoms are of a
single-shot type: each measurement is destructive and gives a
certain density profile $\rho(\bf r).$ A single image contains
information not only about the two point correlation function
$\langle \rho({\bf r})\rho(\bf r') \rangle$, but about higher
order correlations as well. The most important information for our
purposes is contained in the interference pattern at wave vector
${\bf Q}$. Essentially each interference pattern constitutes a
classical measurement of the quantum mechanical operator
$\rho_{\bf Q}$.  We remind the readers that we expect the phase of
$\rho_{\bf Q}$ to be random, so the quantity of interest will  be
$|\rho_{\bf Q}|^2$. By performing measurements several times we
will find not only the average value of this operator, but also
its higher moments. Ultimately we should be able to reconstruct
the entire distribution function for $|\rho_{\bf Q}|^2$.  So the
simplified and idealized procedure that we analyze is the
following: one performs interference experiments many times. Each
experiment is analyzed by doing a Fourier transform of the density
to extract $\rho_{\bf Q}$. The histogram of the measured values of
$|\rho_{\bf Q}|^2$ will be the main subject of these lecture
notes. We will demonstrate the wealth of information that can be
extracted from analysis of such histograms.

As a passing note, we mention that the setup considered in Fig. \ref{fartistic} is not the only possible  configuration for
interference experiments. In another common setup one makes several copies of the same cloud using Bragg pulses, and observes an interference
between them \cite{Hagley,Simsarian,Hellweg,Cacciapuoti, Bongs_review,Miller,Hugbart05}.

It is useful to put our work in the general perspective of noise
analysis in physical systems. Understanding photon fluctuations is
at the heart of modern quantum optics and provides a basis for
creation, detection, and manipulation of non-classical states of
light \cite{Walls}. The field of  quantum optics
 has a long and fruitful tradition of
using the higher-order correlation functions as well as the shot
noise to characterize the quantum states of light. The notion of
higher-order degree of coherence was first introduced by R. Glauber
in 1963 \cite{Glauber63}, also by  Klauder and Sudarshan
\cite{KlaudSud} and by Mandel and Wolf \cite{MaWo}. The knowledge
of photoelectron counting distribution function reveals such
non-classical features of light as antibunching \cite{third},
sub-Poissonian statistics and probe of violation of Bell
inequalities. In particular, third-order correlations provide a
test for distinguishing between quantum and hidden variable theories
in a way analogous to that provided by the
Greenberger-Horne-Zeilinger test of local hidden variable theories
\cite{Kinsler}. Interference of {\it independent} laser beams was first observed in Ref. \cite{MaMa} and stimulated a number of theoretical
studies (for reviews see Refs. \cite{MaWo,mandel,Perina}).


 In condensed matter physics, noise
analysis was also suggested as a powerful approach for analyzing
electron systems \cite{levitov}. It was demonstrated
theoretically that in certain mesoscopic systems current
fluctuations should contain more information than the average
current itself. In particular, the third and higher
moments contain important quantum information on interaction
effects, entanglement and relaxation processes (see e.g. Refs.
\cite{third-meso,sukho}). Specific proposals exist for detecting statistics of
quasi-particles \cite{52}, understanding transmission properties of
small conductors \cite{meso} and observing entanglement between
electrons \cite{beenakker}.  Perhaps the most spectacular
experimental success of the noise analysis in electronic systems has
been the demonstration of the fractional charge of quasiparticles in
the fractional quantum Hall regime \cite{fqhe_noise}. However
generally the noise analysis in condensed matter systems did not
become the detection tool of the same prominence as in quantum
optics. The main reason for this is the excruciating difficulty of
the noise measurements in solid state experiments. One often needs
to measure a signal which is only a part in a million of the
unwanted technical noise.

In the field of ultracold atoms
experiments analyzing quantum noise are only starting. However we
have already seen spectacular successes in several recent
experiments. Analysis of noise correlations in the time of flight
experiments \cite{pra_noise} was used to demonstrate fermionic
pairing \cite{greiner_noise} as well as HBT type correlations for
atoms in optical lattices
\cite{bloch_noise, Spielman_noise, bloch_noise_fermions}(see also I. Bloch's lectures
in this school). Single atom detectors have been used to demonstrate
HBT noise correlations for cold atoms
\cite{Esslinger_g2,Aspect,Aspect2}.  Strongly interacting systems of
cold atoms are expected to realize analogues of important models of
condensed matter systems
\cite{Ketterlereview,ZollerJakschreview,Lewenstein_review}.  Being able to
study noise in such systems should provide an important new
perspective on strongly correlated states of matter and have a
profound effect on many areas of physics.  We hope that these
lecture notes will stimulate more experimental work in analyzing
noise in interference experiments. The first success in this
direction was the recent observation of the
Berezinskii-Kosterlitz-Thouless (BKT) transition
\cite{Berezinskii,KT} by Hadzibabic et al. in Ref. \cite{zoran_KT}.

\subsection{Fundamental sources of noise in interference
experiments with matter}

Two fundamental sources of fluctuations in the amplitude of
interference fringes are the shot noise and the order
parameter fluctuations within individual condensates.

{\it Shot noise} comes from the finite number of atoms used in the
experiments. Let us discuss limiting cases first. Consider an
interference experiment with one atom. Before the expansion the
atom is in a perfect superposition of being between the two wells.
After the expansion we get a perfectly periodic wave function
$\psi({\bf r}) = 2 C \cos (\frac{{\bf Q}{\bf r}+ \phi}{2})$ (for a
more detailed discussion see section \ref{3dinterference}). The expectation value of
the density operator is $\langle \rho ({\bf r}) \rangle = \langle |
\psi ({\bf r})|^2 \rangle =2 |C|^2 (\cos ({\bf Q}{\bf r}+ \phi)+1).$
However this average value will not be measured in a single shot.
A single measurement finds the atom at a single point. The
expectation value of the density determines probabilities with
which we can find atom at any given point, but in a single
measurement we collapse a quantum mechanical wave function and
observe the atom at one point only.  Can one reconstruct the
entire amplitude of the interference pattern $\rho_{\bf Q} = |C|^2
e^{-i \phi}$ from a single measurement? Obviously the answer is no.
In the opposite case of a very large number of atoms  in the same single particle state one should be
able to reconstruct a complete interference pattern already from a
single measurement, since the measurement of positions of many atoms
performs a statistical averaging implicit in quantum mechanics.  In the general case of experiments with a
finite number of atoms, the question arises how well one can
determine the amplitude of interference fringes from doing a
single shot measurement. Formulated more accurately the problem is
to determine probabilities of finding a certain amplitude of
interference fringes, $|\rho_{\bf Q}|^2$, in a single measurement.

\begin{figure}
\psfig{file=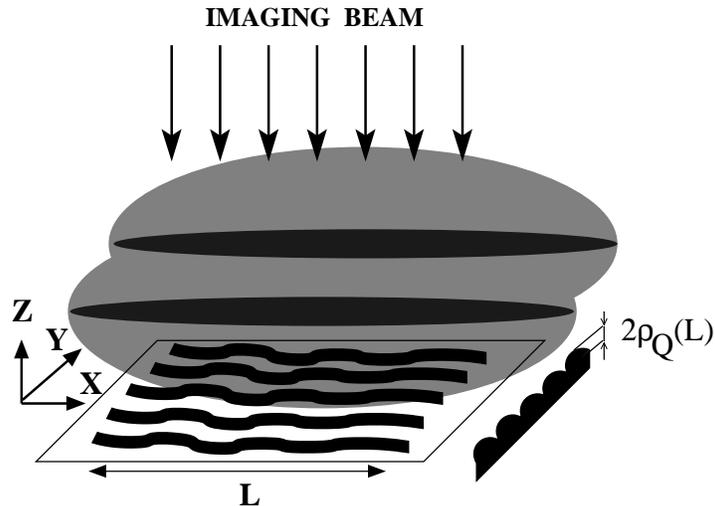} \caption{\label{setup_1D} Simplified
setup of interference experiments with 1D Bose liquids (see e.g. Ref. \cite{Schumm}). Two parallel condensates are extended in the $x$ direction. After atoms are released from the trap,  clouds are imaged by the laser beam propagating along the $z$ axis. Meandering structure of the interference pattern arises from  phase fluctuations along the
condensates. The net interference amplitude $\rho_{\bf Q} (L)$ is defined from the density integrated along the section of length $L.$ }
\end{figure}

{\it Fluctuations of the order parameter} are particularly important
for low dimensional systems. If the condensates are confined in one
\cite{1dexp,gorlitz} or two \cite{stock,zoran_KT,flatland,2dconds,gorlitz} dimensions, then
the true long-range order may not exist. Rigorous theorems forbid
true long range order in two dimensional systems at finite
temperature and in one dimension even at zero temperature
\cite{MerminWagner,Hohenberg,Coleman}. What this means is that low
dimensional condensates can not be characterized by a single phase
and we need to take into account spatial fluctuations of the order
parameter. Effects of such fluctuations on interference experiments
are illustrated schematically in Fig. \ref{setup_1D}. Two one
dimensional clouds expand in the transverse direction. Each point
along the condensates has a local interference pattern, but in the
presence of phase fluctuations (either thermal or quantum), these
patterns are not in phase with each other. It is natural to define
the net interference amplitude from the density integrated over the
axis of the system (the so called columnar density). In many
experiments such integration is done by the measurement procedure
itself. For example, systems such as shown in Fig. \ref{setup_1D}
originally had imaging done along the axis of the interferometer
\cite{Schumm}. Then the laser beam integrates the atomic densities
within the imaging length.
Integrating over local interference patterns which are not in phase
with each other leads to a reduced contrast of the net interference
fringes.

In earlier literature smearing of interference patterns by
fluctuations was considered an unwanted effect \cite{andrews}. The point of view
presented in these notes is quite the opposite. Suppression of the
fringe contrast is an interesting effect which tells us about
important phenomena in Bose condensates, such as
thermal and quantum fluctuations of the order parameter.
By analyzing such suppression we can extract non-trivial information
about the system. In particular,
it has been shown in Ref. \cite{pnas} that the
scaling of the average interference signal with the observation area
contains information about the two-point correlation functions within each cloud.
Recent experiments \cite{zoran_KT} by Hadzibabic et al. used this approach to observe the BKT transition in two
dimensional condensates (see discussion in section \ref{lowdim}). We also note that such experiments can be
used to extract information which is difficult to obtain by other
means. For example, in section \ref{lowdim} we discuss that analysis of the
high moments of the contrast tells us about high order
correlations within individual clouds.

These lecture notes are organized as follows.
In section \ref{3dinterference} we discuss how interference fringes
appear for ideal non interacting 3D BEC's at zero temperature.
In section \ref{Snoise}
we analyze the shot noise for ideal condensates.
In these lecture notes ideal condensates are understood as clouds
of non-interacting atoms which before the expansion occupy a single mode within each of the traps. The problem of interference of independent condensates of
ideal bosons has been extensively analyzed in the literature before
\cite{Glauber, Javanainen, Castin, Zoller,Pethick,Laloe,Wong,Horak,Molmer,Naraschewski,Rohrl}. In particular in an important recent paper
\cite{shotnoise} , Polkovnikov showed that the variance of the fringe
amplitude decreases as an inverse power of $N$, with a non-universal
coefficient which contains information about the state of each cloud
(e.g. coherent states vs Fock states). In this paper we develop a
general formalism for calculating the {\it full distribution functions}
of the fringe amplitudes in interference experiments with ideal condensates with a finite number of atoms.  We apply this formalism to obtain distribution
functions for several experimentally relevant cases such as states
with a well defined phase difference between the two clouds and Fock
states of atoms. Effects of the order parameter fluctuations are
discussed in section \ref{lowdim}. We obtain distribution functions
for both one and two dimensional condensates in the limit when the
number of particles is large and the shot noise can be neglected. We
also discuss intriguing mathematical connections between these
distribution functions and a number of important problems in
physics, such as the quantum impurity problem in a low dimensional interacting electron system \cite{kanefisher} or the distribution of
roughness in systems with $1/f$ noise \cite{1fnoise}.  In these lecture notes we do not
address the issue of technical noise which is obviously important for
understanding real experiments. In the concluding section \ref{conclusions}, however, we
comment on the experimental requirements for observing some of the
phenomena discussed in these lecture notes.

\section{Interference of ideal condensates}

\label{3dinterference}

 In this section we discuss why interference fringes
appear for ideal non interacting BEC's at zero temperature. We follow Ref. \cite{Pethick}, and introduce notations for subsequent sections. First
we consider the case of two clouds with a well defined relative phase,
where appearance of interference fringes can be understood at a
single particle level. Then we show that almost ideal
interference fringes appear even when two expanding clouds are
uncorrelated, provided that the number of particles in each cloud is
large.

\subsection{Interference of condensates with a well defined relative phase}

\subsubsection{Basics of interference experiments. First quantized representation}

 To illustrate how interference fringes arise, let us
start by considering a simple case of two BEC clouds
with a well defined relative phase. Here we neglect interactions between
atoms, so initially all atoms are assumed to be  in the same
single-particle state (single mode approximation).  After the confining potential is removed, the single-particle state evolves with time, but many body wave function remains in the product state. The interference appears
as a result of single particle wave function evolution, which can be
studied in detail.

Normalized single particle wave functions for two clouds will be denoted as
$\psi_1({\bf r},t)$ and $\psi_2({\bf r},t),$ and the initial relative phase is $\varphi.$
 If the total number of particles equals $N,$ then the complete wave function of the system in the
first quantized notations at any moment of time is given  by
\begin{eqnarray}
\Psi({\bf r}_1,...,{\bf
r}_N,t)=\prod_{n=1}^{N}\frac1{\sqrt2}(\psi_1({\bf
r}_n,t)e^{i\varphi/2}+\psi_2({\bf r}_n,t)e^{-i\varphi/2}).
\label{evolwavefunc}
\end{eqnarray}
This wave function satisfies the proper symmetry requirements for
permutations of ${\bf r}_i$ and ${\bf r}_j,$ and evolution of
$\psi_1({\bf r},t)$ and $\psi_2({\bf r},t)$ is controlled by the
single particle Schr\"odinger equation. Initial overlap of  the states
$\psi_1({\bf r})\equiv \psi_1({\bf r},0)$ and $\psi_2({\bf
r})\equiv\psi_2({\bf r},0)$ is assumed to be negligible:
\begin{equation}
\int d{\bf r} \psi_1^*({\bf r})\psi_2({\bf r})\approx 0.
\end{equation}

The expectation value of the
total density corresponding to the wave function (\ref{evolwavefunc})
is
\begin{eqnarray}
\langle \rho({\bf r},t) \rangle =\frac{N}{2}\left(|\psi_1({\bf
r},t)|^2+|\psi_2({\bf
r},t)|^2+2\mbox{Re}\left[e^{i\varphi}\psi_1({\bf
r},t)\psi_2^{*}({\bf r},t)\right]\right). \label{nrt}
\end{eqnarray}
The expectation value of the total density displays an interference pattern due to the last term of Eq. (\ref{nrt}). As a simple example, let us assume that
$\psi_1({\bf r},t)$ and $\psi_2({\bf r},t)$  are initially in the
Gaussian states   centered at points $\pm {\bf d}/2,$ and their
widths are $R_0\ll d.$  Then the evolution of the single particle wave
functions can be simply calculated, and the result is
\begin{eqnarray}
\psi_1({\bf r},t)=\frac{1}{(\pi R_t^2)^{3/4}}e^{-\frac{({\bf r - d}/2)^2(1+i\hbar t/m R_0^2)}{2R_t^2}}, \label{psi1rt}\\
\psi_2({\bf r},t)=\frac{1}{(\pi R_t^2)^{3/4}}e^{-\frac{({\bf r +
d}/2)^2(1+i\hbar t/m R_0^2)}{2R_t^2}}, \label{psi2rt}
\end{eqnarray}
where the widths of the wave packets, $R_t,$ at time $t$ are given by
\begin{eqnarray}
R_t^2=R_0^2+\left(\frac{\hbar t}{m R_0}\right)^2. \label{Rtdef}
\end{eqnarray}
We will be interested in the regime, when the sizes of the clouds $R_t$ are much larger than
original distance between the clouds, that is
\begin{eqnarray}
R_t\gg d \gg R_0.\label{ltl}
\end{eqnarray}
In this regime, the clouds  overlap strongly, and the real parts of
the exponents in Eqs. (\ref{psi1rt}) and (\ref{psi2rt}) are
responsible for the broad overall density profile. Imaginary
parts in the same exponents give rise to interference effects in
the last term of Eq. (\ref{nrt}). Thus the interference part of the density is equal to
\begin{equation}
\frac{N}{(\pi R_t^2)^{3/2}}e^{-\frac{{\bf r^2 + d^2/4}}{R_t^2}}
\cos{\left(\frac{\hbar}{m}\frac{{\bf r d}}{R_0^2R_t^2}t+\varphi
\right)}.
\end{equation}
For sufficiently large $t,$ one can substitute $R_t\approx\hbar
t/m R_0,$ and obtain oscillations of the density at wave vector
${\bf Q}=m {\bf d}/\hbar t,$ with positions of the minima and the maxima controlled by
the relative phase $\varphi$. The Fourier transform of the density at
wave vector $\bf Q$ is
\begin{equation}
\langle \rho_{\bf Q} \rangle =\langle\int d{\bf r} \rho({\bf r}
,t) e^{i \bf Q r} \rangle \approx \frac{N}2 e^{-i\varphi}.
\label{rhoQfq}
\end{equation}

\begin{figure}
\psfig{file=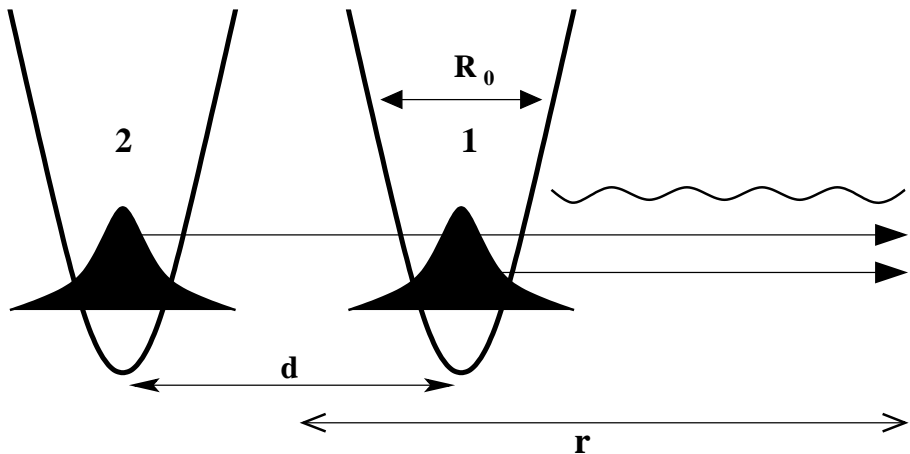} \caption{\label{3dscheme}  Schematic view of the interference
experiment with 3D condensates.}
\end{figure}

Physically $\bf Q$ can be understood as the momentum difference
of the two particles which have been released from the two traps
and arrive simultaneously at the detection point {\bf r}. This can be seen
from the following quasiclassical argument. A particle released from the condensate one and detected at time $t$ at  point {\bf r} has momentum
\begin{eqnarray}
{\bf Q}_1=m({\bf r-d}/2)/(\hbar t). \label{Q1}
\end{eqnarray}
During the expansion this particle picks up a phase ${\bf Q}_1 ({\bf r-d}/2).$  Analogously a particle originating from the condensate two  has momentum
\begin{eqnarray}
{\bf Q}_2=m({\bf r+d}/2)/(\hbar t) \label{Q2}
\end{eqnarray}
and picks up a phase ${\bf Q}_2 ({\bf r+d}/2).$
The interference pattern arises from the oscillating structure of the
phase difference with the wave vector of oscillations
\begin{eqnarray}
{\bf Q}={\bf Q}_1-{\bf Q}_2=m {\bf d}/\hbar t. \label{Q}
\end{eqnarray}
 This simple argument shows that as long as the original sizes of the clouds are much smaller than the distance between them, after sufficient
expansion one should observe oscillations of the density at the wave
vector $\bf Q$ determined by the distance between the clouds.

The interference patterns which we introduced up to this point
appear as a result  of the time  evolution of single particle
states. The many-body nature of the state comes into Eq.
(\ref{nrt}) only as a prefactor $N.$
 In principle, one could have done the same experiment with only one particle. The same result as Eq. (\ref{nrt}) can be obtained in this case by doing experiments many times and averaging over individual experiments: in each particular realization the particle is
observed at some random point ${\bf r}$ with probability
$\langle \rho({\bf r},t) \rangle.$ In experiments with a  large number of atoms, $N,$ used in each shot, each absorption
image is a result of $N$ measurements of  single
particle wave functions. This performs the statistical averaging implicit in quantum mechanics, and leads to the density profile close to Eq.
(\ref{nrt}) in each shot.

\subsubsection{Second quantized representation}
\label{secquant}

To set up the stage for later we will now present the discussion
leading to Eq. (\ref{rhoQfq}) using the second-quantized formalism.
Wave function (\ref{evolwavefunc}) in the second quantization at $t=0$
can be written as
\begin{eqnarray}
|\varphi,N\rangle=\frac{1}{(2^N
N!)^{1/2}}(a_1^{\dagger}e^{i\varphi/2}+a_2^{\dagger}e^{-i\varphi/2})^N|0\rangle.
\label{phasestatesc}
\end{eqnarray}
Here $a_1^{\dagger}$ and $a_2^{\dagger}$ are creation operators  for
clouds one  and  two:
\begin{equation}
a_i^{\dagger}=\int d{\bf r} \psi_i({\bf r})\hat
\psi^{\dagger}({\bf r }).
\end{equation}
$\hat \psi({\bf r })$ is the second quantized operator for
the boson field, which satisfies the usual commutation relations
\begin{eqnarray}
\left[\hat \psi({\bf r' }),\hat \psi^{\dagger}({\bf r})\right]=\delta({\bf r'-r }),
\left[\hat \psi({\bf r' }),\hat \psi({\bf r})\right]=
\left[\hat \psi^{\dagger}({\bf r' }),\hat \psi^{\dagger}({\bf r})\right]=0.    \label{psicomm}
\end{eqnarray}
Operators $a_1, a_2$ and their conjugates satisfy the canonical boson
commutation relations:
\begin{eqnarray}
\left[a_i,a_j^{\dagger}\right]=\delta_{ij},\;
\left[a_i,a_j\right]=\left[a_i^{\dagger},a_j^{\dagger}\right]=0.
 \label{acomm}
\end{eqnarray}
Different initial states can be simply written using the Fock basis
of operators $a^{\dagger}_1$ and $a^{\dagger}_2.$ For example, the
initial state for two independent condensates with $N_1$ and $N_2$
particles in clouds $1$ and $2$ can be conveniently written as
\begin{eqnarray}
|N_1,N_2\rangle=\frac1{\sqrt{N_1!N_2!}}(a_1^{\dagger})^{N_1}(a_2^{\dagger})^{N_2}|0\rangle.
\label{indepinitial}
\end{eqnarray}
Initial state written in the Fock basis of operators $a^{\dagger}_1$
and $a^{\dagger}_2$ contains all information about properties
of the interference amplitudes. During the free expansion
the occupation numbers of states one and two do not change, and only
the single particle wave functions $\psi_1({\bf r},t)$ and
$\psi_2({\bf r},t)$ evolve. After the expansion, the many-body wave
function at time $t$ can be obtained from the initial state written
in the Fock basis of operators $a^{\dagger}_1$ and $a^{\dagger}_2$ using
substitutions
\begin{eqnarray}
a^{\dagger}_1 \rightarrow \int d{\bf r}\psi_1({\bf r},t)\hat
\psi^{\dagger}({\bf r}), \label{a1rule}\\
 a^{\dagger}_2 \rightarrow \int d{\bf r}\psi_2({\bf r},t)\hat
\psi^{\dagger}({\bf r}).\label{a2rule}
\end{eqnarray}
For example, substituting Eqs. (\ref{a1rule})-(\ref{a2rule}) into
Eq. (\ref{phasestatesc}), the wave function (\ref{evolwavefunc})
considered earlier is written as
\begin{eqnarray}
|\varphi,N,t\rangle=\frac{1}{(2^N N!)^{1/2}}\left(\int d{\bf r}
(\psi_1({\bf r},t)e^{i\varphi/2}+\psi_2({\bf
r},t)e^{-i\varphi/2})\hat \psi({\bf r})^{\dagger}\right)^N
|0\rangle. \label{phasestatesc2}
\end{eqnarray}

In the long time limit (\ref{ltl}) considered earlier,
single-particle wave functions $\psi_1({\bf r},t), \psi_2({\bf
r},t)$ can be written as
\begin{eqnarray}
\psi_1({\bf r},t)=u_1({\bf r},t)e^{i {\bf Q}_1 {\bf r}},\\
\psi_2({\bf r},t)=u_2({\bf r},t)e^{i {\bf Q}_2 {\bf r}},
\end{eqnarray}
where ${\bf Q}_1, {\bf Q}_2$ are defined by Eqs.
(\ref{Q1})-(\ref{Q2}) and $u_1({\bf r},t),u_2({\bf r},t)$ are
slowly varying real functions, which determine the overall
density profiles. Since clouds overlap strongly after the expansion, these functions
are normalized according to
\begin{eqnarray}
\int u_1({\bf r},t)^2 d{\bf r}=1,\; \int u_2({\bf r},t)^2 d{\bf
r}=1, \\
\int u_1({\bf r},t) u_2({\bf r},t)  d{\bf r}\approx 1.
\label{unorm}
\end{eqnarray}

The operator, which corresponds to the amplitude of density oscillation at
wave vector $\bf Q$ is written in the  second quantized notations as
\begin{eqnarray}
\hat \rho_{\bf Q}=\int d{\bf r} \hat \rho({\bf r}) e^{i  {\bf Q
r}}= \int d{\bf r} \hat \psi^{\dagger}({\bf r})\hat \psi({\bf r})
e^{i {\bf Q r}}. \label{rhoQdef}
\end{eqnarray}

To find out the {\it statistical} average  of the amplitude of
density oscillations for state (\ref{phasestatesc2}), we need to
evaluate the following matrix element:
\begin{eqnarray}
\langle \rho_{\bf Q}\rangle=\langle\varphi,N,t|\hat \rho_{\bf
Q}|\varphi,N,t\rangle=\langle\varphi,N,t|\int d{\bf r} \hat
\psi^{\dagger}({\bf r}) \hat \psi({\bf r}) e^{i {\bf Q
r}}|\varphi,N,t\rangle.\label{rhoQint}
\end{eqnarray}
To evaluate such matrix elements, first we need to know how
annihilation operator $\hat \psi({\bf r})$ acts on a state
$|\varphi,N,t\rangle.$ Since $|\varphi,N,t\rangle$ is obtained
from $|\varphi,N\rangle$ by substitutions
(\ref{a1rule})-(\ref{a2rule}), it is easy to see that
\begin{eqnarray}
\langle \varphi,N,t| \hat \psi^{\dagger}({\bf r})\hat \psi({\bf
r}) |\varphi,N,t\rangle=\langle \varphi,N|\left( a^{\dagger}_1
u_1({\bf r},t)e^{-i {\bf Q}_1 {\bf r}}+ a^{\dagger}_2 u_2({\bf
r},t)e^{-i {\bf Q}_2 {\bf r}}\right)\left( a_1 u_1({\bf r},t)e^{i
{\bf Q}_1 {\bf r}}+ a_2 u_2({\bf r},t)e^{i {\bf Q}_2 {\bf
r}}\right)|\varphi,N\rangle, \nonumber
\end{eqnarray}
where $|\varphi,N \rangle$ is defined in Eq. (\ref{phasestatesc})
and is written only in terms of $a^{\dagger}_1$ and $a^{\dagger}_2.$
Integration over $d\bf r$ in (\ref{rhoQint}) can be done using
normalization (\ref{unorm}), and assuming that  $u_i({\bf r},t)$
vary at scales much larger than $1/Q.$ Since ${\bf Q}={\bf Q}_1 -
{\bf Q}_2,$ evaluation of Fourier transform picks only the
product of the first term in the first parentheses and of the
second term in the second parentheses of the equation above.
 We obtain
\begin{eqnarray}
\langle \rho_{\bf Q}\rangle=\langle\varphi,N|a_1^{\dagger}a_2\int
u_1({\bf r},t) u_2({\bf r},t)  d{\bf r}|\varphi,N\rangle=\nonumber
\\ \langle\varphi,N|a_1^{\dagger}a_2|\varphi,N\rangle=
\langle\varphi,N-1|\left(\sqrt{\frac{N}2}e^{-i\varphi/2}\right)^2|\varphi,N-1\rangle=\frac{N}2e^{-i\varphi},
\label{rhoQsq}
\end{eqnarray}
which is the same as obtained from a single particle discussion in Eq.
(\ref{rhoQfq}).

The example above illustrates how the matrix elements of many-particle operators at time $t$ can be evaluated
using initial states written in the Fock basis of operators $a^{\dagger}_1$ and $a^{\dagger}_2.$ In general, when one needs to
evaluate an expectation value of some {\bf normal ordered} combination of operators $\hat \psi({\bf r}),$ and $\hat
\psi^{\dagger}({\bf r})$ over the final state at time $t,$ one needs to make
substitutions
\begin{eqnarray}
\hat \psi({\bf r})\rightarrow a_1 u_1({\bf r},t)e^{i {\bf Q}_1
{\bf r}}+ a_2 u_2({\bf r},t)e^{i {\bf Q}_2 {\bf r}},
\label{subst1} \\
\hat \psi^{\dagger}({\bf r})\rightarrow a^{\dagger}_1 u_1({\bf
r},t)e^{-i {\bf Q}_1 {\bf r}}+ a^{\dagger}_2 u_2({\bf r},t)e^{-i
{\bf Q}_2 {\bf r}}, \label{subst2}
\end{eqnarray}
and evaluate matrix elements over the $t=0$ state, written in the Fock basis of operators $a^{\dagger}_1$ and $a^{\dagger}_2.$
It is important that the expression needs to be normal ordered using commutation relations (\ref{psicomm}) before making
 substitutions (\ref{subst1})-(\ref{subst2}), since
 after substitutions (\ref{subst1})-(\ref{subst2}) fields $\hat \psi({\bf r}), \hat \psi^{\dagger}({\bf r})$ do not satisfy the exact
 commutation relations (\ref{psicomm}).

Another way of seeing this is to realize that in Heisenberg representation (see e.g. chapter 6 of Ref. \cite{AGD}) substitutions (\ref{subst1})-(\ref{subst2}) perform the time evolution
of a product of boson annihilation operators
\begin{eqnarray}
\hat \psi({\bf r}_1,t) ... \hat \psi({\bf r}_n,t)=e^{i\hat H t}\hat \psi({\bf r}_1) ... \hat \psi({\bf r}_n)e^{-i\hat H t}
\end{eqnarray}
only when this product acts
on states of the form
\begin{eqnarray}
|\Psi_{0}\rangle=\frac1{\sqrt{N_1!N_2!}}(a_1^{\dagger})^{N_1}(a_2^{\dagger})^{N_2}|0\rangle.
\end{eqnarray}
Thus for calculating the expectation values of the form
\begin{eqnarray}
\langle \Psi_{0}|e^{i\hat H t}\hat \rho({\bf r}_1)...\hat \rho({\bf r}_n)e^{-i\hat H t}|\Psi_{0}\rangle
\end{eqnarray}
one first needs to normal order $\hat \rho({\bf r}_1)...\hat \rho({\bf r}_n)$ using commutation relations (\ref{psicomm}) as
\begin{eqnarray}
\hat \rho({\bf r}_1)...\hat \rho({\bf r}_n)=\sum_{m \leq n}f_m({\bf r}_1,...,{\bf r}_n) \psi^{\dagger}({\bf r}_1)...\hat \psi^{\dagger}({\bf r}_m)\psi({\bf r}_m)...\hat \psi({\bf r}_1),
\end{eqnarray}
and only then use substitutions (\ref{subst1})-(\ref{subst2}) to evaluate matrix elements:
\begin{eqnarray}
\langle \Psi_{0}|e^{i\hat H t}\hat \rho({\bf r}_1)...\hat \rho({\bf r}_n)e^{-i\hat H t}|\Psi_{0}\rangle=\underset{m\leq n}{\sum}f_m({\bf r}_1,...,{\bf r}_n) \langle \Psi_{0}|e^{i\hat H t}\hat \psi^{\dagger}({\bf r}_1) ... \hat \psi^{\dagger}({\bf r}_m)e^{-i\hat H t}e^{i\hat H t}\hat \psi({\bf r}_m) ... \hat \psi({\bf r}_1)e^{-i\hat H t}|\Psi_{0}\rangle=\nonumber\\
\underset{m\leq n}{\sum}f_m({\bf r}_1,...,{\bf r}_n) \langle \Psi_{0}|
\left(a_1^{\dagger} u_1({\bf r}_1,t)e^{-i {\bf Q}_1
{\bf r}_1}+ a_2^{\dagger} u_2({\bf r}_1,t)e^{-i {\bf Q}_2 {\bf r}_1}\right) ...\left(a_1^{\dagger} u_1({\bf r}_m,t)e^{-i {\bf Q}_1
{\bf r}_m}+ a_2^{\dagger} u_2({\bf r}_m,t)e^{-i {\bf Q}_2 {\bf r}_m}\right)\times\nonumber \\
\left(a_1 u_1({\bf r}_m,t)e^{i {\bf Q}_1
{\bf r}_m}+ a_2 u_2({\bf r}_m,t)e^{i {\bf Q}_2 {\bf r}_m}\right) ...\left(a_1 u_1({\bf r}_1,t)e^{i {\bf Q}_1
{\bf r}_1}+ a_2 u_2({\bf r}_1,t)e^{-i {\bf Q}_2 {\bf r}_1}\right)|\Psi_{0}\rangle.
\end{eqnarray}

\subsection{Interference of independent clouds}
The surprising phenomenon which was observed in Ref.
\cite{andrews} is the appearance of interference fringes
in the case when condensates are completely independent. To
illustrate how interference fringes appear for independent clouds,
let us now discuss the case when the numbers of particles in each of the  clouds, $N_1$ and $N_2,$  are fixed, hence the phase difference between the two clouds is not well defined.  Initial state in the Fock basis
in this case is given by (\ref{indepinitial}):
\begin{eqnarray}
|N_1,N_2\rangle=\frac1{\sqrt{N_1!N_2!}}(a_1^{\dagger})^{N_1}(a_2^{\dagger})^{N_2}|0\rangle.
\end{eqnarray}
Using the formalism of second quantization explained earlier, one
can evaluate $\langle \rho_{\bf Q}\rangle$ by analogy to Eq.
(\ref{rhoQsq}):
\begin{eqnarray}
\langle \rho_{\bf Q}\rangle=\langle N_1,N_2|a_1^{\dagger}a_2
|N_1,N_2\rangle=0.
\end{eqnarray}
However, $\langle \rho_{\bf Q}\rangle=0$ doesn't imply that there are no interference
effects for independent condensates. Indeed, $\langle \rho_{\bf
Q}\rangle$ gives only the {\it statistical} average over many
experiments, according to the usual interpretation of expectation
values of operators in quantum mechanics. Being a quantum
operator, $\hat \rho_{\bf Q}$ has non vanishing quantum
fluctuations. In each particular realization of experiment,
complex number $\rho_{\bf Q}$ can have a nonzero value. To show
this, let us evaluate $\langle |\rho_{\bf Q}|^2\rangle,$ which is
the density-density correlation function at wave vector $\bf Q:$
\begin{eqnarray}
\langle |\rho_{\bf Q}|^2\rangle =\langle \hat  \rho_{\bf Q}\hat
\rho_{-\bf Q}\rangle= \langle \int d{\bf r}d{\bf r'} \hat
\psi^{\dagger}({\bf r})\hat \psi({\bf r})\hat \psi^{\dagger}({\bf
r'})\hat \psi({\bf r'}) e^{i {\bf Q (r-r')}}\rangle=\nonumber\\
\langle \int d{\bf r}d{\bf r'} \hat \psi^{\dagger}({\bf r})\hat
\psi^{\dagger}({\bf r'})\hat \psi({\bf r})\hat \psi({\bf r'})
e^{i {\bf Q (r-r')}}+\int d{\bf r} \psi^{\dagger}({\bf r})\hat
\psi({\bf r})\rangle. \label{rhoQ2def}
\end{eqnarray}
These matrix elements can be evaluated using the second
quantization prescription of the previous section, and the result
is
\begin{eqnarray}
\langle |\rho_{\bf Q}|^2\rangle =\langle
N_1,N_2|a_1^{\dagger}a^{\dagger}_2a_1a_2 +
a^{\dagger}_1a_1+a^{\dagger}_2a_2 |N_1,N_2\rangle=N_1 N_2 + N_1 +
N_2.
\end{eqnarray}
In the limit of large $N_1=N_2=N/2,$ the leading contribution to
$\langle |\rho_{\bf Q}|^2\rangle$ is the same as for the state
with the fixed phase. Information about the full distribution of
the quantum operator $\rho_{\bf Q}$ is contained in higher moments of
the distribution. If one considers higher moments of the type
$\langle |\rho_{\bf Q}|^{2n}\rangle=\langle \rho_{\bf
Q}^{n}\rho_{-\bf Q}^{n}\rangle$, the leading contribution in the
limit of large $N_1$ and $N_2$ will again have the form
\begin{eqnarray}
\langle |\rho_{\bf Q}|^{2n}\rangle=(N_1N_2)^n
\left(1+O(\frac1{N_1})+O(\frac1{N_2})\right). \label{rhoQ2n}
\end{eqnarray}
Corrections which appear because of the normal ordering result in subleading
terms which are denoted by $O(1/N_1)+O(1/N_2).$ The leading term implies that in the limit of large $N_1$ and $N_2,$ the
distribution function of $|\rho_{\bf Q}|^{2}$ is highly peaked near the value $N_1 N_2,$
with the relative width which is proportional to the inverse square root of number of particles. Any operator of the form $\rho_{\bf
Q}^{n}\rho_{-\bf Q}^{m}$ will have zero expectation value for
$m\neq n$ similar to $\rho_{\bf Q},$ which means that the phase
of the complex number $\rho_{\bf Q}$ is uniformly distributed
from $0$ to $2\pi.$ The expectation value of any operator which
depends on the phase of $\rho_{\bf Q}$ becomes zero due to
the averaging over the phase.

The physical picture which emerges from the calculations  is
the following \cite{Pethick,Javanainen,Castin}: for two
independent ideal clouds in the limit of large $N$ the  absolute
value of interference fringe amplitude is the same as for the
state with a fixed relative phase, but the position of the intensity
minima fluctuates from shot to shot. The state with a fixed
number of particles is a superposition of states with fixed
relative phases. For example,
\begin{eqnarray}
|N/2,N/2\rangle=\left(\frac{\pi N}{2}\right)^{1/4}\int_0^{2\pi}
\frac{d\varphi}{2\pi}|\varphi,N\rangle.
\end{eqnarray}
In the limit of large $N$ the phase states are almost orthogonal,
and the measurement picks some value of the relative phase. Since
the relative phase is not well defined for independent clouds, in
each particular experiment the positions of the minima will
fluctuate from shot to shot. To distinguish independent clouds
from states which have correlated relative phases, one needs to
do a series of experiments and measure not only the absolute magnitude of interference
fringes, but also the positions of the minima. Experiments which
distinguish states with a fixed relative phase from some other
many body states are  already being done, and can be used i.e. to
measure the temperature \cite{Gati} or to study the dynamical
evolution \cite{phasedyn, Schumm, Hofferberth, MIT_squeezing,Jo2}
 of the relative phase.

\section{Full counting statistics of shot noise}

\label{Snoise} As has been explained in the previous section, for
experiments with  independent clouds  the average interference amplitude depends
only on the number of particles per cloud.
In this section we consider not only the average interference
amplitude, but also its shot to shot fluctuations  due to a finite
number of atoms in the clouds. We will demonstrate that  while the
average value of $|\rho_{\bf Q}|^2$ depends only on the number of
particles per cloud, the full distribution function of the variable
\begin{eqnarray}
R=|\rho_{\bf Q}|^2
\label{Eq38}
\end{eqnarray}
contains information about the states of individual clouds. Our
analysis is motivated by the earlier work of Polkovnikov
\cite{shotnoise}, who showed   that the variance of the fringe
amplitude decreases as the inverse power of the number of particles
per cloud, $N,$ with a non-universal coefficient which contains
information about the state of the clouds. Experimental observation
of the effects discussed in this section requires systems with a
small number of atoms. This may be realized with micro-BEC's on
chips \cite{phasedyn, Shin2,Schumm, Hofferberth,
MIT_squeezing,Wang,Jo2,Chip_review}.


The shot noise for finite $N$ has a fundamental nature, which
stems from the probabilistic nature of quantum mechanics.
Distributions of $R$ obtained below correspond to the following
"idealized" experimental procedure: release the confining
potential and take an absorption image of the columnar density on an ideal CCD camera with $100\%$ efficiency (photon shot noise is ignored). To obtain the amplitude of interference fringes, $\rho_{\bf Q},$ extract a Fourier
component of the density at wave vector $\bf Q$ from each image separately.
The results of many experiments give the histogram $W(R)$ of the values
of $R=|\rho_{\bf Q}|^2.$
We note that the quantum observable $\hat R$  defined in such way is
a {\it many-body} operator, and calculation of its full distribution
is a non-trivial task, even when all atoms are in the same state,
such as for the case of a well defined  relative phase between atoms
in the two wells. In this section we develop a general method to
find distribution functions of $R$ analytically.

We note that in our idealized setting we find interference patterns
at a well defined wave vector ${\bf Q}.$ We expect that the finite
size of the systems in transverse direction after expansion and
collisions during the initial stage of expansion broaden the peak in
the Fourier space to a finite, but small range of wave vectors
around ${\bf Q}$ \cite{Bloch_reply}. Hence a one should consider
$R=|\rho_{\bf Q}|^2$ as an integral over the peak in the Fourier
image of the density. Also in this paper we will discuss the
amplitude of the interference fringes whereas experimental papers
typically discuss visibility of the interference patterns. The two
quantities differ only by the trivial rescaling.

\begin{figure}
\psfig{file=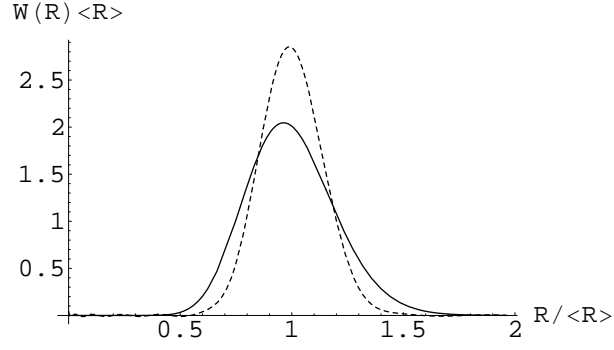} \caption{\label{shotnoisefig2} Rescaled distribution functions of  $R=|\rho_{\bf Q}|^2$ for independent
clouds in the coherent states (solid) or in the states with a well defined
numbers of atoms (dashed). Here $N_1=N_2=100.$ }
\end{figure}

\begin{figure}
\psfig{file=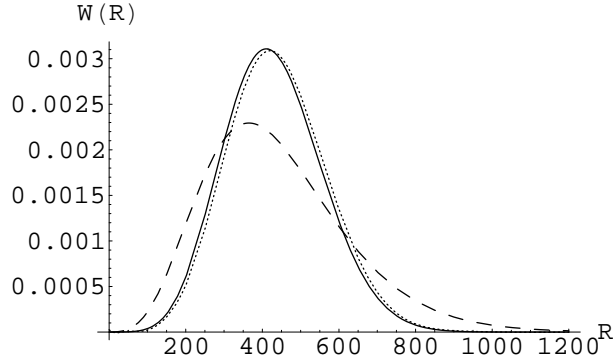} \caption{\label{shotnoisefig} Distribution functions of  $R=|\rho_{\bf Q}|^2$ for independent
clouds in the coherent (dashed) and in the number
(dotted) states with $N_1=N_2=20.$ Solid line is a distribution function of  $R$ for clouds with a fixed relative phase  with total number $N=N_1 + N_2=40.$ Distribution functions for states with a fixed relative phase and with fixed numbers are very close, and become indistinguishable when $R$ is normalized by its
average value $\langle R \rangle.$ }
\end{figure}

Results for independent clouds in coherent (solid) and number
(dashed) states for $N_1=N_2=100$ are presented in Fig.
\ref{shotnoisefig2}. One can see, that even for a relatively large
number of atoms, $N=100,$ fluctuations due to shot noise are
appreciable . In Fig. \ref{shotnoisefig} we compare the full
counting statistics of  $R=|\rho_{\bf Q}|^2$ for independent clouds
in coherent (dashed) and  number (dotted) states with $N_1=N_2=20,$
and for clouds with a fixed relative phase (solid) with total number
of atoms $N=N_1 + N_2=40.$ Distribution functions for the cases of
(i) well defined relative phase between the clouds and (ii) fixed
number of atoms in each cloud  are very close. They become
indistinguishable when $R$ is rescaled by its average value $\langle
R \rangle,$ although each of them differs considerably from the
Gaussian distribution. The distribution function of $R$ is wider for
coherent states compared to number states, as was suggested in Ref.
\cite{shotnoise} based on the study of the variance of the two
distributions. Hence the conclusion is that coherent and number
states can be easily distinguished based on the statistics of
fluctuations of $R$ relative to its average value.

In principle, when the two clouds are prepared with the same relative
phase over many experiments, it is possible to distinguish
independent condensates in number states from states with a fixed
relative phase using a set of several interference experiments: minima positions are uniformly distributed for independent clouds, while for states with  a fixed relative phase  positions  of the interference minima are always at the same points in space. However one can imagine the situation when the clouds
are prepared in  state with a fixed relative phase but the relative
phase itself is random from realization to realization. Our results show that it is practically impossible to distinguish such states from number
states by looking at the distribution of the amplitude of
interference fringes. Our calculations below provide additional
support to the physical interpretation
\cite{Pethick,Javanainen,Zoller,Castin} presented in the previous
section. Fluctuations of the absolute value of the interference
amplitude are the same for the cases when clouds have random relative phase and when clouds are prepared in number states so the random relative phase is "measurement induced".


The method developed in this section can be generalized to a variety
of experimental situations, i.e. several independent condensates.
Different squeezed states within individual condensates can be
considered, and measurement of full counting statistics of shot
noise can be used as an experimental probe to
distinguish between different correlated states.


As noted earlier, we will be interested in the full distribution function
of the positive definite quantum observable
\begin{eqnarray}
\hat R=|\hat \rho_{\bf Q}|^2=\hat \rho_{\bf Q}\hat \rho_{-\bf Q},
\end{eqnarray}
defined by Eq. (\ref{rhoQ2def}). To calculate its full
distribution function, $W(R),$ one needs to know expressions for
higher moments $\langle \hat R^n \rangle.$ After that, one has to
solve the "problem of moments", i.e. to recover the distribution
function on the $(0,\infty)$ interval using all moments. In general,
this procedure is numerically hard and unstable, unless higher
moments have a certain analytical form. If the expression for higher
moments is known analytically, then one can sometimes avoid the
"problem of moments" by calculating the so-called characteristic function, $\chi(\lambda),$  which is the Laplace transform of
$W(R):$
\begin{eqnarray}
\chi(\lambda)=\int_0^{\infty} e^{-\lambda R}W(R)
dR=\int_0^{\infty}\sum^{\infty}_{i=0} \frac{(-\lambda
R)^n}{n!}W(R)dR=\sum^{\infty}_{i=0}\frac{(-\lambda)^n \langle \hat
R^n\rangle }{n!}
\end{eqnarray}
If $\chi(\lambda)$ can be calculated analytically, then $W(R)$ can
be recovered by the inverse Laplace transform. In our case it is
more practical to calculate not the characteristic function, but the
analog of the Hankel transformation \cite{hankel,korn} of $W(R),$
given by
\begin{eqnarray}
Z(i\lambda)=\sum_{n=0}^{n=\infty}
\frac{(i\lambda)^{2n}}{(n!)^2}\langle \hat R^n\rangle.
\label{Zilamdef}
\end{eqnarray}
Using the expansion of the zeroth order Bessel function, one can write
\begin{eqnarray}
Z(i\lambda)=\int_0^{\infty} W(R) J_0(2 \lambda \sqrt{R}) dR.\\
\end{eqnarray}
The inversion of the transformation can be found using the orthogonality condition for the zeroth order Bessel  functions
$\int_0^\infty J_0(\lambda x) J_0(\lambda y) |x|\lambda
d\lambda=\delta(|x|-|y|),$ which gives
\begin{eqnarray}
W(R)=2\int_0^{\infty} Z(i\lambda)J_0(2 \lambda \sqrt{R}) \lambda
d\lambda.\label{wA}
\end{eqnarray}
By the end of this section we will provide analytical expressions
for $Z(i\lambda)$ for certain cases (see Eqs. (\ref{cohfinal2}), (\ref{fockfinal}), (\ref{gegenbauer})), from which $W(R)$ can be
obtained by simple numerical integration according to Eq. (\ref{wA}).

To proceed we note that $\hat \rho_{\bf Q}$ and $\hat \rho_{\bf -Q}$  commute with each other:
\begin{eqnarray}
[\hat \rho_{\bf Q},\hat \rho_{\bf -Q}]=0.
\end{eqnarray}
Operators $\hat \rho_{\bf Q}$ are understood as in Eq. (\ref{rhoQdef}),
without the projection on single particle states $\psi_1({\bf r},t), \psi_2({\bf r},t)$ as in Eqs. (\ref{subst1})-(\ref{subst2}). The latter
substitutions can be only done after normal ordering (see discussion in section \ref{secquant}). Hence we find
\begin{eqnarray}
Z(i\lambda)=\langle \int_0^{2\pi}\frac{d
\varphi}{2\pi}e^{i\lambda\left(\hat \rho_{\bf Q}e^{i\varphi}+\hat
\rho_{\bf -Q}e^{-i\varphi}\right)}\rangle \label{zil}.
\end{eqnarray}
Indeed, after expanding the exponent and integration,
only even degrees of $i \lambda$ survive, and non vanishing  terms are
exactly what is needed for Eq. (\ref{Zilamdef}).

The normal ordering of Eq. (\ref{zil}) can be done using the following identity:
\begin{eqnarray}
e^{  \int  f({\bf r}) \hat \psi^{\dagger}({\bf r})\hat \psi({\bf r})d{\bf r}}=
:e^{\int(e^{f({\bf r})}-1)\hat \psi^{\dagger}({\bf r})\hat \psi({\bf r})d{\bf r}}:. \label{normalordering}
\end{eqnarray}
Here we have assumed that operators $\psi^{\dagger}({\bf r}), \psi({\bf r}),$ have the canonical commutation relations given by Eq.(\ref{psicomm}). Eq. (\ref{normalordering}) is a generalization of the simpler identity \cite{Blasiak}:
\begin{eqnarray}
e^{\lambda a^{\dagger} a}=:e^{(e^{\lambda}-1)a^{\dagger}a}: \label{aordering}
\end{eqnarray}
for operators which obey the commutation relations
\begin{eqnarray}
[a,a^{\dagger}]=1.
\end{eqnarray}
The normal ordering signs $:\;\;:$ mean that all creation operators should be put to the left of annihilation operators in  Taylor expansion of expressions being ordered.  To illustrate the meaning of Eq. (\ref{aordering}), let us consider the expansions of left and right side up to $\lambda^2.$ The left hand side is
\begin{eqnarray}
1+\lambda a^{\dagger} a+\frac{\lambda^2}2a^{\dagger} aa^{\dagger} a+O({\lambda}^3)=
1+(\lambda + \frac{\lambda^2}2)a^{\dagger} a+\frac{\lambda^2}2a^{\dagger} a^{\dagger}a a+O({\lambda}^3).
\end{eqnarray}
The right hand side is
\begin{eqnarray}
1+(e^\lambda-1): a^{\dagger} a:+\frac{(e^\lambda-1)^2}2
:a^{\dagger} aa^{\dagger} a:+ O({\lambda}^3)=\nonumber \\
1+(\lambda + \frac{\lambda^2}2+ O({\lambda}^3)):a^{\dagger}
a:+\frac{\lambda^2+O({\lambda}^3)}2:a^{\dagger}a
a^{\dagger}a:+O({\lambda}^3)=\nonumber\\ 1+(\lambda +
\frac{\lambda^2}2)a^{\dagger} a+\frac{\lambda^2}2a^{\dagger}
a^{\dagger}a a+ O({\lambda}^3).\nonumber
\end{eqnarray}
Eq. (\ref{aordering}) holds not only up to $\lambda^2,$ but to all orders in $\lambda$ and plays an important role in quantum optics.

Using the definition of $\rho_{\bf Q}$ given by Eq. (\ref{rhoQdef}),
 one can apply Eq. (\ref{normalordering}) with $f({\bf r})=2 i \lambda \cos{({\bf Q r} +\varphi)}$
and rewrite $Z(i\lambda)$ in Eq. (\ref{zil}) as

\begin{eqnarray}
Z(i\lambda)=\langle : \int_0^{2\pi}\frac{d \varphi}{2\pi}e^{\int(e^{2 i \lambda \cos{({\bf Q r} +\varphi)}}-1)\hat \psi^{\dagger}({\bf r})\hat \psi({\bf r})d{\bf r}}:\rangle.
\label{zilordered}
\end{eqnarray}

\subsection{Interference of two independent coherent condensates}
Let us first explain how to evaluate $Z(i\lambda)$ for independent
clouds in coherent states \cite{Glauber63,OrlandNegele} of operators $a_1$
and $a_2$ with eigenvalues $\sqrt N_1 e^{i\psi_1}$ and $\sqrt N_2
e^{i\psi_2}.$ Since coherent states  form a complete basis, any
initial state can be expanded in this basis, and thus the problem
of calculating of $W(R)$ is essentially solved for arbitrary initial
states. Coherent states are convenient, since they are the eigen
states of the annihilation operator, and the annihilation operator acts on
them as a $c-$number. Hence after making substitutions
(\ref{subst1})-(\ref{subst2}) into the normal ordered expression, one
can substitute operators $a_i, a^{\dagger}_i$ by numbers
$\sqrt N_i e^{i\psi_i},\sqrt  N_i e^{-i\psi_i}.$ Since the normal
ordered expression is obtained by the normal ordering of the Taylor
expansion of Eq. (\ref{zilordered}), one needs to collect the Taylor
series back. For coherent states, the whole procedure is
equivalent to removing the normal ordering signs in Eq.
(\ref{zilordered}), making substitutions
(\ref{subst1})-(\ref{subst2}) and treating operators $a_i,
a^{\dagger}_i$ as numbers $\sqrt N_i e^{i\psi_i},\sqrt  N_i
e^{-i\psi_i}.$ Thus we obtain
\begin{eqnarray}
Z(i\lambda; \sqrt  N_1 e^{i\psi_1},\sqrt N_2 e^{i\psi_2})=
 \int_0^{2\pi}\frac{d \varphi}{2\pi}e^{
 \int\left(e^{2 i \lambda \cos{({\bf Q r} +\varphi)}}-1\right)
\left(N_1u_1({\bf r},t)^2 +N_2u_2({\bf r},t)^2 +2 \sqrt{N_1 N_2}
\cos{(\psi_1-\psi_2+ {\bf Q r})u_1({\bf r},t)u_2({\bf r},t)}
\right)  d{\bf r} }.\label{zilcoh}
\end{eqnarray}
Similar to section \ref{secquant} we assume the that normalized functions $u_1({\bf r},t)$ and  $u_2({\bf r},t)$ strongly overlap
and vary at scales much larger than $1/Q,$ which is equivalent to
\begin{eqnarray}
\int e^{i n\bf Q r}u_\alpha({\bf r},t)u_\beta({\bf r},t) d{\bf r}=\delta_{n0}.
\end{eqnarray}
Then integration over $d \bf r$ in the exponent of Eq. (\ref{zilcoh}) can be done using the following equations:
\begin{eqnarray}
\int\left(e^{2 i \lambda \cos{({\bf Q r} +\varphi)}}-1\right)(N_1 u_1({\bf r},t)^2 +N_2u_1({\bf r},t)^2) d{\bf r}=\nonumber\\
\sum_{m=1}^{m=\infty}\frac{(i\lambda)^{2m}}{(2m)!}\int (N_1 u_1({\bf r},t)^2 +N_2u_1({\bf r},t)^2)(e^{i ({\bf Q r} +\varphi)}+e^{-i ({\bf Q r} +\varphi)})^{2m}d{\bf r} =\nonumber\\
\sum_{m=1}^{m=\infty}(N_1 +N_2)\frac{(i\lambda)^{2m}}{(2m)!} \frac{(2m)!}{m!m!}=(J_0(2\lambda)-1)(N_1 +N_2);\label{j0}\\
2\sqrt{N_1 N_2}\int\left(e^{2 i \lambda \cos{({\bf Q r} +\varphi)}}-1\right) \cos{(\psi_1-\psi_2 +{\bf Q r})}u_1({\bf r},t)u_2({\bf r},t)d{\bf r}=\nonumber \\
\sqrt{N_1 N_2}\sum_{m=0}^{m=\infty}\frac{(i\lambda)^{2m+1}}{(2m+1)!}\int(e^{i ({\bf Q r} +\varphi)}+e^{-i ({\bf Q r} +\varphi)})^{2m+1}\left(e^{i (\psi_1-\psi_2+ {\bf Q r})}+e^{-i (\psi_1-\psi_2+ {\bf Q r})}\right) u_1({\bf r},t)u_2({\bf r},t)d{\bf r}=\nonumber \\
\sqrt{N_1 N_2}\sum_{m=0}^{m=\infty}\frac{(i\lambda)^{2m+1}}{(2m+1)!}\frac{(2m+1)!}{m!(m+1)!}
\left(e^{i (\psi_1-\psi_2 -\varphi)}+e^{-i (\psi_1-\psi_2 -\varphi)}\right)=2 i \sqrt{N_1 N_2}J_1(2\lambda) \cos{(\psi_1-\psi_2 -\varphi)}.\label{j1}
\end{eqnarray}

Substituting (\ref{j0}) and (\ref{j1}) into (\ref{zilcoh}), and
doing the integral  over $\varphi,$ we finally obtain the central
result of this section:
\begin{eqnarray}
Z(i\lambda;\sqrt  N_1 e^{i\psi_1},\sqrt N_2
e^{i\psi_2})=\int_0^{2\pi}\frac{d
\varphi}{2\pi}e^{(J_0(2\lambda)-1)(N_1 +N_2)+2 i \sqrt{N_1 N_2}
J_1(2\lambda) \cos{(\psi_1-\psi_2 -\varphi)}}=\label{cohfinal1}\\
e^{(J_0(2\lambda)-1)(N_1 +N_2)} J_0\left(2 \sqrt{N_1
N_2}J_1(2\lambda)\right).\label{cohfinal2}
\end{eqnarray}

\subsection{Interference of independent clouds in number states}

Let us now explain how to calculate $Z^f(i\lambda, N_1, N_2)$ for
the Fock states with the number of particles equal to $N_1$ and $N_2.$
First, we need to expand the Fock states $|N_1,N_2\rangle$ using
the coherent states basis. Since the basis of coherent states is
overcomplete \cite{OrlandNegele}, there are many ways to do a
decomposition. For our purposes it is convenient to use
\begin{eqnarray}
|N_1,N_2\rangle=\frac{(a_1^\dagger)^{N_1}}{\sqrt{N_1!}}\frac{(a_2^\dagger)^{N_2}}{\sqrt{N_2!}}|0\rangle=
\sqrt{N_1!
N_2!}\alpha^{-N_1-N_2}e^{\alpha^2}\int_0^{2\pi}\int_0^{2\pi}\frac{d\varphi_{1R}d\varphi_{2R}}{(2\pi)^2}
e^{-i N_1 \varphi_{1R}-i N_2 \varphi_{2R}} |\alpha
e^{i\varphi_{1R}},\alpha e^{i\varphi_{2R}}\rangle, \label{ketexp}
\end{eqnarray}
where $\alpha$ is an arbitrary real positive number. Coherent
states are given by
\begin{eqnarray}
|\alpha e^{i\varphi_{1R}},\alpha
e^{i\varphi_{2R}}\rangle=e^{-\alpha^2+\alpha e^{i \varphi_{1R}}
a_1^\dagger+\alpha e^{i \varphi_{2R}} a_2^\dagger}|0\rangle, \label{cohdef}
\end{eqnarray}
and the overlap between them equals
\begin{equation}
\langle \alpha e^{i\varphi_{1L}},\alpha e^{i\varphi_{2L}}|\alpha
e^{i\varphi_{1R}},\alpha
e^{i\varphi_{2R}}\rangle=e^{\alpha^2(e^{i\varphi_{1R}-i\varphi_{1L}}+e^{i\varphi_{2R}-i\varphi_{2L}}-2)}.\label{cohoverlap}
\end{equation}
 One can also expand the bra
state $\langle N_1,N_2|$ similarly to (\ref{ketexp}) by
introducing integration variables $\varphi_{1L},\varphi_{2L}.$ For
any given values of
$\varphi,\varphi_{1L},\varphi_{2L},\varphi_{1R},$ and $\varphi_{2R},$
matrix elements between coherent states can be evaluated as in
previous section, and by simple modification of Eq.
(\ref{cohfinal1}) we obtain
\begin{eqnarray}
Z^f(i\lambda, N_1, N_2)=\int\int\int\int\int\frac{d\varphi
d\varphi_{1L} d\varphi_{2L} d\varphi_{1R} d\varphi_{2R}}{(2\pi)^5}
N_1! N_2!\alpha^{-2N_1-2N_2}e^{2\alpha^2} e^{i N_1
(\varphi_{1L}-\varphi_{1R}) + i N_2 (\varphi_{2L}-\varphi_{2R})}
\times \nonumber\\
 \langle \alpha e^{i\varphi_{1L}},\alpha
e^{i\varphi_{2L}}|\alpha e^{i\varphi_{1R}},\alpha
e^{i\varphi_{2R}}\rangle\times \nonumber\\
e^{(J_0(2\lambda)-1)\alpha^2(e^{i\varphi_{1R}-i\varphi_{1L}}
+e^{i\varphi_{2R}-i\varphi_{2L}})}e^{i \alpha^2 J_1(2\lambda)
\left(e^{i(\varphi-\varphi_{1L}+\varphi_{2R})} +
e^{-i(\varphi-\varphi_{1R}+\varphi_{2L})}\right)}.\label{Fockbig}
\end{eqnarray}
One can now substitute Eq.(\ref{cohoverlap}) into the equation above,
introduce variables
\begin{eqnarray}
\varphi_1=\varphi_{1L}- \varphi_{1R} \; \mbox{and} \;
\varphi_2=\varphi_{2L}- \varphi_{2R},
\end{eqnarray}
and integrate over $\varphi,  \varphi_{1R}, \varphi_{2R}.$
Multiple cancellations occur, and eventually we obtain
\begin{eqnarray}
Z^f(i\lambda, N_1, N_2)=\int\int\frac{d\varphi_{1} d\varphi_{2}
}{(2\pi)^2}N_1! N_2!\alpha^{-2N_1-2N_2}e^{i N_1 \varphi_{1} + i
N_2 \varphi_{2}} e^{J_0(2\lambda)\alpha^2(e^{-i\varphi_1}
+e^{-i\varphi_2})} J_0\left(2 \alpha^2 J_1(2\lambda)
e^{-i(\varphi_1+\varphi_2)/2} \right).
\end{eqnarray}
Both integrations in the equation above can be done in a closed
form for arbitrary positive integer $N_1$ and $N_2$ using
hypergeometric functions.  Here we will present the results only
for $N_1=N_2=N.$ One needs to expand the last exponent and $J_0(
2\alpha^2 J_1(2\lambda) e^{-i(\varphi_1+\varphi_2)/2})$ using
Taylor series. After integration over $d\varphi_1$ and
$d\varphi_2$ dependence on $\alpha$ disappears, and we obtain the
final result for Fock states:
\begin{eqnarray}
Z^f(i\lambda, N, N)=\sum_{k=0}^{N}(N!)^2
\frac{J_0(2\lambda)^{2k}}{k!^2}\frac{\left(i
J_1(2\lambda)\right)^{2(N-k)}}{(N-k)!^2}=_2F_1\left(-n,-n;1;-\frac{J_0(2\lambda)^{2}}{J_1(2\lambda)^{2}}\right)
\left(i J_1(2\lambda)\right)^{2N}, \label{fockfinal}
\end{eqnarray}
where $_2F_1\left(a,b;c;x\right)$ in a hypergeometric function
defined by
\begin{equation}
_2F_1\left(a,b;c;x\right)=1+\frac{ab}{c}\frac{x}{1!}+\frac{a(a+1)b(b+1)}{c(c+1)}\frac{x^2}{2!}+...
\end{equation}

\subsection{Clouds with a well defined relative phase}

Let us now consider the case of clouds with a fixed relative phase,
when the initial state $|\varphi_0,N\rangle$ at $t=0$ is given by
Eq. (\ref{phasestatesc}). This state can be expanded using the  coherent states basis as
\begin{eqnarray}
|\varphi_0,N\rangle=\frac{1}{(2^N
N!)^{1/2}}(a_1^{\dagger}e^{i\varphi_0/2}+a_2^{\dagger}e^{-i\varphi_0/2})^N|0\rangle=
\sqrt{N!}(\sqrt{2}\alpha)^{-N} e^{\alpha^2}\int_0^{2\pi}\frac{d\varphi_{R}}{2\pi}
e^{-i N \varphi_{R}} |\alpha
e^{i\varphi_{R}+ i\varphi_{0}/2},\alpha e^{i\varphi_{R}- i\varphi_{0}/2}\rangle.
 \nonumber
\end{eqnarray}
A similar expansion can be written for bra- vector $\langle \varphi_0,N|$ using the phase variable $\varphi_{L}.$ Coherent states and their overlaps are given by Eqs. (\ref{cohdef})-(\ref{cohoverlap}), and one obtains an expression for the generating function
$Z(i\lambda,N)$ similar to Eq. (\ref{Fockbig}):
\begin{eqnarray}
Z(i\lambda,N)=\int\int\int\frac{d\varphi
d\varphi_{L} d\varphi_{R}}{(2\pi)^3}
N!(\sqrt2\alpha)^{-2N} e^{2\alpha^2} e^{i N
(\varphi_{L}-\varphi_{R})}
\langle \alpha e^{i\varphi_{L}+i\frac{\varphi_{0}}{2}},\alpha
e^{i\varphi_{L}-i\frac{\varphi_{0}}{2}}|\alpha e^{i\varphi_{R}+i\frac{\varphi_{0}}{2}},\alpha
e^{i\varphi_{R}-i\frac{\varphi_{0}}{2}}\rangle \nonumber\\
\times
e^{2(J_0(2\lambda)-1)\alpha^2e^{i\varphi_{R}-i\varphi_{L}}
}e^{i \alpha^2 J_1(2\lambda)
\left(e^{i(\varphi-\varphi_0+\varphi_{R}-\varphi_{L})} +
e^{-i(\varphi-\varphi_0+\varphi_{L}-\varphi_{R})}\right)}.\nonumber
\end{eqnarray}
The integrand depends only on the difference $\Delta \varphi=\varphi_{L}-\varphi_{R},$ and the integral over $\varphi$ can be done analytically. Dependence of $Z(i\lambda,N)$
on $\varphi_0$ drops out, as expected:
\begin{eqnarray}
Z(i\lambda,N)=\int\int\frac{d\Delta\varphi
}{2\pi}N! \alpha^{-2N} 2^{-N}e^{i N \Delta\varphi} e^{2 J_0(2\lambda)\alpha^2e^{-i\Delta\varphi}} J_0\left(2 \alpha^2 J_1(2\lambda)
e^{-i\Delta\varphi} \right).
\end{eqnarray}

 Expanding the last exponent and $J_0\left(2 \alpha^2 J_1(2\lambda)
e^{-i\Delta\varphi} \right)$ in the expression above and integrating over $\Delta \varphi$, we obtain the final expression for even $N:$
\begin{eqnarray}
Z(i\lambda,N)=N!2^{-N}\sum_{k=0}^{N/2}\frac{\left(i J_1(2\lambda)\right)^{2k}}{(k!)^2}
\frac{\left(2 J_0(2\lambda)\right)^{N-2k}}{(N-2k)!}=2^{-N}\left(i J_1(2\lambda)\right)^N C_N^{-N}\left(\frac{J_0(2\lambda)}{iJ_1(2\lambda)}\right), \label{gegenbauer}
\end{eqnarray}
where $C^{\alpha}_n(x)$ is a  Gegenbauer polynomial \cite{korn}.

\section{Interference of fluctuating low-dimensional gases}
\label{lowdim}

\begin{figure}
\psfig{file=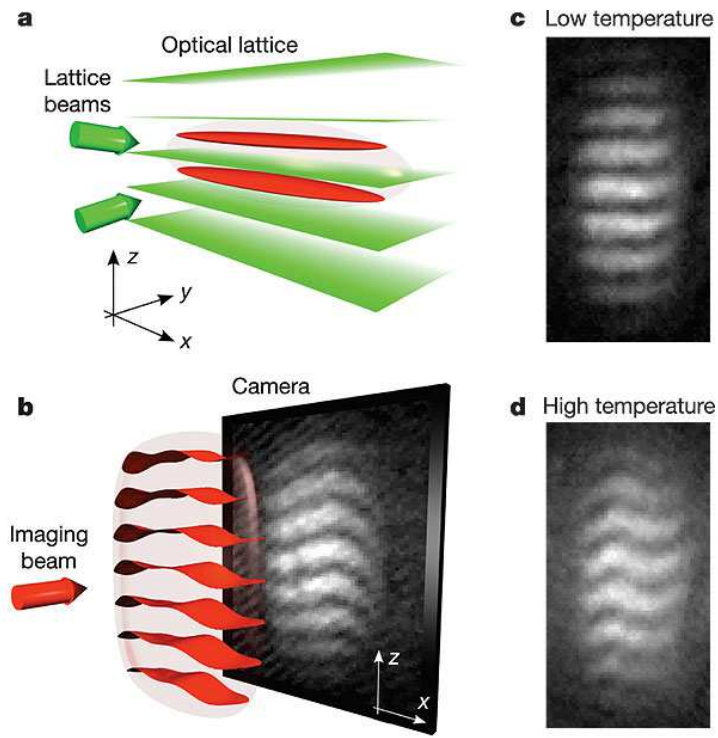} \caption{\label{setup_2D} Experimental setup
for interference of 2D gases. Note that the interference patterns are straight at low temperatures  indicating suppressed phase fluctuations. Meandering patterns at high temperatures come from strong phase fluctuations. Taken from Ref. \cite{zoran_KT}.}
\end{figure}

As we discussed in previous sections, for macroscopic
three dimensional Bose-Einstein Condensates the long-range phase
coherence manifests itself in the nearly perfect interference fringes
between two independent condensates \cite{andrews}. For low
dimensional Bose gases, the situation is different, since phase
fluctuations are very effective in destroying the long-range order.
In one dimension, long-range coherence is prohibited even at zero
temperature \cite{Coleman}, while in two dimensions any nonzero
temperature destroys long-range order \cite{Hohenberg}. In
addition,  the Berezinskii-Kosterlitz-Thouless (BKT) phase transition
occurs \cite{Berezinskii,KT}, which separates the low temperature phase
with power-law correlations from the high temperature phase with short-range correlations. Phase
fluctuations reduce the average visibility of the interference
fringes, and result in the shot to shot fluctuations of the visibility.

In this section, we discuss how measurements of interference
fringes can reveal information about spatial correlations within
individual condensates. The typical experimental setups are shown in
Figs. \ref{setup_1D} and \ref{setup_2D}. They correspond to the so-called
open boundary conditions (OBC). Essentially the OBC mean that the imaged area is cut out of a larger system. As a theoretical model one can also
consider a one-dimensional condensate with periodic boundary
conditions (PBC), which corresponds to interference experiments
with two coaxial rings lying in two parallel $xy$ planes. While this model is somewhat artificial from the experimental point of view (see however Ref. \cite{Stamper-Kurn_toroid}), it allows a very elegant theoretical analysis, hence we will discuss it in these lecture notes as well.

The confining potential is highly anisotropic, and after it is switched
off, the  clouds predominantly expand  in the transverse direction, while
no significant expansion occurs in the axial (for 1D gases) or in-plane
(for 2D gases) directions. For low-dimensional gases the phase of
the condensate doesn't have a long-range order due to quantum or
thermal fluctuations. Locally the phase determines the positions of
the minima of the absorption intensity, and fluctuations of the
phase lead to fluctuations of the interference fringe positions
along the condensates, as shown in Fig. \ref{setup_1D}. Fluctuations
of the fringe positions contain information about the original phase
fluctuations present in the system, which are preserved during
expansion.

To extract information about fringe position fluctuations for the 1D
case, we will integrate the intensity along the axes of the clouds.
 Fluctuations of the relative phase result in fluctuations of the
minima positions for different $x.$ For each $y,$ the image can be
integrated along the $x$ direction to give the integrated fringe
amplitude $\rho_{\bf Q}(L)$ (see Fig. \ref{setup_1D}).  Note that the integrated
fringe amplitude depends on the integration length $L.$ One experimental image can
be used to extract information for different values of $L.$ Many
images are still required to obtain distribution functions for each
$L.$ For 2D gases, the setup is analogous and is shown in Fig
\ref{setup_2D}. Here part of the integration is performed by the
imaging beam itself. The size of the integration area along the
direction of the imaging beam can be controlled by applying magnetic
field gradients, so that only a specified section of the cloud is
resonant with the probe light.

The operator which corresponds to the fringe amplitude $\rho_{\bf Q}(L),$  illustrated in
Fig. \ref{setup_1D}, is the same as $\rho_{\bf Q}$ defined by Eq.
(\ref{rhoQdef}), where the integration along $x-$ dimension is
limited to the section of length $L.$ Let us first consider the
expectation value $\langle |\rho_{\bf Q}(L)|^2\rangle$ (expectation values of
operators which depend on the phase of $\rho_{\bf Q}(L)$ vanish, similar to 3D
case, since two clouds are assumed to be independent). One has to
use modified formulas (\ref{subst1})-(\ref{subst2}), where operators
$a^\dagger_i, a_i $ are now allowed to have $x-$dependence. In the
limit when the number of particles in the section of size $L$ is large,
the average value of $\langle |\rho_{\bf Q}(L)|^2\rangle$ is given
 by \cite{pnas}
\begin{eqnarray}\label{a2}
\langle |\rho_{\bf Q}(L)|^2\rangle=\int_0^L \int_0^L dx_1 dx_1' \langle
a^\dagger_1(x_1)a^\dagger_2(x_1') a_1(x_1') a_2(x_1)\rangle.
\end{eqnarray}
Note that in Eq. (\ref{a2}) we used the normal ordered form of the operators which means that we neglect the shot noise considered in section
\ref{Snoise}. This is justified for long condensates as we discuss below.
>From now on, we will concentrate on the case when independent
clouds are identical and have the same density of particles
with equal interaction strengths. Then
\begin{eqnarray}
\langle |\rho_{\bf Q}(L)|^2\rangle=\int_0^L \int_0^L dx_1 dx_1' \langle
a^\dagger(x_1)a(x_1')\rangle^2. \label{Asquared}
\end{eqnarray}

To gain intuition into the physical meaning of the average amplitude
of interference fringes, we address two limiting cases. First,
consider the situation when $\langle a^\dagger (x) a(0)\rangle$
decays exponentially with distance and the correlation length is given by
$\xi<<L$. Then Eq. (\ref{Asquared}) implies that $|\rho_{\bf Q}(L)|\propto
\sqrt{L \xi}$, which has a simple physical interpretation. Since the
phase is only coherent over a length $\xi,$ the system is
effectively equivalent to  a series of  $L/\xi$ pairs of
independent condensates. Each pair contributes interference
fringes with a constant amplitude proportional to $\xi$ and a
random phase. The total amplitude $\rho_{\bf Q}(L)$ is the result of
adding $L/\xi$ independent vectors of constant length $\xi$ and
random direction. Adding random uncorrelated vectors gives a zero average except for a typical square root fluctuation. Thus scaling of the
absolute value of the net interference amplitude is $\sqrt{L \xi}.$
 This observation is similar in spirit to that made in Ref. \cite{zoran}
of interference between $30$ independent condensates in a chain.
Fringes can be seen, though their average amplitude is suppressed
by a factor of $\sqrt{30}$ compared to the interference between
two condensates. Now consider the opposite limit of perfect
condensates, for which $\langle{a^{\dagger}(x)a(0)}\rangle$ is
constant. In this case Eq. (\ref{Asquared}) implies that $|\rho_{\bf Q}(L)| \propto L$. Pictorially this is the result of adding vectors which are all aligned, resulting in a fringe amplitude absolute value of which scales as the total size of the system.

Methods developed in this section for analyzing $|\rho_{\bf Q}(L)|^2$ can be applied to condensates with either uniform and non uniform densities.
For simplicity, we concentrate on the case
when $L$ is much smaller than the size of the clouds, so the
change in the atomic density along the clouds can be ignored. In this case correlation functions for 1D gases are described by the Luttinger liquid theory \cite{Haldane, Cazalilla}. For OBC at zero temperature two-point correlation functions are given by
\begin{eqnarray}
\langle a^{\dagger}(x)a(y)\rangle \sim \rho
\left(\xi_h/|x-y|\right)^{1/2K}. \label{1dcor}
\end{eqnarray}
Here $\rho$ is the particle density, $\xi_h$ is the healing
length, which also serves as the short range cutoff, and $K$ is
the so-called Luttinger parameter, which characterizes the strength of interactions.  For bosons with a repulsive short-range potential, $K$ ranges between $1$ and $\infty$, with $K=1$
corresponding to strong interactions, or ``impenetrable'' bosons,
while $K\to \infty$ for weakly interacting bosons. Substituting
Eq.~(\ref{1dcor}) into Eq.~(\ref{Asquared}) and assuming that
$L\gg \xi_h,$ we obtain \cite{pnas}
\begin{eqnarray}
\langle |\rho_{\bf Q}(L)|^2 \rangle = \tilde C\rho^2 L^2\left(\frac{\xi_h}{
L}\right)^{1/K}, \label{1dT0}
\end{eqnarray}
where $\tilde C$ is a constant of order unity. We see that the
amplitude of the interference fringes $\langle |\rho_{\bf Q}(L)|\rangle$ scales as a non trivial power of the imaging length. In the non
interacting limit ($K\to\infty$) the scaling is linear $\langle
|\rho_{\bf Q}(L)|\rangle \sim L,$ as expected for a fully coherent system.
Interestingly, $\langle |\rho_{\bf Q}(L)|\rangle \sim \sqrt{L}$ appears in the hard core limit ($K=1$), as in systems with short range correlations which were discussed above.


One may be concerned that Eq. (\ref{1dT0}) gives only the long distance
asymptotic behavior of the correlation functions, and does not describe the short distance behavior. From Eq. (\ref{Asquared}) one finds that the contribution of the short distance part of the correlation functions to
$\langle |\rho_{\bf Q}(L)|^2 \rangle$ scales as $L.$ In the physically
relevant case of $K>1$ and in the limit of large $L$ this contribution is
smaller than Eq. (\ref{1dT0}). We note that in principle one can use
the exact Bethe ansatz solution of the Lieb-Liniger model \cite{LL} to
obtain correlation functions valid at all distances \cite{Caux}. Another
contribution which has been neglected is the shot noise. The shot noise contribution to $\langle |\rho_{\bf Q}(L)|^2 \rangle$ comes from the normal ordering of operators, and equals
\begin{eqnarray}
\int_0^L \int_0^L dx_1 dx_1' \langle
a^\dagger_1(x_1)a_2(x_1) a^\dagger_2(x_1') a_1(x_1')\rangle
-\int_0^L \int_0^L dx_1 dx_1' \langle
a^\dagger_1(x_1)a^\dagger_2(x_1') a_1(x_1') a_2(x_1)\rangle=\nonumber \\
\int_0^L\int_0^L dx_1 dx_1'\delta(x_1 -x_1')\langle
a^\dagger_1(x_1) a_1(x_1')\rangle =n_{1D} L.
\end{eqnarray}
In the limit of large $L$ and $K>1$ this is again a subleading contribution and can be neglected.

For 2D, one can use similar approach to describe the contrast
distribution at finite temperature below the BKT transition.
We note that we assume that the temperature is small enough
such that 2D gas is in a quasicondensate regime \cite{2dshlyapnikov, Shlyapnikov_review},
when only phase fluctuations are present. In this
case, correlation functions are given by \cite{Berezinskii, KT, Tsvelik}
\begin{eqnarray}
 \langle
a^{\dagger}(r)a(0)\rangle \sim \rho
\left(\frac{\xi_h}{r}\right)^{\eta(T)}, \label{2dcorrfunc}
\end{eqnarray}
  where $\eta(T)=m
T/(2\pi\hbar^2 \rho_s(T))$
 depends on the temperature and the superfluid
density $\rho_s(T).$ The BKT transition takes place at the universal value
$\eta_c(T_c)=1/4.$ To keep connection to the 1D case, we introduce
\begin{eqnarray}
K=1/(2\eta(T)),\label{Keta}
\end{eqnarray}
and restrict our attention to $K>K_c=2.$ For temperatures above the
BKT transition Eq. (\ref{2dcorrfunc}) doesn't hold, and
correlations decay exponentially. This means that the integrated interference amplitude will only increase as the square root of the integration area \cite{pnas}.

Fig. \ref{setup_2D} illustrates experiments performed with 2D gases to
identify the BKT transition \cite{zoran_KT}. Two independent 2D
condensates are confined in transverse directions using an optical lattice potential. After the optical potential is switched off and clouds expand, the density is imaged on a CCD camera. When temperature is small, interference fringes are straight lines. As the temperature is increased, the fringes start to meander due to spatial fluctuations of
the phase. Integrating the image along the section of variable
length $L$ in $x-$ direction gives $L-$ dependent fringe amplitude
$\langle |\rho_{\bf Q}(L)|\rangle.$ Scaling of this amplitude with $L$ contains information about $\eta(T),$ which is expected to
have a universal value $\eta_c(T_c)=1/4$ at  the BKT transition.

\begin{figure}
\psfig{file=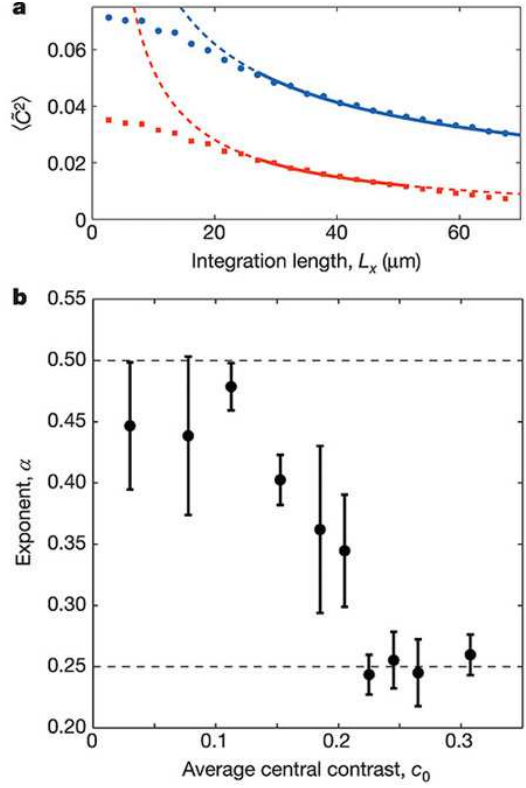} \caption{\label{zoranfig2}
Emergence of quasi-long-range order in a 2D gas. {\bf a,} Examples of average integrated interference contrasts $\langle \tilde C^2\rangle$
 are shown for low (blue circles) and high (red squares) temperatures; $L_x$ is the integration length along $x-$ direction. The lines are fits to the data using the power-law functions $1/(L_x)^{2\alpha}.$ {\bf b,} Exponent $\alpha$ as a function of the central contrast $c_0$. Central contrast $c_0$ serves as a "thermometer", such that smaller values of $c_0$ correspond to higher temperatures. Dashed lines indicate  theoretically expected values of $\alpha$ above and below the BKT transition in a uniform system. Taken from Ref. \cite{zoran_KT}.}
\end{figure}

Fig. \ref{zoranfig2} illustrates the procedure used to extract
scaling exponents in experiments. $\langle |\rho_{\bf
Q}(L)|^2\rangle$ (denoted as $\langle \tilde C^2\rangle$ in
Fig.~\ref{zoranfig2}A) is plotted as a function of $L,$ and its
scaling with $L$ in a certain range (see Ref. \cite{zoran_KT} for
more details) is used to extract the exponent $\eta(T)$ (denoted
as $\alpha$ in Fig.~\ref{zoranfig2}B). The average central
contrast $c_0$ serves as a "thermometer", such that smaller
values of $c_0$ correspond to higher temperatures. Above the BKT
temperature the value of $\eta(T)$ extracted from interference
experiments is expected \cite{pnas} to be equal to $0.5,$ while
at the transition point it is equal to  $\eta_c(T_c)=1/4.$
Fig.~\ref{zoranfig2}B shows a sudden change of the exponent in  a
relatively narrow range of temperatures. This change is
reminiscent of the universal jump in  the superfluid density for
2D helium films \cite{BishopReppy}. Of course, since experiments
are done with finite systems and imaging done along the
$y-$direction performs averaging over the inhomogeneous density
profile, one shouldn't expect the universal jump, but rather a
crossover. The presence of the trap also affects  parameters of
the BKT transition \cite{BKTtrap,BKTtrap2}.

The BKT transition is an example of a topological transition
which is driven by the unbinding of vortices, and the remarkable
feature of experiments in Ref. \cite{zoran_KT} is the ability to
independently resolve the vortices. When only one of the condensates has a vortex, the interference pattern will have a disclination-like structure \cite{stock}. It was shown experimentally in Ref. \cite{zoran_KT} that proliferation of
vortices occurs at the same point in the parameter space as the jump
in the scaling exponent. In following sections we will show that
in uniform systems not only the scaling exponent, but also the  full distribution function of the fringe contrast has a universal form at the BKT transition.

\subsection{Interference amplitudes: from high moments to full distribution functions}
\label{highermomentssec}
Measuring atom density to obtain an interference pattern is a
classical measurement on a quantum mechanical wave function.
The process of the measurement itself introduces an intrinsic
quantum mechanical noise. Said differently, from shot to shot we will
not get precisely the same value of the amplitude of interference fringes.
Expressions for $\langle |\rho_{\bf Q}(L)|^2\rangle$ which we derived in Eqs. (\ref{a2})-(\ref{1dT0}) correspond to averaging over many shots.
For example, data points from Fig.~\ref{zoranfig2}A correspond to averaging over approximately one hundred measurements \cite{flatland}. However each individual shot will give the value of  $ |\rho_{\bf Q}(L)|^2$ which may be different from its average value.

An interesting question to consider is how this amplitude
fluctuates from one experimental run to another. To address this
question we need to consider higher moments of the operator
$|\rho_{\bf Q}(L)|^2$. Generalizing the argument which lead to
Eq. (\ref{Asquared}), we obtain
\begin{eqnarray}
\langle |\rho_{\bf Q}(L)|^{2n} \rangle = \int_0^L\dots\int_0^L dx_1 \dots dx_n
dx_1' \dots dx_n' | \langle a^\dagger(x_1) \dots a^\dagger(x_n)
a(x_1') \dots a(x_n') \rangle |^2. \label{A2n}
\end{eqnarray}

In Eq. (\ref{A2n}) we used a normal ordered correlation function similar to Eq. (\ref{Asquared}). One can calculate \cite{shotnoise} corrections due to normal ordering for higher moments of $|\rho_{\bf Q}(L)|^{2},$
and show that in the limit of large $L$ and $K>1$ they can be neglected.

>From  Eq. (\ref{Asquared}) we observe that $\langle
|\rho_{\bf Q}(L)|^{2} \rangle$ contains information about two-point correlation
functions of individual clouds. Eq. (\ref{A2n}) shows that higher
moments of $|\rho_{\bf Q}(L)|^{2}$ contain information about higher order
correlation functions. The full distribution of the fluctuating
variable $|\rho_{\bf Q}(L)|^2$  contains information about all high order
correlation functions.

In the Luttinger liquid theory fluctuations of the phase
are described by the Gaussian action. For Gaussian  actions higher
order correlation functions are simply related to two-point
correlation functions (see e.g. Ref. \cite{Tsvelik}):
\begin{eqnarray}
\langle a^\dagger(x_1) \dots a^\dagger(x_n) a(x_1') \dots a(x_n')
\rangle = \frac{\underset{ij}{\prod}\langle a^\dagger(x_i)
a(x_j')\rangle }{\underset{i<j}{\prod}\langle
a^\dagger(x_i)a(x_j)\rangle\underset{i<j}{\prod}\langle
a^\dagger(x_i') a(x_j')\rangle}.
\end{eqnarray}
Using this formula together with Eq. (\ref{1dcor}), the higher moments of fringe amplitudes can be written as
\begin{eqnarray}
\langle |\rho_{\bf Q}(L)|^{2n}\rangle =A_0^{2n}  Z_{2n}, \; \mbox{where} \;
A_0=\sqrt{ C \rho^2 \xi_h^{1/K} L^{2-1/K}},\label{A0}
\end{eqnarray}
$C$ is a constant of the order of unity and for OBC in 1D
\begin{eqnarray}
 Z_{2n}=\int_{0}^1
...\int_{0}^1 du_1 ...  dv_n \left|\frac{\underset{i<j}{\prod}|u_i
-u_j| \underset{i<j}{\prod}|v_i -v_j|}{\underset{ij}{\prod}
|u_i-v_j| } \right|^{\frac1K}
 =\int_0^1
...\int_0^1 du_1 ... dv_n
e^{\frac1K\left(\underset{i<j}{\sum}G(u_i,u_j) +
\underset{i<j}{\sum}G(v_i ,v_j)- \underset{ij}{\sum}G(u_i,v_j)
\right)}. \label{Z2n}
\end{eqnarray}
Here for OBC $G(x,y)$ is given by
\begin{eqnarray}
G(x,y)=\log{|x-y|}. \label{Gxydef}
\end{eqnarray}
Integrals similar to Eq. (\ref{Z2n}) appeared in the
literature before \cite{Jack,geneuler}, but they are not easy to compute.

The Gaussian model possesses a powerful conformal symmetry
\cite{Cazalilla, Tsvelik}, which dictates the form of the
correlation functions for periodic boundary conditions or nonzero
temperatures.  For PBC with circumference of the condensates
equal to the imaging length, the change in the correlation
functions leads to \cite{Cazalilla}
\begin{equation}
G^{per}(x,y)=\log{\frac1\pi\sin{\pi|x-y|}}. \label{corrper}
\end{equation}
For nonzero temperature, $Z_{2n}$ depends on $K$ and the thermal
length $\xi_T=\hbar v_s/(k_B T),$ where $v_s$ is the sound
velocity:
\begin{equation}
G(x,y,\frac{\xi_T}{L})=\log{\left(\frac{\xi_T}{\pi
L}\sinh{\frac{\pi|x-y|L}{\xi_T}}\right)}.
\end{equation}
The analysis in this section (except for section \ref{BSG})  doesn't depend on the particular form of $G(x,y).$ The only general restriction is
\begin{eqnarray}
G(x,y)=G(y,x).
\end{eqnarray}

Eq. (\ref{Z2n}) is also valid for 2D case below the BKT transition
temperature, with $u_i$ and $v_i$ being 2D variables on a
rectangle with $G(\vec x,\vec y)=\log{|\vec x-\vec y|},$ and with a properly redefined $A_0.$

In what follows, we will be interested in
distribution functions $W(\alpha)$ of the variable
$\alpha=|\rho_{\bf Q}(L)|^{2}/A_0^2 $ and  of its normalized version $\tilde
\alpha=|\rho_{\bf Q}(L)|^{2}/\langle \rho_{\bf Q}(L)^2\rangle.$
By the definition of higher moments
\begin{eqnarray}
Z_{2n}=\int_0^\infty W(\alpha) \alpha^n d\alpha,\; \mbox{and} \; \frac{Z_{2n}}{Z_2^n}=\int_0^\infty \tilde W(\tilde \alpha)\tilde \alpha^n d\tilde \alpha.\label{walpha}
\end{eqnarray}

The general problem we consider now
is  how to construct distribution functions
$W(\alpha)$ given by Eq. (\ref{walpha}) with $Z_{2n}$ given by Eq.
(\ref{Z2n}). $G(x,y)$ can be  an {\it arbitrary} symmetric
function of $x$ and $y,$ and not necessarily the function of their
difference. This allows us to study systems with
non-uniform density in external traps, where one needs to use  modified correlation functions \cite{Shlyapnikov_review}. In section \ref{SGsection} we will show that $W(\alpha)$ is connected to the partition functions of various  Sine-Gordon models and Coulomb gases, and methods developed in this section provide a new non perturbative tool for calculating partition functions of such models.

One can think of $Z_{2n}$ in Eq. (\ref{Z2n}) as  partition
functions of a classical two-component gas of fictitious charged
"particles" in a microcanonical ensemble, with $K$ being the
"temperature". We  first briefly comment on the limiting cases $K\gg 1,$
and $ K\rightarrow 1.$ For $K\gg 1,$ expansion of $Z_{2n}$ gives $Z_{2n}\approx 1$ in the zeroth
order. This means that $\tilde W(\tilde
\alpha)\approx\delta(\tilde \alpha-1),$ i.e. a very narrow
function peaked near its average value. Higher order terms in
the expansion give $\tilde W(\tilde \alpha)$ a small width of the
order of $1/K,$ and are studied in detail in Appendix
\ref{largeKexpansion}. For 1D gases at zero temperature with
$K\rightarrow1$ or nonzero temperatures with $\xi_T K/L\ll 1$ the
distribution function $\tilde W(\tilde\alpha)$ is Poissonian
irrespective of OBC or PBC: $\tilde W(\tilde\alpha)= e^{-\tilde
\alpha}$. To demonstrate this we need to verify that
$Z_{2n}/Z_2^n=\int_0^{\infty}\tilde\alpha^n
e^{-\tilde\alpha}d\tilde\alpha=n!$. This can be shown  using
 the classical gas analogy: as $K\rightarrow 1,
Z_2=\int1/|x-y|^{1/K} dx dy$  starts to diverge for $x\rightarrow
y,$ and the main contribution to $Z_{2n}$ comes from "molecular"
states of the two-component gas, i.e. from the parts of the configuration
space in which each "particle" has a "particle" of the opposite
charge in its neighborhood; $n!$ is just the number of ways to
form such pairs. In this language $K\gg1$ corresponds to the
"plasma" phase of the classical charged gas, and the evolution of
the distribution function can be understood as a formation of
"molecules" out of the "plasma" phase as the "temperature" (i.e. the Luttinger parameter $K$) is lowered.
For finite $\xi_T,$ the main contribution to $Z_{2n}$ comes from
distances $\sim \xi_T K/L,$ and if this parameter is much
smaller than 1, then again "molecular" contributions dominate.

There is a simple physical interpretation to $W(\alpha)$ for both
$K\gg1$ and $K\rightarrow 1.$ When bosons do not interact and $K=\infty,$
there should be no phase fluctuations within individual condensates.
Hence in each experiment we should find a perfect interference pattern although positions of the density minima are unpredictable (see discussion in section \ref{3dinterference}). Alternatively in the regime of strong interactions when $K\rightarrow 1$ we can think of the  net interference as a result of adding many random uncorrelated two dimensional vectors (see discussion after Eq. (\ref{Asquared})). Earlier we used the fact that for a random walk the net displacement is proportional to the square root of the number of steps. But we also know that for 2D random walks the distribution function of the square of the net displacement is Poissonian, which is what we find for $\tilde W (\tilde \alpha).$

As $K$ varies from $1$ to $\infty,$ the distribution function
$\tilde W(\tilde \alpha)$ should evolve from being a very broad Poissonian
function to a narrow delta function. The evolution of the
distribution in the intermediate regimes will be studied in detail below.
In Appendix \ref{largeKexpansion} we develop a systematic expansion of
$Z_{2n}$ in powers of $1/K$ for large $K$ which works for
different dimensions and boundary conditions, and investigate the
distribution functions in this limit.
The rest of this section is organized as follows.
 In section  \ref{SGsection} we discuss the connection of our problem
to the Sine-Gordon models, which describe various physical
problems, ranging from flux line in superconductors \cite{AHNS} to string
theory \cite{Witten}. In section  \ref{BSG} we obtain full distributions for the 1D case with PBC using the exact solution of the boundary  Sine-Gordon model on a circle \cite{Gritsev}. In
section \ref{Exactsolution} we present a novel non perturbative
solution which is applicable for any value of $K$ and for various dimensions and boundary conditions. We discuss the connection of the distribution functions of
fringe visibilities to the statistics of random surfaces \cite{fcsshort}, and prove that  for 1D case with  periodic boundary conditions in the
limit of large $K$ the distribution of fringe visibilities is
given by the Gumbel distribution, one of the extreme value statistical
distributions \cite{gumbelbooks}.

\subsection{Connection of the fringe visibility distribution functions to the partition functions of Sine-Gordon models}
\label{SGsection}
To illustrate the connection of the fringe visibility distribution functions to the partition functions of Sine-Gordon models we start from a 1D system with
periodic boundary conditions. Shortly we will demonstrate that this example provides
a unique possibility to make an excursion into very
different subjects of physics and mathematics. As shown earlier,
for a one-dimensional ring condensate with PBC and circumference  equal to the imaging length, the
correlation functions lead to
\begin{equation}
G^{per}(x,y)=\log|\frac{1}{\pi}\sin(\pi(x-y))|.
\end{equation}
This function describes a two-component gas of
particles interacting via a 1D Coulomb potential, which is a logarithmic
in the interparticle separation. For PBC the distance is a chord function
$\sin(\pi|x-y|)/\pi.$ It is common to define the grand partition function
of such Coulomb gas as
\begin{eqnarray} \label{86}
Z_{0}(K,g)=\sum_{n=0}^{\infty} \frac{g^{2n}}{(n!)^2}
Z^{per}_{2n},
\end{eqnarray}
where $Z^{per}_{2n}$ is given by Eq. (\ref{Z2n}) with $G(x,y)$ substituted by
$G^{per}(x,y).$

Several different physical problems can be related to the partition
function given by Eq. (\ref{86}). Here we mention only a few examples: a) the anisotropic Kondo model \cite{YA}; b) the quantum impurity model in
one dimensional interacting electron systems introduced in Ref. \cite{kanefisher}, which has been  extensively studied in the context of edge states in Quantum Hall systems \cite{Saleur_notes}; c) the background-independent string theory and the model of strings attached
to a D-brane originally introduced in Ref. \cite{Witten} (see Ref. \cite{Sen} for a recent review); d)
Calogero-Sutherland model \cite{Sutherlandbook}, which has
numerous application as an effective model; e) flux line pinning in
superconductors \cite{AHNS}; f) quantum tunneling  in the
presence of dissipation within Caldeira-Leggett model
\cite{caldeira-leggett}; g) interference of two one-dimensional
condensates \cite{pnas,Gritsev,fcsshort}.

%

We now briefly describe how the expression given by Eq. (\ref{86}) appears as a partition function in physical systems. We consider the imaginary time action
\begin{eqnarray}\label{impact}
S_{per}[g]=\frac{1}{2}\int_{-\infty}^{\infty}dx\int_{0}^{1}d\tau[(\partial_{\tau}\phi)^{2}+(\partial_{x}\phi)^{2}]+
2g\int_{0}^{1}d\tau\cos[\beta\phi(x=0,\tau)+2\pi p\tau]
\end{eqnarray}
known as the boundary Sine-Gordon model \cite{boundarySG}. The
quantum field $\phi(x,\tau)$ in Eq. (\ref{impact}) is defined on
an infinite line in $x$ direction and is assumed to be periodic
along $\tau$ direction: $\phi(x,\tau)\equiv \phi(x,\tau+1).$ The
interaction term is present only at $x=0.$ A typical physical
system which is described by action (\ref{impact}) is an
interacting 1D electron liquid scattered by an impurity
\cite{kanefisher}. In this case the cos-term describes
backscattering of electrons within the Luttinger liquid
formalism. In Eq.~(\ref{impact}) we also added a $p-$dependent
phase winding term. For a quantum impurity problem this  is
somewhat reminiscent of having a finite voltage
\cite{Zamolodchikov_voltage}, while in interference experiments
this term corresponds to a relative momentum along the
condensates \cite{Gritsev} (except for section \ref{BSG} we only
consider the case with $p=0$). Partition function of Eq.
(\ref{impact}) is defined as \cite{Jack}

\begin{eqnarray}\label{zp} Z_{p}(K,g)&=&\frac{{\cal Z}_{p}(K,g)}{{\cal
Z}_{p}(K,g=0)},\qquad {\cal Z}_{p}(K,g)=\int{\cal D}\phi e^{-S_{per}[g]}.
\end{eqnarray}
 One can expand $Z_{p}(K,g)$  in Taylor series of the coupling $g.$
Nonvanishing contributions come from combinations which have equal number of $\exp(+i\beta\phi)$ and $\exp(-i\beta\phi)$ terms.
 This is essentially a charge neutrality condition for the Coulomb gas. If we identify
\begin{eqnarray} \frac{\beta^{2}}{2\pi}=\frac{1}{K},\;\;\;
x=g\left(2\pi\right)^{\frac{1}{2K}} , \end{eqnarray}  we obtain \begin{eqnarray}
Z_{p}(K,g)=\sum_{n=0}^{\infty} \frac{x^{2n}}{(n!)^2} Z_{2n}^{(p)}.
\label{zpert} \end{eqnarray} Here the microcanonical partition functions
$Z_{2n}^{(p)}$ are represented by the following integrals

\begin{eqnarray} &
&Z_{2n}^{(p)}=\int_0^{2\pi}\!\dots\!\int_0^{2\pi}\prod_{i=1}^{n}\frac{du_i}{
2\pi}\frac{dv_i}{
2\pi}e^{i2p\sum_{i}(u_{i}-v_{i})}\left|\frac{\prod_{i<j}
2\sin\left(\frac{u_i-u_j}{2}\right)\prod_{k<l}
2\sin\left(\frac{v_k -v_l}{ 2}\right)}{
\prod_{i,k}2\sin\left(\frac{u_i-v_k}{
2}\right)}\right|^{\frac{1}{K}}. \label{z2n} \end{eqnarray}


$Z_{2n}^{(p)}$  for $p=0$  is related to $Z^{per}_{2n}$
defined above by
\begin{equation}
Z_{2n}^{(0)}=Z^{per}_{2n} (2\pi)^{-n/K}.
\end{equation}
Then
\begin{eqnarray} Z_{0}(K,g)=\sum_{n=0}^{\infty} \frac{g^{2n}}{(n!)^2}
Z^{per}_{2n}= \sum_{n=0}^{\infty} \frac{g^{2n}}{(n!)^2} \langle \alpha^n
\rangle= \sum_{n=0}^{\infty} \frac{g^{2n}}{(n!)^2}
\int_0^{\infty}W(\alpha)\alpha^n d \alpha = \int_0^\infty
W(\alpha)\, I_0(2g \sqrt{\alpha})\, d\alpha. \label{zpKgdef}
\end{eqnarray}

$Z_{0}(K,g)$ is essentially a Hankel transformation of
$W(\alpha),$ and inverting Eq.~(\ref{zpKgdef}) we can express the
probability $W(\alpha)$ through the partition function
$Z_{0}(K,g)$. Noting that $I_0(ix)=J_0(x)$ and using the
completeness relation for Bessel functions, $\int_0^\infty
J_0(\lambda x) J_0(\lambda y) |x|\lambda
d\lambda=\delta(|x|-|y|)$, we obtain
\begin{equation}
W(\alpha)=2\int_0^\infty Z_{0}(K, i g)J_0(2 g \sqrt{\alpha}) g dg.
\label{wz} \end{equation} It is important that the last equation has the
partition function at imaginary value of the coupling constant.
This should be understood as an analytic continuation of
$Z_{0}(K,g).$ For PBC in 1D, partition functions $Z_{p}(K,g)$  can be
evaluated using the exact solution of the boundary Sine-Gordon model
with periodic boundary conditions, and this approach will be presented in
section \ref{BSG}.

For open boundary conditions one can also write the grand canonical partition
function
\begin{eqnarray}
Z(g)=\sum_{n=0}^{\infty} \frac{g^{2n}}{(n!)^2}
Z_{2n}, \label{zg1}
\end{eqnarray}
where $Z_{2n}$ are given by Eqs. (\ref{Z2n}),(\ref{Gxydef}). We can express Eq. (\ref{zg1}) as a partition function of a certain Sine-Gordon model:
\begin{eqnarray}
Z(g)=\frac{\int {\cal D}\phi e^{-S(g)}}{\int {\cal D}\phi
e^{-S(0)}}, \label{zg2}
\end{eqnarray}
 where
\begin{eqnarray}
S(g)=\pi K \int_{-\infty}^{\infty}dx\int_{-\infty}^{\infty}d\tau[(\partial_{\tau}\phi)^{2}+(\partial_{x}\phi)^{2}]+
2g\int_{0}^{1}d\tau \cos[2\pi\phi(x=0,\tau)].
\label{sgaction}
\end{eqnarray}
Note different limits of the $\tau$ integration in the first and the second terms. One can see that Eqs. (\ref{zg1}) and (\ref{zg2})
 define the same $Z(g)$ by observing that
$-\log\left(|x-y|\right)/(2\pi)$ is a free propagator of the Gaussian
action on a plane.


So far we established the relation between the distribution
functions of the amplitude of interference fringes of 1D
condensates and the boundary Sine-Gordon models (\ref{impact}),
(\ref{sgaction}). Generalization of this argument to the case of
2D condensates is straightforward. Higher moments of the
interference amplitude are given by integrals of type
(\ref{Z2n}), but each $u$ and $v$ is now a two-dimensional
coordinate.
Distribution function of fringe amplitudes is then related to the partition function of the {\it bulk} Sine-Gordon model
\begin{eqnarray}
S_{2D}(g)=\pi K \int_{-\infty}^{\infty}dx\int_{-\infty}^{\infty}d\tau[(\partial_{\tau}\phi)^{2}+(\partial_{x}\phi)^{2}]+
2g\int_{0}^{1}\int_{0}^{1}d\tau dx \cos[2\pi\phi(x,\tau)].
\label{bulksgaction}
\end{eqnarray}
Expanding the partition function corresponding to action (\ref{bulksgaction}) in powers of $g,$ one finds the expression identical to Eq. (\ref{zpKgdef}), but with $W(\alpha)$ corresponding to interference of 2D condensates.

When we describe interference of systems with OBC, we use Eqs. (\ref{sgaction}) and (\ref{bulksgaction}), in which  the field $\phi(x,\tau)$ is defined  on a whole plane, and is not periodic in $\tau.$ In both these cases the interaction is present only in some part of the system, so translational invariance is lost in both $x$ and $\tau$ dimensions. Thus to calculate $W(\alpha)$ using Eq. (\ref{wz})  one needs to calculate partition functions of inhomogeneous Sine-Gordon models. Exact solution of Sine-Gordon model is available only for periodic boundary conditions, so to treat open boundary conditions one needs to develop alternative methods. Conversely, if one has a solution for $W(\alpha),$
this provides a tool for calculating partition functions of inhomogeneous
Sine-Gordon models and Coulomb gases using Eq. (\ref{zpKgdef}).
In section \ref{Exactsolution} we present a novel mapping of $W(\alpha)$
for arbitrary $G(x,y)$ to the statistical properties of random surfaces, which provides a new tool for calculating partition functions of a wide class of Sine-Gordon models and Coulomb gases. In general, partition functions with inhomogeneous $g(x,\tau)$ can be evaluated using this mapping as well. In this case the function $g(x,\tau)$ will appear in the integrand
of Eq.~(\ref{bulksgaction}). We point out that the mapping which we use is not related to the existence of the exact solution of Sine-Gordon models, but relies only on the structure of the correlation functions in the absence of interactions. We also note that a suitable extension of our method can be used to compute correlation functions of Coulomb-gas models in equilibrium and non-equilibrium situations.

\subsection{Distribution functions for 1D gas with periodic boundary conditions}
\label{BSG}

There are three natural ways to compute partition function
$Z_{0}(K,g)$ for complex $g,$ which can be used to construct
distribution functions for 1D gas with PBC using Eq. (\ref{wz}).
The first one is related to the theory of symmetric polynomials
and is described in Appendix~\ref{Jack}. The second one relies on
the integrability of the quantum impurity model  defined by the
action (\ref{impact}), and  makes use of the thermodynamic Bethe
ansatz (TBA). This approach is discussed in Appendix~\ref{TBA}.
These two approaches have been presented in Refs. \cite{Jack},
\cite{FLS2}. They are difficult to implement directly and are
included for completeness. The third approach described below in
section ~\ref{thirdway} leads to the most transparent answer.
Although it is intrinsically related to the first two, it has a
broader applicability and will be studied in detail.

At imaginary $g$ the theory  (\ref{impact}) is apparently
non-Hermitian, but as we will see later the model belongs to a
special class of non-Hermitian field theories, which have real
spectrum. We describe an interesting connection with ${\cal
PT}$-symmetric quantum mechanics and conformal field theories
with negative central charges.
 Here we also  make close contact with the impurity problem in a Luttinger
liquid \cite{kanefisher}. This connection is based on the following observation:
when we consider Taylor expansion of the quantum impurity partition function in powers of $g,$ we need to take multi-point correlation functions at the same $x$ but at different times. To calculate higher
moments of the interference amplitude signal we need to use correlation
functions for equal time, but at different points in space. The two are the same because of the relativistic invariance of the Luttinger liquid \cite{Cazalilla,Tsvelik}.


\subsubsection{Mapping to integrable structure of CFT and singular anharmonic
oscillator}\label{thirdway}


To introduce our method we need to make several formal remarks
regarding one dimensional conformal field theories.
In a recent series of papers Bazhanov, Lukyanov, and Zamolodchikov
explored an integrable structure of conformal field theories
focusing on connections to solvable problems on lattices
\cite{BLZ13}. The key ingredients of the solvability of lattice models are
the so-called transfer matrix operators ${\bf T}(\lambda).$
These operators contain information about all  integrals of motion
as well as excitation spectra of the system. Transfer matrices are
defined as a function of the so-called spectral parameter
$\lambda$ (in the continuum limit $\lambda$ corresponds to
rapidity) and commute for different values of $\lambda$. The
latter property is a direct manifestation of the existence of
an infinite number of commuting integrals of motion. In his studies
of the 8-vertex model, Baxter \cite{Baxter} introduced the operator
${\bf Q}(\lambda)$ which helps to find eigenvalues of ${\bf T}$.
Operators ${\bf T}$ and ${\bf Q}_{\pm}$ satisfy a set of
commutation relations, in particular \cite{BLZ13}
\begin{eqnarray}\label{BOE}
{\bf T}(\lambda){\bf Q}_{\pm}(\lambda) = {\bf
Q}_{\pm}(q\lambda)+{\bf Q}_{\pm}(q^{-1}\lambda),
\end{eqnarray}
where $q=\exp(i\pi/2K)$.  So $\bf T$ matrices can be obtained
explicitly when one knows the $\bf Q$ operators.

By construction, these operators act in the representation space
of Virasoro algebra which can be constructed from the Fock space
of free bosonic operators $a_{\pm n}$ defined such that
\begin{eqnarray}
a_{n}|p\rangle &=&0, \quad \mbox{for} \quad n>0,\quad
P|p\rangle=p|p\rangle, \quad
~[a_{n},a_{m}]=\frac{n}{2}\frac{2\pi}{K}\delta_{n+m,0}.
\end{eqnarray}
Here $|p\rangle$ denotes the vacuum vector and $p$ is the zero
mode of the corresponding bosonic field $\phi(u)$:
\begin{eqnarray}
\phi (u) = iQ +iPu+\sum_{n\neq 0}\frac{a_{-n}}{n}e^{inu}.
\end{eqnarray}
Operators ${\bf A}_{\pm}(\lambda)={\bf
Q}_{\pm}(\lambda)\lambda^{\mp 4p/K}$ act in the representation
space of Virasoro algebra, which can be constructed from the Fock
space of bosonic operators $a_{\pm n}$ satisfying
$a_{n}|p\rangle=0$, $(n>0)$. The Fock vacuum state $|p\rangle$ is
an eigen state of the momentum operator, $P|p\rangle=p|p\rangle$.
For $p=N/2$ ($N=0,1,2...$) the vacuum eigenvalues of the operator
${\bf A}_{\pm}(\lambda)$, ${\bf
A}_{\pm}(\lambda)|p\rangle=A_{\pm}^{(vac)}|p\rangle$, are given by
(below we consider only the quantities with the $+$ subscript
which correspond to the positive $p$)
\begin{eqnarray}
A^{(vac)}(\lambda)=Z_{p}(K,-ig),
\end{eqnarray}
where
\begin{eqnarray}\label{lambda}
\lambda =\frac{x}{\pi}\sin\left(\frac{\pi}{2K}\right).
\end{eqnarray}
At this point it is not clear what we gained by connecting
the analytically continued partition function of the impurity problem
$Z_{p}(K,ig)$ to the expectation value of the operator $A^{(vac)}(\lambda).$ As we discuss below, a considerable number of
important results have been derived for $A^{(vac)}(\lambda).$ We will
be able to make use of these results to obtain the distribution
functions of the fringe amplitudes.
The function $A^{(vac)}$ has known large-$\lambda$ asymptotics
\cite{BLZ13} \begin{eqnarray}\label{asymp} \log(A^{(vac)})(\lambda)\sim
M(K)\left(-\lambda^{2}\right)^{\frac{K}{2K-1}}, \end{eqnarray} where the
constant $M(K)$ is given by \begin{eqnarray}
M(K)=\frac{\sqrt{\pi}\,\Gamma(\frac{1}{4K-2})[\Gamma(\frac{2K-1}{2K})]
^{\frac{2K}{2K-1}}}{\cos(\frac{\pi}{4K-2})\Gamma(\frac{K}{2K-1})}.
\end{eqnarray} The function $A^{(vac)}(\lambda)$ is entire function for
$K>1$ and is completely determined by its zeros $\lambda_{k}$,
$k=0,1,...$. Therefore $A^{(vac)}(\lambda)$ can be represented by
the convergent product
\begin{eqnarray}\label{entire}
A^{(vac)}(\lambda)=\prod_{k=0}^{\infty}\left(1-\frac{\lambda^{2}}{\lambda_{k}^{2}}\right),
\ & &\ A^{(vac)}(0)=1.
\end{eqnarray}

On the basis of analysis of a certain class of exactly solvable
model, corresponding to the integrable perturbation of the
conformal field theory, it  was conjectured in \cite{DT} that the
so-called $Y$-system and related $T$ system (where
$Y=e^{\epsilon_{r}}$ and $\epsilon_{r}$ are the Bethe-ansatz
energies parametrized by $r$, the nodes of the Dynkin diagrams
(see TBA-section) satisfy the same functional equations and
possess the same analytical structure and asymptotics as the
spectral determinant of the one-dimensional anharmonic oscillator.
Further, the same functional equations, analytical properties
(\ref{entire}) and asymptotics (\ref{asymp}) are satisfied for the
vacuum eigenvalues of ${\bf Q}$-operator for special values of $p$
and the latter are given by the spectral determinant of the
following Schr\"{o}dinger equation \begin{eqnarray}
(-\partial_{x}^{2}+x^{2\alpha})\Psi(x)=E\Psi(x). \end{eqnarray} The spectral
determinant is defined as
\begin{eqnarray}\label{specdet}
D(E)=\prod_{n=1}^{\infty}\left(1-\frac{E}{E_{n}}\right).
\end{eqnarray}
Soon after, in Ref.~[\onlinecite{BLZ}], this conjecture has been
extended to all values of $p$ :
\begin{eqnarray}
A^{(vac)}(\lambda, p)= D(\rho\lambda^{2}),
\end{eqnarray}
where now $D(E)$ is the spectral determinant of the
Schr\"{o}dinger equation
\begin{equation}
-\partial_{x}^{2}\Psi(x)+\left(x^{4K-2}+\frac{l(l+1)}{x^{2}}\right)\Psi(x)=E\Psi(x),
\label{schrod}
\end{equation}
with $l=4pK-1/2$. Here $\rho =(4K)^{2-1/K}[\Gamma(1-1/(2K))]^{2}$.
In relation to the interference problem, $p=\left(m d/\hbar
t\right)\tan\theta$, where $m$ is the atom's mass, $d$ is the
separation between the two condensates and $t$ is the time when
the measurement was done after the free expansion started.

For some values of parameters the Eq.(\ref{schrod}) can be solved
exactly:
\begin{itemize}

 \item[a)] $K=1$. In this case, corresponding to a singular
harmonic oscillator $E_{n}=4n+2l-1$, $n=1,2,...$ and
\begin{equation}
D(E,l)=\frac{\Gamma(\frac{3}{4}+\frac{l}{2})e^{-CE}}{\Gamma(\frac{3}{4}+\frac{l}{2}-\frac{E}{4})},
\end{equation}
where $C$ is nonuniversal renormalization constant.

\item[b)] The case
$K\rightarrow\infty$ is recovered by the rigid well potential for
which the eigen energies are given by the zeroes of the Bessel
function, and therefore
\begin{equation}
D(E,l)=\Gamma(l+3/2)(\sqrt{E}/2)^{-l-1/2}J_{l+1/2}(\sqrt{E}).
\end{equation}
\item[c)] For $K=3/4$ and $l=0$  the result is expressed in terms of Airy function.

\end{itemize}
For generic values of $K$ and $l$ the Schr\"{o}dinger equation can
be solved numerically with a very good precision with subsequent
computation of the spectral determinant.  Alternatively, for
$n\geq 5-10$ the spectrum of the equation (\ref{schrod}) is very
well approximated by the standard WKB expression (see e.g.
\cite{QR}) \begin{eqnarray}\label{WKB}
E_{n}=\epsilon(K)(n-\gamma_{l}(K))^{\frac{2K-1}{K}}, \label{wkb1}
\end{eqnarray} where $n=1,2,..$. Here $\gamma_{l}(K)$ is the Maslov index.
For $1/2<K<\infty$, $\gamma_l(K)=\frac{1}{4}-\frac{l}{2}$, for
$K=\infty$, $\gamma_{l}(K)=-l/2$  and for $0<K<\frac{1}{2}$,
$\gamma_l(K)=\frac{4K-2l-1}{8K}$. The function $\epsilon(K)$ in
Eq.~(\ref{wkb1}) reads \begin{eqnarray}
\epsilon(K)=\left[\frac{2\sqrt{\pi}\,\Gamma(\frac{3}{2}+\frac{1}{4K-2})}
{\Gamma(1+\frac{1}{4K-2})}\right]^{\frac{2K-1}{K}}. \end{eqnarray} In
principle, the function $\gamma_l(K)$ can be  a smooth function
interpolating between limiting values given above and can be
considered as a noninteger Maslov index~\cite{FT}. This
interpolation allows to (approximately) evaluate the partition
function and the distribution function in many cases. In the
limiting cases $K\to 1$ and $K\to \infty$ the WKB approximation
gives the exact spectrum. We
point out  however, that using the approximate WKB spectrum carelessly can
result in non-physical results, e.g. negative values of the
distribution function $W(\alpha)$, and thus have to be used with
care.
WE emphasize that  the solution of the ODE (\ref{schrod}) for
several potentials gives us {\it already analytically continued}
function $Z_p(K,ig)$. It seems that this approach to finding the
solution is easier and more elegant than the solution of TBA
equations with subsequent analytical continuation. Moreover a
variety of approximate methods are available for solving
Schr\"{o}dinger-like 1D equations.

Note that Eq. (\ref{schrod}) has an interesting duality
symmetry which generalizes the Coulomb-harmonic oscillator duality
and which allows to relate the $K>1/2$ and $0<K<1/2$ sectors.
Making the transformation $x\rightarrow y^{1/A},
\Psi(x)\rightarrow y^{\lambda}\phi(y)$ with
$\lambda=-(1/2)+1/(2A)$, and $A=2K$ and then rescaling
$y\rightarrow\sqrt{\alpha}y$ with $\alpha=(-4K^{2}/E)^{2K}$ we
obtain an equation of the same form with parameters $K', l', E'$
given by \begin{eqnarray} 4K'=\frac{1}{K},\quad l'=2p-1/2,\quad
E'=\frac{1}{4K^{2}}\left(\frac{4K^{2}}{E}\right)^{2K}. \end{eqnarray} The
point $K=1/2$ is a self-dual point of this transformation.
Presumably this symmetry is the origin of the Seiberg-Witten
type duality in the impurity problem observed in Ref. \cite{FSd}.

\subsubsection{PT-symmetric quantum mechanics }

The surprising link between the ordinary differential equation
(\ref{schrod}) and the exactly solvable problems related to
conformal field theories with negative central charge and/or
two-color Coulomb gas on a circle is not limited to the
examples considered above. There is a deep connection between these
problems and a non Hermitian ${\cal PT}$-symmetric quantum
mechanics. The latter has been formulated and studied by C. Bender
in various contexts \cite{Bender}. This story goes back to the
(unpublished) work of Bessis and Zinn-Justin where, inspired by
studies of Lee-Yang model, they conjectured that the 1D
Schr\"{o}dinger equation with potential $V(x)=(ix)^{3}$ has a {\it
real} spectrum. Later in Ref. \cite{BB} this statement was generalized
for arbitrary power-law potentials, $V(x)=(ix)^{n}$ for real $n.$
It was also conjectured that the Schr\"{o}dinger equation is
invariant with respect to combined action of the ${\cal PT}$ symmetry:
${\cal P}$ (parity inversion): $x\rightarrow -x,$ $p\rightarrow
-p;$ ${\cal T}:$ $x\rightarrow x,$ $p\rightarrow -p,$
$i\rightarrow -i.$

As we have seen, the partition function of the boundary Sine-Gordon
model (as well as of the other related models) is real for
the imaginary value of the coupling constant. This is of course not
a generic property of field theory models. Therefore the boundary
sine-Gordon model with imaginary coupling also belongs to the
class of ${\cal PT}$-symmetric systems. It is then not
surprising why the ODE (\ref{schrod}) appears in such a theory.

This intriguing relation between ordinary differential equations
and various integrable models has been recently extended to a
large class of models (for recent review and references see
\cite{DT2}) by making a close link to the ${\cal PT}$-symmetric
systems. These relationships have been used \cite{Lu} for the
so-called circular brane model which in certain limit describes
the so-called Ambegaokar-Eckern-Sch\"{o}n model \cite{AES}. It is  therefore reasonable to expect other new interesting applications of these findings.

\subsubsection{Analysis of distribution functions}

Using exact analytic expressions for $Z_p(K,ig)$ at $K=1$ after
the integral transform we obtain the  Poissonian distribution for
$W_p(\alpha)$. The case $K\rightarrow\infty$ is recovered easily
as well. One sees that when $p=0$, the distribution function is a
delta function in the leading order (corrections to this result
will be considered in Appendix \ref{largeKexpansion}). As $p$
increases, $W_p(\alpha)$ rapidly broadens. In the limit of large
$K$, the function $W_p(\alpha)$ depends only on the product $K p$
and takes a simple form: \begin{eqnarray}\label{weibul} W_p(\alpha)\approx
\left\{
                                   \begin{array}{ll}
                                   4Kp (1-\alpha)^{4Kp-1}, & \alpha <1 \\
                                     0, & \alpha >1.
                                     \end{array}
                                     \right.
\end{eqnarray}

This function is peaked at $\alpha=1$ for $K p<1/4$, it becomes a
step function exactly at $K p=1/4$, and for $K p>1/4$.
$W_p(\alpha)$ is a monotonically decreasing function of $\alpha$.
When the product $K p$ becomes large $K p\gtrsim 1$ the function
$W_p$ becomes Poissonian: $W_p(\alpha)\approx 4 K p \exp(-4 K P
\alpha)$. In general, the tendency of broadening of the
distribution function remains true for all values of $K$.
%

The distribution given in (\ref{weibul}) is a particular
representative of class of  extreme-value distributions called
Weibull distributions. At $p=0$ the limit of $K\rightarrow\infty$
it will be analytically proven in section \ref{gumbelproofsection}
that the normalized amplitude $\tilde \alpha$ is characterized by  the
universal Gumbel distribution, for which
\begin{eqnarray}
\tilde W(\tilde \alpha)=K e^{K(\tilde \alpha-1) -\gamma -
e^{K(\tilde \alpha-1) - \gamma}},
\label{Gumbeldef}
\end{eqnarray}
where $\gamma\approx 0.577$ is the Euler gamma-constant. The
Gumbel distribution, which also belongs to a class of extreme
value statistics frequently appears in various problems (see e.g.
books \cite{gumbelbooks} for historic introduction), including
number theory \cite{EL}, $1/f$ noise \cite{1fnoise},
Kardar-Parisi-Zhang growth \cite{KPZ}, free Bose gas \cite{CLM}, etc.
Probably the whole distribution function in our interference
context can in general be characterized as a certain extreme-value
statistics.

\subsection{Non perturbative solution for the general case}
\label{Exactsolution}


In this section we
present a novel approach for calculating $W(\alpha),$ which is  based on
a mapping to the statistics of random surfaces \cite{fcsshort}.
We use this method to evaluate $W(\alpha)$ numerically for a variety of situations.
We point out that our method is not related to
the existence of the exact solution of Sine-Gordon models, but relies
only  on the structure of the multi-point correlation functions in the absence of interactions.  As has been discussed in section \ref{SGsection}, $W(\alpha)$ is connected to the partition
functions of Sine-Gordon models and the partition functions of Coulomb gases. Our mapping provides a new non perturbative tool for calculating partition functions of such models. We expect that there should be numerous applications of our new method to other physical problems which can be related to Sine-Gordon and Coulomb gas models. As a concrete application of our approach we prove analytically  that  for periodic boundary conditions in
the limit of large $K$ the distribution of fringe visibilities is
given by the Gumbel distribution, one of the extreme value statistical
distributions \cite{gumbelbooks}.

\subsubsection{Mapping to the statistics of random surfaces}

We start by observing that $G(x,y)$ is real and symmetric. Hence
it can be diagonalized on $(0,1)$ by solving the eigenvalue equations
\begin{equation}
 \int_{0}^{1}G(x,y) \Psi_{f}(y) dy=G(f) \Psi_{f}(x). \label{eigeneq}
\end{equation}
Here $f$ is an integer index, which goes from $1$ to $\infty.$
$\Psi_{f}(x)$ can be chosen to be real and normalized according to
\begin{eqnarray}
\int_{0}^{1}\Psi_{f}(x)\Psi_{k}(x) dx=\delta(f,k).
\end{eqnarray}
Then, $G(x,y)$ is given by
\begin{eqnarray}
 G(x,y)=\sum_{f=1}^{f=\infty}G(f)\Psi_{f}(x) \Psi_{f}(y). \label{decomp}
\end{eqnarray}
Decomposition given by Eqs. (\ref{eigeneq})-(\ref{decomp}) is
similar to diagonalization of a symmetric matrix by finding its
eigenvectors and eigenvalues. Now we can write $Z_{2n}$ from Eq.
(\ref{Z2n}) as
\begin{eqnarray}
Z_{2n}=\int_{0}^1 ...\int_{0}^1 du_1 ... du_n dv_1... dv_n
e^{\frac1K\left(\sum_{i<j}G(u_i, u_j) +
\sum_{i<j}G(v_i,v_j)-\sum_{ij} G(u_i,v_j)  \right)}= \nonumber \\
 \int_{0}^{1}...\int_{0}^{1}du_1 ... dv_n
e^{\sum_f\frac{G(f)}K\left[\frac{\left(\sum_{i}\Psi_{f}(u_i)\right)^2+\left(\sum_{i}
\Psi_{f}(v_i)\right)^2-\sum_{i}\Psi_{f}(u_i)^2-\sum_{i}\Psi_{f}(v_i)^2}{2}
-\left(\sum_{i}\Psi_{f}(u_i)\right)\left(\sum_{i}\Psi_{f}(v_i)\right)
\right]}=\nonumber\\  \int_{0}^1...\int_{0}^1 du_1 ... dv_n
e^{\sum_f\frac{G(f)}{2K}\left[\left(\sum_{i=1}^{i=n}\Psi_{f}(u_i)-\Psi_{f}(v_i)\right)^2
-\sum_{i=1}^{i=n}\left(\Psi_{f}(u_i)^2+\Psi_{f}(v_i)^2
\right)\right]}.
\end{eqnarray}

Square of the first $i-$sum in the last line above can be
decoupled by introducing Hubbard-Stratonovich integrations over
auxiliary variables $t_f:$
\begin{eqnarray}
Z_{2n}=\int_{0}^{1}...\int_{0}^1 du_1 ...
dv_n\prod_{f=1}^{f=\infty} \frac{ \int_{-\infty}^{\infty}d
t_{f}e^{-\frac{t_{f}^2}2} e^{\sum_{i}
t_f\sqrt{\frac{G(f)}{K}}\left(\Psi_{f}(u_i)-\Psi_{f}(v_i)
\right)-\frac{G(f)}{2K}\left(\Psi_{f}(u_i)^2+\Psi_{f}(v_i)^2
\right)}} {\sqrt{2\pi}}. \label{185}
\end{eqnarray}
Now we can simply integrate over $du_1, ..., dv_n,$ since all
integrals over $u-$variables are the same (integrals over $v-$
variables are also identical):
\begin{eqnarray}
Z_{2n}=\left(\prod_{f=1}^{f=\infty}\frac{\int_{-\infty}^{\infty}
e^{-\frac{t_{f}^2}2}d t_{f}}{\sqrt{2\pi}}\right)
g(\{t_f\})^ng(\{-t_f\})^n, \label{10}
\end{eqnarray}
where
\begin{eqnarray}
g(\{t_f\})=\int_{0}^{1}dx\; e^{\sum_f
t_f\sqrt{\frac{G(f)}{K}}\Psi_{f}(x)-\frac{G(f)}{2K}\Psi_{f}(x)^2}.\label{gt}
\end{eqnarray}
If all eigenvalues $G(f)$ are negative, then
\begin{eqnarray}
g(-\{t_f\})= g(\{t_f\})^*,\;\;
g(\{t_f\})g(\{-t_f\})=|g(\{t_f\})|^2.\nonumber
\end{eqnarray}
>From comparison of Eqs. (\ref{walpha}) and (\ref{10}) we obtain
the central result of this section
\begin{eqnarray}
W(\alpha)=\prod_{f=1}^{f=\infty}\frac{\int_{-\infty}^{\infty}
e^{-\frac{t_{f}^2}2}d
t_{f}}{\sqrt{2\pi}}\delta\left[\alpha-|g(\{t_f\})|^2\right].
\label{Walpha}
\end{eqnarray}

\begin{figure}
\psfig{file=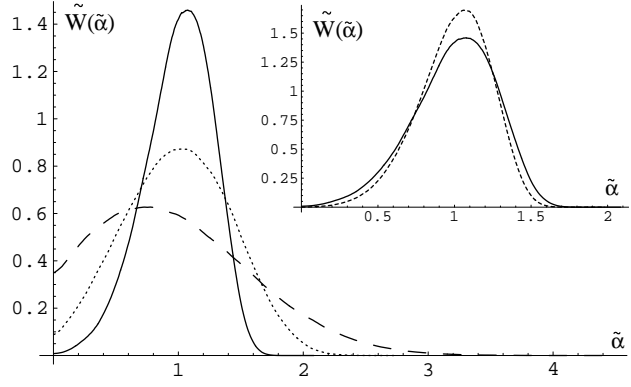} \caption{\label{1dzeroT} Scaled
distribution functions of  the normalized interference amplitude $\tilde W(\tilde
\alpha)$ at $T=0$ for 1D gases with open boundary conditions,
shown for Luttinger parameters $K=2\;\mbox{(dashed)},
K=3\;\mbox{(dotted)}, $ and $ K=5 \;\mbox{(solid)}.$ The inset
shows a comparison between open (solid) and periodic (dashed)
boundary conditions for $K=5$. The figure is taken from Ref.
\cite{fcsshort}.}
\end{figure}

Eq.~(\ref{Walpha}) can be used to simulate distributions
$W(\alpha)$ using Monte-Carlo approach. First, one needs to solve
integral Eqs. (\ref{eigeneq})-(\ref{decomp}) numerically to obtain
eigen functions and eigen vectors. Then one needs to choose random
numbers $\{t_f\}$ from the Gaussian ensemble, and plot the histogram
of the results for $|g(\{t_f\})|^2.$ In what follows we perform
simulations of $W(\alpha)$ with up to $N=10^6-10^7$ realizations
of $\{t_f\}$ and smooth the data. We use a finite value of
$f_{max}$ and check for convergence with $f_{max},$ typically
$\sim 30.$ $\langle \alpha\rangle $ is always kept within $1\%$
from its expected value. For most of the presented results, all
eigenvalues $G(f)$ are negative, and Eq.~(\ref{Walpha}) can be
directly applied. Special care should be taken for 1D case with
nonzero temperature, where one of the eigenvalues $G(1)$ can be
positive. This situation can be handled by subtracting
sufficiently large positive constant $C$ from
$G(x,y,\frac{\xi_T}{L}),$ which makes all eigenvalues negative.
According to (\ref{Z2n}), this leads to rescaling of $\alpha$ by a
factor $e^{-C},$ which can be easily taken into account.

\begin{figure}
\psfig{file=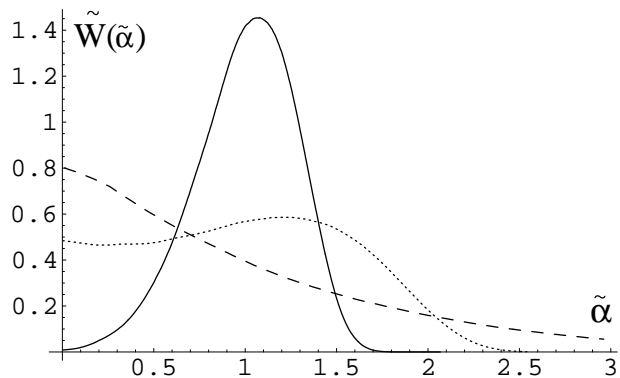} \caption{\label{1dnztfig} Scaled
distribution functions of the normalized interference amplitude $\tilde W(\tilde
\alpha)$ for a 1D Bose gas with open boundary conditions at nonzero
temperature and $K=5.$ Different curves correspond to ratios $K
\xi_T/L =\infty\;\mbox{(solid)}, K \xi_T/L=1\;\mbox{(dotted)},$
and $K \xi_T/L=0.25 \;\mbox{(dashed)}.$ $\xi_T$ is the thermal
correlation length, $K$ is the Luttinger parameter and $L$ is the
imaging length. The  figure is taken from Ref. \cite{fcsshort}.}
\end{figure}

\begin{figure}
\psfig{file=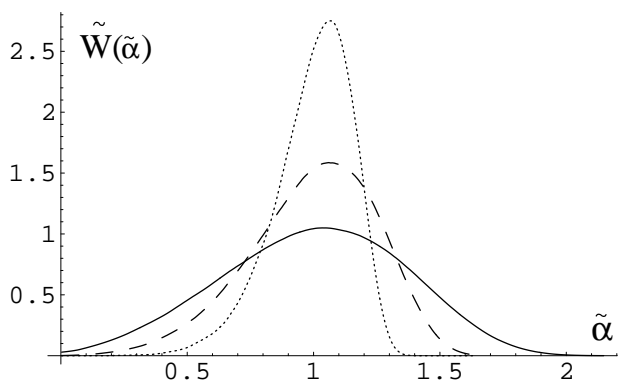} \caption{\label{2dfig}Scaled distribution
functions of the normalized interference amplitude $\tilde W(\tilde
\alpha)$ for a two dimensional Bose gas with the aspect ratio of the
imaging area equal to unity and open boundary conditions.
Temperature is below the Berezinskii-Kosterlitz-Thouless (BKT)
transition temperature. Different curves correspond to $T=T_{BKT}$ (
BKT transition point, solid), $T=(2/3) T_{BKT}$ (dashed line) and
$T=(2/5)T_{BKT}$ ( dotted line). Above the BKT transition
temperature the distribution function is Poissonian. The figure is
taken from Ref. \cite{fcsshort}.}
\end{figure}

In Fig.~\ref{1dzeroT} we show scaled distribution of contrast
$\tilde W(\tilde \alpha)$ at $T=0$  for 1D gases with OBC for
various $K.$ The inset shows a comparison between OBC and PBC for
$K=5.$ In Fig. \ref{1dnztfig} we show the scaled distribution of
contrast for 1D gas with OBC at nonzero temperature and $K=5.$  As
has been discussed earlier, for $\xi_T K/L\ll1$ distribution is
Poissonian and wide, while for $K\gg1$ and  $\xi_T K/L\gg1$ it is very
narrow. Evolution of the full distribution function of the
visibilities while $L$ is varied can be used to  measure
the thermal length $,\xi_T=\hbar v_s/(k_B T),$ precisely and to extract the temperature.   As seen in Fig. \ref{1dnztfig}, at $T\neq 0$ the
distribution  function has characteristic features, i.e. it is
generally non-symmetric and can have a {\it minimum}. These
features can be used to distinguish the noise due to fluctuations
of the phase from technical noise. Finally, in Fig.~\ref{2dfig} we
show scaled distribution of contrast for 2D gas with unity aspect
ratio of imaging area and OBC below BKT temperature. Above BKT
temperature distribution functions become Poissonian for $L\gg
\xi,$ where $\xi$ is the correlation length. In 2D one cannot
describe the crossover at $L\sim \xi$ similar to 1D, since the
action which describes the fluctuations of the phase is not
quadratic in this region, and Eq. (\ref{Z2n}) doesn't hold.

Result (\ref{Walpha}) can be interpreted as the mapping of our
problem to the statistics of random surfaces (strings in 1D),
subject to classical noise. In this mapping, $\Psi_{f}(x)$ are the
eigen modes of the surface vibrations, $t_f$ are the fluctuating
mode amplitudes, and $|G(f)|$ is the noise power. This mapping holds
as long as all eigenvalues $G(f)$ are negative, as discussed
earlier. Infinite dimensional integral over $\{t_f\}$ variables can
be understood as an averaging over fluctuations of the surface. For
particular realization of noise variables $\{t_f\},$ complex valued
surface coordinate at point $x$ is given by
\begin{equation}
h(x;\{t_f\})=\sum_f
t_f\sqrt{\frac{G(f)}{K}}\Psi_{f}(x)-\frac{G(f)}{2K}\Psi_{f}(x)^2.
\label{hxdef}
\end{equation}
$g(\{t_f\})$ is a number defined for each realization of a random
surface $\{t_f\},$ given by Eq. (\ref{gt}). Fringe amplitude
$\alpha$ for each realization of $\{t_f\}$ variables is given by
\begin{equation}
\alpha=|g(\{t_f\})|^2=\left|\int_0^1 dx e^{h(x;\{t_f\})}\right|^2.
\label{toexpand}
\end{equation}

\subsubsection{From interference of 1D  Bose liquids of  weakly interacting atoms to extreme value statistics}
\label{gumbelproofsection}

To illustrate the power of the interpretation of the interference fringe amplitudes  in terms of random
surfaces, we now prove analytically that for periodic
boundary conditions in 1D and  in the limit of large $K,$ the
normalized distribution $\tilde W(\tilde \alpha)$ is given  by the Gumbel distribution (\ref{Gumbeldef}), one of the extreme value statistical
distributions. We note that to prove such result using $1/K$ expansion described in Appendix \ref{largeKexpansion}, one needs to go to the infinite
order of the perturbation theory, so essentially our result is non-perturbative in $K.$

The Gumbel function
\begin{equation}
P_G(x)=e^{-(x+\gamma)-e^{-(x+\gamma)}} \label{gumbelfuncdef}
\end{equation}
plays the same role in the extreme value statistics \cite{gumbelbooks}
as the usual Gaussian distribution plays in the statistics of the
average value. According to the central limit theorem, the average
value of $N$ numbers taken from the same ensemble in the limit of
large $N$ is distributed according to Gaussian function. One can
prove similar theorems for the distribution of the extreme, i.e.
largest value of $N\gg1$ numbers taken from the same ensemble.
Gumbel function (\ref{gumbelfuncdef}) is one of the universal
functions describing ensembles  which decay faster than any
algebraic function at infinity.  The Gumbel distribution often
appears in applied mathematics, including problems in finance or climate studies, where it is used to describe rare events such as stock market crashes or earthquakes.

For the periodic case corresponding to Eq. (\ref{corrper}),
the eigenvalue problem (\ref{eigeneq})-(\ref{decomp}) can be solved
analytically, and the noise has $1/f$ power spectrum:
\begin{equation}
G^{per}(x,y)=\log{\frac1\pi\sin{\pi|x-y|}}=-\log{2\pi}-\sum_{f=1}^{f=\infty}\frac1{2f}\left[(\sqrt{2}\cos{2\pi
f x})(\sqrt{2}\cos{2\pi f y})+(\sqrt{2}\sin{2\pi f
x})(\sqrt{2}\sin{2\pi f y})\right].
\end{equation}
Eigen modes are given by simple harmonic functions, and all of
them, except one, are doubly degenerate: we will denote
corresponding noise variables as $t_{1,f}, t_{2,f}$ and eigen
values as $G(f)=-1/(2f).$  Simplification in the limit $K\gg1$ stems
from the fact that exponent in Eq. (\ref{toexpand}) can be expanded
in Taylor series, since  $h(x;\{t_f\})$ is small for large $K$
according to its definition (\ref{hxdef}). In this case it can be
shown that the distribution of $\alpha$ is linearly related to
roughness, or mean square fluctuation of the surface, as defined
in Ref.~\cite{1fnoise}. It has been shown in  Ref.~\cite{1fnoise}
that $1/f$ noise  results in the Gumbel distribution of the
roughness, which has been interpreted in terms of extreme value
statistics in Ref. \cite{BertinClusel}.

  Terms of the
order of $1/\sqrt{K}$ vanish in the Taylor expansion of
Eq. (\ref{toexpand}), since the average values of $\cos{(2\pi f x)}$
and $\sin{(2\pi f x)}$ on interval $(0,1)$ are equal to 0. To
order $1/K,$ we obtain
\begin{equation}
\alpha \approx
1+\frac1K\log{2\pi}-\sum_{f=1}^{\infty}\frac{1}{2fK} (t_{1,f}^2
+t_{2,f}^2-2).
\end{equation}
The constant term  $-\log{2\pi}$ in $G^{per}(x,y)$ gives a
constant rescaling of the distribution function of $\alpha,$ and doesn't
show up in the normalized fringe amplitude $\tilde \alpha:$
\begin{equation}
\tilde \alpha \approx 1-\sum_{f=1}^{\infty}\frac{1}{2fK}
(t_{1,f}^2 +t_{2,f}^2-2).
\end{equation}
Thus to the leading order in $1/K$ expansion, the distribution $Y(x)$ of
the rescaled variable
\begin{equation}
x=-K (\tilde \alpha-1)= \sum_{f=1}^{\infty}\frac{1}{2f} (t_{1,f}^2
+t_{2,f}^2-2)
\end{equation}
equals
\begin{eqnarray}
Y(x)=\prod_{f=1}^{f=\infty}\int_{-\infty}^\infty
\int_{-\infty}^\infty \frac{d t_{1,f} d
t_{2,f}}{2\pi}e^{-\sum_{f=1}^{f=\infty}\frac{t_{1,f}^2+t_{2,f}^2}2}\delta\left(x-\sum_{f=1}^{f=\infty}\frac1{2f}(t_{1,f}^2
+t_{2,f}^2-2)\right). \label{Wx}
\end{eqnarray}
To prove Eq. (\ref{Gumbeldef}), we need to show $Y(x)=P_G(x),$ where
Gumbel function $P_G(x)$ is given by Eq. (\ref{gumbelfuncdef}). Let
us introduce new positive variables
\begin{equation}
u_f= \frac{(t_{1,f}^2 + t_{2,f}^2)}{2f}>0 ,
\end{equation}
then
\begin{eqnarray}
Y(x)=\prod_{f=1}^{f=\infty}\int_{0}^\infty f d
u_{f}e^{-\sum_{f=1}^{f=\infty}  f u_f
}\delta\left(x-\sum_{f=1}^{f=\infty}(u_f -1/f)\right).
\label{Wxnew}
\end{eqnarray}
To prove that $Y(x)=P_G(x)$ we will calculate their Fourier
transforms:
\begin{eqnarray}
Y(is)=\int_{-\infty}^{\infty} dx
Y(x)e^{isx}=\prod_{f=1}^{f=\infty} f d u_{f}e^{-  f u_f}
e^{is(u_f-1/f)}=\prod_{f=1}^{f=\infty}\frac{f
e^{-is/f}}{f-is}=e^{- i\gamma s }\Gamma[1-i s],\\
P_G(is)=\int_{-\infty}^{\infty} dx
P_G(x)e^{isx}=\int_{-\infty}^{\infty}
dxe^{-(x+\gamma)-e^{-(x+\gamma)}}e^{isx}=e^{-i\gamma s }\Gamma[1-i
s].
\end{eqnarray}
The proof above doesn't illustrate the meaning of $Y(x)$ as of a
distribution of extreme value. Here we will  follow the method of
Ref. \cite{BertinClusel} and explicitly construct the variable,
extreme value of which generates Gumbel distribution. Let us
impose a finite cutoff $f_{max}=N,$ and at the end of calculation
we will send $N$ to infinity. If one identifies
\begin{eqnarray}
z_1=u_{N}\\ z_2=u_{N-1}+u_{N},\\ ...\\ z_{N}=u_1+u_2+...+u_{N},
\end{eqnarray}
then $\{z_1,...,z_N\}$ is an ordered set (since $u_i>0$)  of
outcomes taken from Poissonian distribution, since
\begin{eqnarray}
e^{-\sum_{f=1}^{f=N}f u_f}=e^{-\sum_{f=1}^{f=N}z_f},
\end{eqnarray}
and Jacobian of transformation from  variables $\{u_f\}$ to
$\{z_f\}$ variables is  unity. Then
\begin{eqnarray}
Y_N(x)=N!\int_0^\infty e^{-z_1} dz_1 \int_{z_1}^\infty e^{-z_2}
dz_2 ... \int_{z_{N-1}}^\infty e^{-z_N} dz_N
\delta\left(x-\left(z_N-\sum_{f=1}^N\frac1f\right)\right)
\end{eqnarray}
is nothing but the shifted distribution of the largest  of $N$
numbers taken from the Poissonian distribution, and in the limit
of large $N$ this distribution converges to Gumbel function.

 One can understand the appearance of the Gumbel distribution by noting that for $K\gg 1$ the
distribution function of the interference amplitude is dominated
by rare fluctuations of the random periodic 1D string, which are
spatially well localized. The Gumbel distribution was introduced
precisely to describe similar rare events such as stock market
crashes or earthquakes. For open boundary conditions the universal
distribution for large $K$ is slightly different from Gumbel
function, similar to $1/f$ noise in other systems \cite{1fnoise}.
But the main properties, like the presence of asymmetry or the
asymptotic form of the tails are preserved.

\section{Conclusions}
\label{conclusions}
\subsection{Summary}

When we discuss interference experiments with ultracold atoms, the
conventional idea of the particle-wave duality takes a new meaning.
On the one hand, these experiments probe phase coherence which is
typically associated with coherent non-interacting waves.  On the
other hand, one can use powerful tools of atomic physics to control
interactions between atoms in a wide range and to reach the regime of
strong correlations. One can also prepare atomic systems in states
which would be difficult if not impossible to obtain in optics, e.g. low dimensional condensates with strong thermal or quantum fluctuations.  This
remarkable combination places interference experiments with ultracold
atoms in a unique position: they can address a problem of how the
interactions, correlations, and fluctuations affect the coherent
properties of matter. This question appears in many areas of physics,
including high energy and condensed matter physics, nonlinear quantum
optics, and quantum information. While the naive answer that
interactions suppress interference turns out to be correct in most
cases, the goal of these lecture notes was to demonstrate that the
{\it quantitative} analysis of this suppression can provide a lot of
nontrivial information about the original  systems.

We discussed two effects which contribute to the reduction of the
interference fringe contrast in matter interferometers. The first
effect is the shot noise arising from a finite number of atoms
used in a single measurement. This analysis is particularly
important for interference experiments with independent
condensates in which the position of interference fringes is
random and averaging over many shots can not be performed.  In
this case one needs to rely on single shot measurements to observe
interference patterns. While interference of independent
condensates has been discussed before \cite{Glauber, Javanainen,
Castin,
Zoller,Pethick,Laloe,Wong,Horak,Molmer,Naraschewski,Rohrl}, to our
knowledge, we provide the first derivation of the full
distribution function of the amplitude of interference fringes.
Another mechanism of the suppression of the amplitude of
interference fringes discussed in these lecture notes is the
quantum and thermal fluctuations of the order parameter in low
dimensional condensates. The motivation for this discussion comes
from the observation that interference experiments between
independent fluctuating condensates can be used to study
correlation functions in such systems \cite{pnas}. For example,
one can use the scaling of the integrated amplitude of
interference patterns to analyze two point correlation functions.
This method has been successfully applied by Hadzibabic et al.
\cite{zoran_KT} to observe the Berezinskii-Kosterlitz-Thouless
transition in two dimensional condensates. One conceptual approach
to understanding interference experiments with independent
condensates is to consider them as analogues of the Hanbury Brown
and  Twiss experiments in optics \cite{HBT}. In the latter
experiments interference between incoherent light sources appears
not in the average signal but in the higher order correlation
function. One important difference however is that matter
interference experiments are of a single shot type and information
is contained not only in the average fringe contrast but also in
the variation of the signal between individual shots. In
particular higher moments of the amplitude of interference fringes
contain information about higher order correlation functions
\cite{pnas}. A complete theoretical description of the fringe
contrast variations is contained in the full distribution
functions of the fringe amplitudes, which we calculate for one and
two dimensional condensates \cite{Gritsev, fcsshort} in the limit
when the number of atoms is large and the shot noise can be
neglected \cite{shotnoise}. An important aspect of these lecture
notes was identifying intriguing mathematical connections which
exist between the problem of calculating distribution functions of
interference fringe amplitudes and several other problems in field
theory and statistical physics, such as the quantum impurity
problem \cite{kanefisher}, tunneling in the presence of the
dissipation \cite{caldeira-leggett}, non-hermitian {\cal
PT}-symmetric quantum mechanics \cite{BB,Bender} and various
conformal field theories. We developed a novel mapping of a wide
class of such problems to the statistics of random surfaces,
which provided a complete {\it non perturbative} solution. In
certain cases we have  analytically proven \cite{fcsshort} the
relationship between the distribution function of fringe
amplitudes and the universal extreme value statistical
distribution \cite{gumbelbooks}.

\subsection{Some experimental issues}

We now comment on a few issues relevant for experimental analysis of
noise in interference experiments.
The amount of information contained in the experimentally measured
distribution function is directly related to the number of cumulants
which can be accurately extracted. This includes the second
cumulant $k_2$, which corresponds to the width of the distribution;
the third cumulant $k_3$, which is related to skewness,
$g_{1}=k_{3}/k_{2}^{3/2}$, and describes the asymmetry of the
distribution function, and so on. In general, the statistical error
in determining the $n^{th}$ order cumulant after $N$ measurements
scales as $\sqrt{A_{n}/N}$, where $A_n$ is a constant which grows
with $n$ and depends on the higher moments of the distribution. For
example, to experimentally distinguish the normal and Gumbel
distributions it is necessary that the statistical error in skewness
is at least a factor of two smaller than the mean skewness, which
for the Gumbel distribution is $g_{1}\approx -1.14$. Thus the
minimal number of measurements required is $N_{min}\approx
24/g_{1}^{2}\approx 20$, where we used $A_3\approx 6$,  appropriate
for the normal distribution~\cite{kendall}. In practice the
required number of measurements may be higher because of the
influence of other possible sources of noise. However, it is  certainly
experimentally feasible.

Another experimentally relevant issue is the effect of the
inhomogeneous density due to the parabolic confining potential.
While the approach discussed in these lecture notes can be extended to include the inhomogeneous density profile, interpretation of the experimental results is more straightforward when density variations can be neglected.
 We note that when the condensate density varies gradually in space, the
power-law decay of the correlation functions is not strongly
affected \cite{KGDS}, except that the exponent may be different in
different parts of the trap (correlation function exponents
typically depend on the density). We expect that qualitatively our
results will not change provided that the power law decay of the
correlation functions is much stronger than the change of the
condensate density in the measured part of the cloud. To be more
precise, the best comparison with theory can be achieved when the
observation region $L$ is much smaller than the size of the
condensate, determined by the effective Thomas-Fermi length
\cite{DLO}, $R_{TF}=(3N\hbar^{2}/(m^{2}\omega^{2}a_{1D}))^{1/3}$
(here $N$ is the number of atoms of mass $m$ and $a_{1D}$ is the
one-dimensional scattering length). As long as $L$ remains much
larger  the healing length, our analysis is valid. In the regime
of weakly interacting atoms, one can show that the ratio between
the effective Luttinger parameter $K$ at the center of the
harmonic trap and at the boundary of the observation region is
given by $1-L^{2}/8R_{TF}^{2}$, thus giving only a small
correction to the distribution function computed in the central
region.  One can also reach similar conclusions in the strongly
interacting regime. It is also worth pointing out that we expect
the limiting case of the Poissonian distribution to be
particularly robust to the inhomogeneous density of atoms. Indeed
the Poissonian distribution is related to the fast $1/\sqrt{x}$
decay of the one-particle correlation functions in the strongly
interacting limit. This scaling is a universal feature of the
Tonks-Girardeau limit of bosons and is not affected by the weak
harmonic trap \cite{Minguzzi-Gangardt}.


\subsection{Outlook}

Before concluding these lecture notes we would like to discuss
questions which still need to be understood in the
context of interference experiments with ultracold atoms. We also
suggest an outlook for  future theoretical work.
\begin{trivlist}

\item {\it Combining shot noise with the order parameter fluctuations}.
\newline
A careful reader has undoubtedly noticed that we discussed {\it
either} the shot noise {\it or} the order parameter fluctuations. At
this point we are still lacking theoretical tools which would allow
to include both effects simultaneously. One of the difficulties is
that such analysis requires the  knowledge of   the correlation
functions for all distances rather than  the long distance
asymptotic form. Indeed, in section \ref{lowdim} we showed` that the short distance part of the correlation functions gives contribution of the same
order as the shot noise. In the particular case of the interference
of 2D condensates, the knowledge of the short distance behavior of
the correlation functions is needed to include the effect of the
vortex excitations below the BKT transition.

\item {\it Stacks of independent condensates}
\newline
In these lecture notes we focused on interference patterns from a
single pair of condensates. However in experiments one often has a
stack of several condensates (see e.g. \cite{stock}). In this
case interference arises from all possible pairs, and the system
provides intrinsic averaging and suppression of the noise. For a finite number of condensates self-averaging is not complete
and one expects finite fluctuations of the fringe contrast. It would
be useful to generalize analysis of the  shot noise and order
parameter fluctuations to such systems.

\item{\it Dynamics of interacting atoms}.
\newline
One of the advantages of the cold atoms systems is the possibility
to study non equilibrium coherent dynamics of interacting systems.
In particular dynamical splitting of a single condensate into a pair
of condensates has been performed in experiments on microchips
\cite{Hofferberth, phasedyn, MIT_squeezing,Jo2} and stimulated theoretical work on
the subject \cite{sanpera,Caz,Altman_decoh,Burkov,gritsev1,gritsev2}. Similar experiments
can also be done using superlattice potentials in optical lattices
which are now available in experiments \cite{mainz_superlattice, NIST_superlattice}. While
analysis of fringe amplitude distribution functions presented in these lecture notes dealt exclusively with
systems in the thermodynamic equilibrium, it would be interesting to
generalize it to systems undergoing nonequilibrium dynamical evolution.

\item{\it Interference experiments with fermions}
\newline
The discussion presented in these notes was limited to the case of
interference of bosons.  Such experiments can also be done
with fermions \cite{fermion_interference}, which are available
experimentally in different dimensions \cite{Demarcojin1999,1dfermions, 2dfermions}.
For fermions, modulation of the density can be
related to fermion antibunching \cite{Henny,Oliver,pra_noise}. Analysis of the noise of the fringe contrast
visibility for fermions would be an interesting problem too.

\item{\it Generalization to other systems}
\newline
We note that mapping of the Coulomb gas into the statistics of
random surfaces introduced in section \ref{Exactsolution} should
have applications beyond calculating the distribution functions
of the interference fringe amplitudes. This is a new non
perturbative tool to calculate partition functions of a variety
of other systems that can be represented as Coulomb gas models.
Examples include quantum impurity-related problems
\cite{kanefisher}, Sine-Gordon models where interaction strength
can depend on position, and many others.  Our mapping is not
related to the existence of the exact solution of Sine-Gordon
models, but relies only on the factorable structure of the
many-point correlation functions in the absence of interactions,
which is a general property of a Gaussian action.

\end{trivlist}

\subsection{Acknowledgements}

Many results presented in this review originally appeared in the
research papers published together with E. Altman and A.
Polkovnikov, to whom we owe a special gratitude. We also thank I.
Bloch, R. Cherng, M. Greiner, M. Lukin, G. Morigi, T. Porto, J.
Schmiedmayer, J. Thywissen, V. Vuletic, and P. Zoller for numerous
enlightening discussions. This work was partially supported by the
NSF Grant No. DMR-0132874, MIT-Harvard CUA and AFOSR. V.G. was also
supported by the Swiss National Science Foundation, grant
PBFR2-110423.

\appendix
\section{Large $K$ expansion}
\label{largeKexpansion}

In this Appendix, we will describe a systematic "diagrammatic
technique" to calculate $Z_{2n}$ or $Z^{per}_{2n}$ as an
expansion in small parameter $1/K.$  It corresponds to the "high
temperature" limit of the classical gas analogy discussed
in section \ref{highermomentssec}. This expansion can be applied both in 1D or 2D, and can be used to study the limiting distribution at large $K,$ which
for PBC in 1D has been conjectured \cite{Gritsev} and proven
\cite{fcsshort} to be the Gumbel distribution \cite{gumbelbooks}.

\subsection{Expansion to order $(1/K)^2$.}
\label{k2} We will start from the 1D case by expanding the exponent
in Eq. (\ref{Z2n}):

\begin{eqnarray}
 Z_{2n}=\int_{0}^1
...\int_{0}^1 du_1 ... du_n dv_1... dv_n
e^{\frac1K\left(\sum_{i<j}G(u_i, u_j) + \sum_{i<j}G(v_i,v_j)-\sum_{ij} G(u_i,v_j)  \right)}= \nonumber \\ \int_{0}^1
...\int_{0}^1 du_1 ... du_n dv_1... dv_n \sum
_{m=0}^{\infty}\frac1{m!K^m}\left(\sum_{i<j}G(u_i,u_j) +
\sum_{i<j}G(v_i,v_j)-\sum_{ij} G(u_i,v_j)  \right)^m.
\label{1Kexp}
\end{eqnarray}

In the  first order of expansion in powers of $1/K,$  $G(x,y)$
dependence comes only through one integral
\begin{eqnarray}
I_0=\int_{0}^1\int_{0}^1 dx dy G(x,y)
\end{eqnarray}
 after the integration in (\ref{1Kexp}).

 The prefactor depends only on $n,$ and can be calculated analytically:
 the total number of $u-u$ terms is $n(n-1)/2,$ the total number of $v-v$ terms is also $n(n-1)/2,$
 and the total number of $u-v$ terms is $n^2.$ The latter come with $-1$ sign, so the total expression for $Z_{2n}$ up to
 $O(1/K^2)$ is

\begin{eqnarray}
Z_{2n}=1-\frac1K n I_0 + O(\frac1{K^2}).
\end{eqnarray}
In second order,
 dependence of $Z_{2n}$ on $G(x,y)$ comes through three different integrals:

\begin{eqnarray}
I_{1,2}=\int_{0}^1\int_{0}^1 \int_{0}^1\int_{0}^1 dx dy  dz dt G(x,y) G(z,t)= \left (\int_{0}^1\int_{0}^1 dx dy G(x,y) \right )^2=I_0^2, \\
I_{2,2}=\int_{0}^1\int_{0}^1 \int_{0}^1 dx dy dz G(x,y) G(y,z), \\
I_{3,2}=\int_{0}^1\int_{0}^1  dx dy G(x,y) G(x,y).
\end{eqnarray}

\begin{figure}
\psfig{file=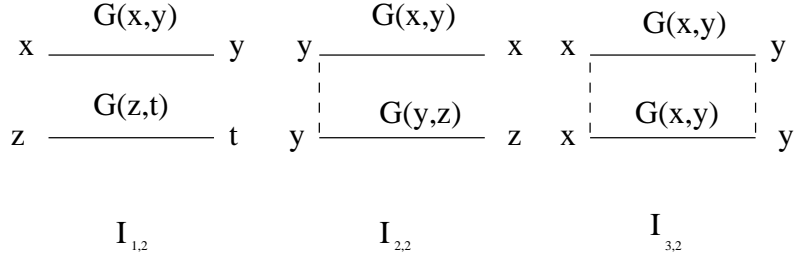} \caption{\label{k2diags} Topologically
inequivalent diagrams, corresponding to $1/K^2$ terms in expansion
of $Z_{2n}$. }
\end{figure}

Pictorially, these expressions can be represented by the
corresponding diagrams, shown in Fig. \ref{k2diags}. There,
horizontal solid line  corresponds to $G(x_i, x_j).$ If two ends
are connected by a dashed line, then in the integral these two
ends should correspond to the same variable. To write  the
expression for $Z_{2n}$ in  $1/K^2$ order, one has to calculate
the total number of terms corresponding to $I_{1,2}, I_{2,2},
I_{3,2}.$ Clearly, it will be some universal polynomial depending
on $n,$ which is determined by combinatorics. Here we describe
how to do this combinatorics in detail.

When parenthesis are multiplied in (\ref{1Kexp}) there are
several types of expressions which appear, and come with different
signs to the integral:

\begin{eqnarray}
\left(\sum_{i<j}G(u_i, u_j) +
\sum_{i<j}G(v_i, v_j)-\sum_{ij} G(u_i,v_j)  \right)^2= \sum_{i1,j1,i2,j2}G(u_{i1},v_{j1})G(u_{i2},v_{j2})+ \nonumber\\
\sum_{i1<j1,i2<j2}\left(G(u_{i1},u_{j1})G(u_{i2},u_{j2})+ G(v_{i1},v_{j1})G(v_{i2},v_{j2})+ 2 G(u_{i1},u_{j1})G(v_{i2},v_{j2})\right)\nonumber  \\
- 2 \sum_{i1<j1,i2,j2} (G(u_{i1},u_{j1})G(u_{i2},v_{j2}) + G(v_{i1},v_{j1})G(u_{i2},v_{j2}) ).
\label{m2exp}
\end{eqnarray}

 These are (signs are indicated)

\begin{eqnarray}
+G(u_{i1},u_{j1})G(u_{i2},u_{j2}), \; \mbox{and equivalent after integration} \; G(v_{i1},v_{j1})G(v_{i2},v_{j2});  \label{c1}\\
+G(u_{i1}, v_{j1})G(u_{i2},v_{j2}); \label{c2} \\
-G(u_{i1}, u_{j1})G(u_{i2},v_{j2}), \; \mbox{and equivalent after integration}\; -G(v_{i1}, v_{j1})G(u_{i2},v_{j2}); \label{c3} \\
+G(u_{i1}, u_{j1})G(v_{i2},v_{j2}); \label{c4}
\end{eqnarray}

For terms which have $G(u_{ik},u_{jk})$ or $G(v_{ik},v_{jk}),  k=\{1,2\}$ there is a restriction that $ik<jk,$ but there is
no such restriction for $G(u_{ik},v_{jk}).$

Lets calculate the total number of $I_{1,2}, I_{2,2}$ and $I_{3,2}$ terms for $u-u-u-u$ expressions of type (\ref{c1}). Total number of $I_{1,2}$ terms compatible
with restrictions is
\begin{eqnarray}
\frac1{2!^2}n(n-1)(n-2)(n-3).
\end{eqnarray}
$1/2!^2$ appears since pairs $\{i1, j1\}$ and $\{i2, j2\}$ are ordered , while $n(n-1)(n-2)(n-3)$ is the total number of ways
to choose a sequence of four different  not ordered numbers out of the set of size $n.$

Total number of $I_{2,2}$ terms compatible with restrictions is

\begin{eqnarray}
\frac1{2!^2}n(n-1)(n-2)*4.
\end{eqnarray}

\begin{figure}
\psfig{file=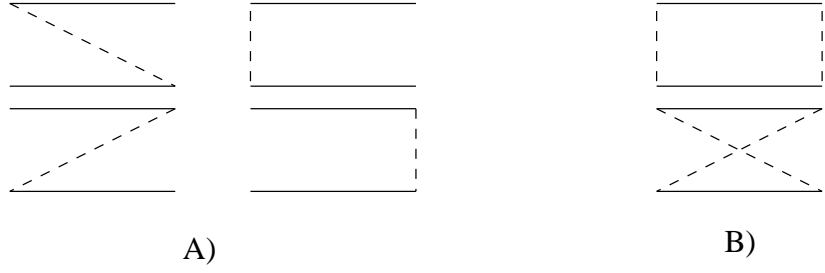} \caption{\label{k2topeqdiags}
Topologically equivalent diagrams, corresponding to a)$I_{2,2}$
and b)$I_{3,2}$. }
\end{figure}

Here the situation is similar to $I_{1,2},$ but there is a "topological" prefactor of $4,$ which corresponds to four topologically equivalent
diagrams for $I_{2,2},$ as shown in Fig. \ref{k2topeqdiags}.a.

Finally, total number of $I_{3,2}$ terms is
 \begin{eqnarray}
\frac1{2!^2}n(n-1)*2.
\end{eqnarray}
Here, $2$ is again a "topological" prefactor, corresponding to two topologically equivalent diagrams shown in Fig. \ref{k2topeqdiags}.b.

Overall, $u-u-u-u$ term of (\ref{c1}) gives
\begin{eqnarray}
\frac1{4}n(n-1)(n-2)(n-3) I_{1,2} + n(n-1)(n-2)I_{2,2} + \frac1{2}n(n-1) I_{3,2}
\end{eqnarray}
As a simple check of combinatorics one can calculate the total numbers of $u-u-u-u$ terms of type (\ref{c1}). It is

\begin{eqnarray}
\frac1{2!^2}n(n-1)(n-2)(n-3)+\frac1{2!^2}n(n-1)(n-2)*4+\frac1{2!^2}n(n-1)*2 =\left(\frac{n(n-1)}{2!}\right)^2,
\end{eqnarray}
as it should be from the calculation which neglects the dashed lines, and doesn't impose any restrictions on the terms
at different horizontal rows.

Analogously, one can calculate polynomials  for expressions (\ref{c2})-(\ref{c4}), and results are as follows, respectively:
\begin{eqnarray}
n^2(n-1)^2 I_{1,2} + 2n^2(n-1) I_{2,2} + n^2 I_{3,2}, \\
-\frac12n^2(n-1)(n-2) I_{1,2} -n^2(n-1) I_{2,2} , \\
\frac1{2^2}n^2(n-1)^2 I_{1,2}.
\end{eqnarray}
Summing all the terms with corresponding prefactors from (\ref{m2exp}), we obtain

\begin{eqnarray}
Z_{2n}= 1-\frac1K n I_0 +\frac1{2!K^2}(P_{1,2} I_{1,2} + P_{2,2}I_{2,2} + P_{3,2} I_{3,2}) + O(\frac1{K^3}) =\nonumber\\
1-\frac1K n I_0 +\frac1{2!K^2}(3n(n-1) I_{1,2} - 4n(n-1)I_{2,2} + (2n-1)n I_{3,2}) + O(\frac1{K^3}),
\label{z2nuptok2}
\end{eqnarray}
where $P_{1,2}=3n(n-1),P_{2,2}= - 4n(n-1),$ and $P_{3,2}=(2n-1)n$ are universal polynomials of $n.$ Notice, that due to sign cancellations of terms in $(\ref{c1})-(\ref{c4}),$ the overall degree of polynomials in $1/K^2$ order is
$m=2,$ compared to naively expected $2m=4.$

Using (\ref{z2nuptok2}), one can calculate $Z_{2n}/Z^n_2$ up to
$O(1/K^3):$

\begin{eqnarray}
\frac{Z_{2n}}{Z_2^{n}}=1+\frac{n(n-1)}{K^2}(I_{1,2}-2I_{2,2}+I_{3,2})+ O(\frac1{K^3})
\end{eqnarray}

Calculated values of $\{I_0, I_{1,2}, I_{2,2}, I_{3,2}\}$ and $\{I_0^{per},I_{1,2}^{per}, I_{2,2}^{per}, I_{3,2}^{per}\}$ are
\begin{eqnarray}
I_0=-\frac32=-1.5, \; I_0^{per}=-\log{2\pi}\approx-1.83788;\\
I_{1,2}=\frac94=2.25, \; I_{1,2}^{per}=(\log{2\pi})^2\approx3.37779;\\
I_{2,2}=\frac{51-\pi^2}{18}\approx2.28502, \; I_{2,2}^{per}=(\log{2\pi})^2\approx3.37779; \\
I_{3,2}=\frac72=3.5, \; I_{3,2}^{per}=\frac{\pi^2+12(\log{2\pi})^2 }{12}\approx4.20026.
\end{eqnarray}
Thus, for $K\rightarrow \infty,$ the limiting ratio of  the widths of the distributions for OBC and PBC are
\begin{eqnarray}
\frac{Z_{4}/(Z_{2})^2-1}{Z^{per}_{4}/(Z^{per}_{2})^2-1}\rightarrow\frac{I_{1,2}-2I_{2,2}+I_{3,2}}
{(I_{1,2}^{per}-2I_{2,2}^{per}+I_{3,2}^{per})}\approx 1.43465.
\end{eqnarray}

\subsection{General properties of $(1/K)^m$ terms,  and expansion to order $(1/K)^5$.}

>From the expansion of the previous subsection, one can formulate the general properties
of the "diagram technique" to calculate terms up to $(1/K)^m:$
\begin{itemize}
\item[a)]
First, one has to draw all possible topologically inequivalent diagrams, which consist of $m$ horizontal solid lines, with some of the ends connected by dashed lines. Each end can have at most two dashed lines coming out of it.
\item[b)]
Ends which are connected by a dashed line correspond to the same variable. Diagrams for which two opposite ends of the horizontal line correspond to the same variable are excluded.
\item[c)]
Expression which corresponds to a diagram is constructed the following way:  if variables at the end of a given solid horizontal line are $x$ and $y,$ then $G(x,y)$ should be put as one of the terms in the product under the integrand. Thus the integrand consists of the product of function $G$ of some variables $m$ times. Diagrams for which the integrands are the same up to relabeling of the variables are considered to be identical.
\item[d)]all free variables should be integrated from $0$ to $1.$
\end{itemize}

Diagrams can be connected or disconnected. For example, in Fig. \ref{k2diags} diagram corresponding
to $I_{1,2}$ is disconnected, and diagrams corresponding to $I_{2,2}$ and $I_{3,2}$ are connected. The integral which corresponds to a disconnected diagram is a product of expressions, corresponding to its parts: for example, $I_{1,2}=I_0 * I_0.$

If the number of topologically inequivalent diagrams of order $m$ is $g(m),$ then the term
of the order $1/K^m$ in $Z_{2n}$ has the following form:

\begin{eqnarray}
\int_{0}^1
...\int_{0}^1 du_1 ... dv_n \frac1{m!K^m}\left(\sum_{i<j}G(u_i,u_j) +
\sum_{i<j}G(v_i,v_j)-\sum_{ij} G(u_i,v_j)  \right)^m=\frac1{m!K^m}(\sum_{r=1}^{g(m)} P_{r,m}(n) I_{r,m}),\label{kmterm}
\end{eqnarray}
 where $P_{r,m}(n)$ are universal polynomials of $n,$  which can be calculated combinatorially, as described in section \ref{k2}.

 Polynomials $P_{r,m}(n)$ should satisfy the following requirements:
 \begin{itemize}
 \item[a)] For positive integer $n,$  the values of $P_{r,m}(n)$ are integer, and
 \begin{eqnarray}
 P_{r,m}(0)=0.
\end{eqnarray}

 \item[b)] If one sets $G(x,y)=1,$ then $I_{r,m}=1,$ and lhs of (\ref{kmterm}) can be trivially calculated. This implies
 \begin{eqnarray}
\sum_{r=1}^{g(m)} P_{r,m}(n)=(-n)^m. \label{itemb}
\end{eqnarray}

\item[c)] Degree of each polynomial $P_{r,m}(n)$ is not larger than $m.$
This has been shown above for $m=1$ and $m=2,$ and has been checked up to $m=5,$ although we didn't succeed in proving it directly. This conjecture is supported by the fact, that it guarantees that for any $G(x,y)$ for $K\rightarrow \infty$
distributions fit on the top of each other after proper rescaling (see next section).

 \end{itemize}

\begin{figure}
\psfig{file=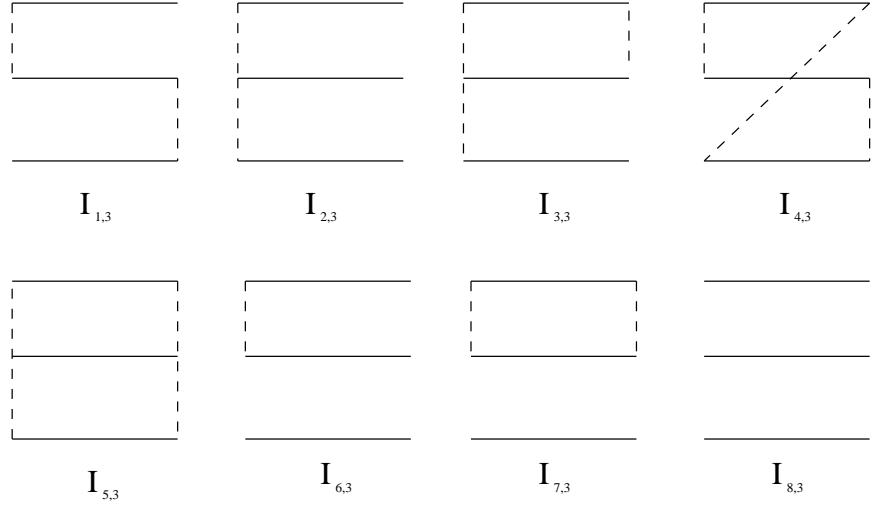} \caption{\label{m3} Topologically inequivalent diagrams for $m=3.$ Diagrams $I_{1,3}-I_{5,3}$ are connected and $I_{6,3}-I_{8,3}$ are disconnected.}
\end{figure}

One can do the combinatorial calculations similar to previous section for $m=3,$ and it takes about a day "by hand". We will only show the results here.
 For $m=3$ there are $g(3)=8$ topologically inequivalent diagrams, which are shown in Fig. \ref{m3}, out of which $5$ diagrams ($I_{1,3}$ to $I_{5,3}$) are connected. Expressions corresponding to each diagram  and universal polynomials are, respectively:

\begin{eqnarray}
I_{1,3}=\int_{0}^1\int_{0}^1 \int_{0}^1\int_{0}^1 dx dy dz dt G(x,y) G(x,z) G(z,t), P_{1,3}(n)=-12n(n-1)(2n-3),\\
I_{2,3}=\int_{0}^1\int_{0}^1 \int_{0}^1 \int_{0}^1 dx dy dz dt G(x,y) G(x,z) G(x,t), P_{2,3}(n)=12n(n-1),\\
I_{3,3}=\int_{0}^1\int_{0}^1 \int_{0}^1 dx dy dz G(x,y)^2 G(x,z), P_{3,3}(n)=-12n(n-1),\\
I_{4,3}=\int_{0}^1\int_{0}^1 \int_{0}^1 dx dy dz G(x,y) G(x,z) G(y,z),P_{4,3}(n)=4n(n-1)(2n-1),\\
I_{5,3}=\int_{0}^1\int_{0}^1 dx dy G(x,y)^3,P_{5,3}(n)=-n,\\
I_{6,3}=I_{2,2}* I_0,P_{6,3}(n)=36n(n-1)(n-2),\\
I_{7,3}=I_{3,2}* I_0,P_{7,3}(n)=-3n(n-1)(2n-3),\\
I_{8,3}=I_{1,2}* I_0= I_0^3, P_{8,3}(n)=-15n(n-1)(n-2).
\end{eqnarray}

\begin{figure}
\psfig{file=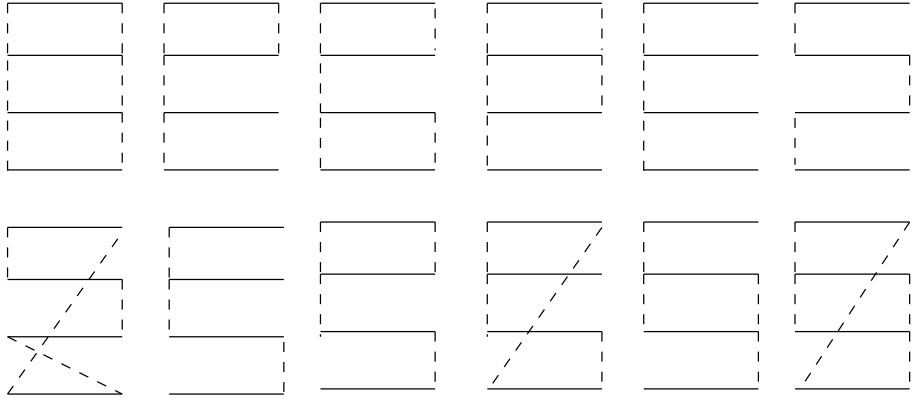} \caption{\label{m4} Irreducible diagrams for $m=4.$}
\end{figure}

Numerically evaluated integrals for OBC and PBC are

\begin{eqnarray}
I_{1,3}=-3.49399, I^{per}_{1,3}=-6.20797;\\
I_{2,3}=-3.5268, I^{per}_{2,3}=-6.20797;\\
I_{3,3}=-5.32704, I^{per}_{3,3}=-7.71956;\\
I_{4,3}=-4.21255, I^{per}_{4,3}=-6.50848;\\
I_{5,3}=-11.25, I^{per}_{5,3}=-12.5458 ;\\
I_{6,3}=-3.42753, I^{per}_{6,3}=-6.20797;\\
I_{7,3}=-5.25, I^{per}_{7,3}=-7.71956;\\
I_{8,3}=-3.375, I^{per}_{8,3}=-6.20797.
\end{eqnarray}
For $m>3$ it becomes too cumbersome to manually calculate
universal polynomials $P_{r,m}.$ We wrote a program in { \it
Mathematica}, which expands $m-\mbox{th}$ term of (\ref{1Kexp})
directly, and calculates the values  $\{P_{r,m}(0), ...,
P_{r,m}(n)\}$ using powerful pattern recognition tools. After
that, $P_{r,m}(n)$ is recovered using Newton's formula. Results
for $m=3$ can be recovered that way. One can also check in each
order that the degree of $P_{r,m}(n)$ is not larger than $m.$ For
each $m$ the program needs as an input all topologically
inequivalent diagrams, and currently the results have been
extended to $m=5.$ For $m=4,$ the overall number of diagrams is
$23,$ out of which $12$ are irreducible, and shown in Fig.
\ref{m4}. For $m=5,$ the overall number of diagrams is $66,$ out
of which $33$ are irreducible. Numerical prefactors of polynomials
$P_{r,m}(n)$ grow with $m,$ while their overall sum has a
prefactor $1,$ as follows from (\ref{itemb}). For example, one of
the polynomials for $m=5$ has a prefactor of $384$. This puts
stringent requirements on the errors in calculation of
$I_{r,m}(n),$  so going beyond $m=5$ will require additional
numerical effort. For $m=5$ results are reliable, since we can
check the numerical accuracy of the calculation of integrals by
comparison with analytically proven \cite{fcsshort} Gumbel
distribution.

While calculation of original $Z_{2n}$ requires
$2n-\mbox{dimensional}$ numerical integration, calculation of the
$I_{r,m}$ requires at most  $m-\mbox{dimensional}$ numerical
integration. This can be  seen from the fact that even for
irreducible diagrams, only the parts which contain "loops" have
to be integrated numerically. Horizontal lines which have a free
end can be "integrated out" analytically, and the dimension of
the numerical integral has at most dimension of the loop. For
example, first and third horizontal bars in $I_{1,3}$ of Fig
\ref{m3} can be integrated out using the following identity:
\begin{equation}
\int_{0}^{1}dy\log{|x - y|}=-1 + \log{(1 - x)} - x \log{(1 - x)} + x \log{x}, \label{simplifyint}
\end{equation}
so one has to do only 2-dimensional integral numerically.
Analogously, diagram $I_{2,3}$ requires only one dimensional
numerical integration. In each order $m$ there is only one
diagram which requires $m-\mbox{dimensional}$ integration, i.e.
$I_{4,3}$ in Fig.\ref{m3} for $m=3.$ All the rest require at most
$(m-1)-\mbox{dimensional}$ integration.

\subsection{Properties of the $K\rightarrow \infty$ distribution}
\label{Kinfty} For $K\rightarrow \infty$ distribution  function
becomes very narrow, and it is an interesting question to
investigate the limiting behavior of the distribution function.
Let us consider the distribution of normalized contrast $\tilde
\alpha,$ defined by (\ref{walpha}). Due to normalization the
following relations hold:

\begin{eqnarray}
\int_0^{\infty} \tilde W(\tilde \alpha) d\tilde \alpha\equiv 1, \int_0^{\infty}  \tilde \alpha \tilde W(\tilde \alpha) d\tilde \alpha\equiv 1.
\end{eqnarray}

For large $K$ distribution function is peaked near $\tilde
\alpha=1,$ and its width is proportional to $1/K.$ To calculate
the properties of the universal function we assume that it decays
relatively fast (exponential decay is enough) away from $\tilde
\alpha=1$ for large $K,$ so we can extend integrations to
infinity. If we define the fluctuation of the normalized contrast
$\tilde \beta=\tilde \alpha -1$ with distribution function
$\tilde W_0(\tilde \beta)=\tilde W(1+\tilde \beta),$ then

\begin{eqnarray}
-1+\frac{Z_{2}}{Z_2}\equiv0=\int_0^{\infty} (\tilde \alpha-1) \tilde W(\tilde \alpha) d\tilde \alpha =\int_{-1}^{\infty}\tilde \beta\tilde W_0(\tilde\beta) d\tilde\beta  \approx \int_{-\infty}^{\infty}\tilde \beta\tilde W_0(\tilde\beta) d\tilde\beta\equiv \tilde M_1,\label{z1tilde}\\
-1+\frac{Z_{4}}{Z_2^2}= \int_0^{\infty} (\tilde \alpha^2-1)\tilde W(\tilde \alpha) d\tilde \alpha   \approx \int_{-\infty}^{\infty} (2\tilde\beta+\tilde\beta^2) \tilde W_0(\tilde\beta) d\tilde\beta\equiv 2 \tilde M_1 + \tilde M_2,\label{z2tilde}\\
-1+\frac{Z_{6}}{Z_2^3}= \int_0^{\infty} (\tilde \alpha^3-1)\tilde W(\tilde \alpha) d\tilde \alpha  \approx \int_{-\infty}^{\infty} (3\tilde \beta+3\tilde \beta^2+\tilde \beta^3) \tilde W_0(\tilde \beta) d\tilde \beta\equiv 3 \tilde M_1 + 3 \tilde M_2+ \tilde M_3 \label{z3tilde},\\
...,
\end{eqnarray}
where $\tilde M_1,\tilde M_2,\tilde M_3 $ are the moments of $\tilde W_0(\tilde\beta).$
Inverting (\ref{z1tilde})-(\ref{z3tilde}), one can find $\tilde
M_2, \tilde M_3, ...$ as series expansion in powers of $1/K.$ If
for large $K$ the distribution functions collapse on the top of
each other after proper rescaling, as has been conjectured in
\cite{Gritsev}, then  the expansion of $\tilde M_m$ in powers of
$1/K$ should start from  $1/K^m.$ Below we will show, that this is
necessarily true for all $n$, if the degrees of universal
polynomials $P_{r,m}(n)$ are not larger than $m$ and $P_{r,m}(0)=0.$

To show this, let us consider an analog of Eqs.
(\ref{z1tilde})-(\ref{z3tilde}) for the fluctuation of
the unnormalized contrast $\beta=\alpha-1$ with distribution
$W_0(\beta)=W(1+\beta)$ (normalized contrast considered in
Appendix \ref{Kinfty} is related to $\alpha$ by simple rescaling).
Then
\begin{eqnarray}
Z_{2n}=\int_0^{\infty} \alpha ^n W(\alpha)d\alpha=\int_{-1}^{\infty}(1+\beta)^n W_0(\beta) d\beta  \approx\int_{-\infty}^{\infty}\sum_{i=0}^{i=n}C^i_n\beta^i W_0(\beta) d\beta =\sum_{i=0}^{i=n}C^i_n M_i,\label{Midef}
\end{eqnarray}
where $M_i$ is the $i-\mbox{th}$ moment of $W_0(\beta).$ We need
to show that expansion of $M_i$ in powers of $1/K$ starts only
from the terms of the order of $1/K^i.$ This can be seen from
expansion (\ref{kmterm}) together with (\ref{Midef}).
If degrees of $P_{r,m}(n)$  are not larger than $m,$ then
(\ref{kmterm}) means that $1/K^m$ terms in $Z_{2n}$ grow at most
as  $n^m$  for large $n.$ On the other hand,
$C^i_n=\frac{n!}{i!(n-i)!}$  grows as  $n^i$  for large $n,$ and
if expansion of $M_i$ in powers of $1/K$ starts before $1/K^i,$
this will contradict the previous sentence.

To find the first nontrivial
contribution to $\tilde M_m,$ one has to go to order $1/K^m$ in
the expansion. Using results for $m=5$ calculated above, one can calculate for $K \rightarrow \infty$ limiting behavior of  $K^2 \tilde M_2,...,K^5\tilde M_5$ for periodic and non-periodic cases, and compare it with the result of the Gumbel distribution
\begin{eqnarray}
\tilde W_G(\tilde \beta)=K e^{K\tilde \beta -\gamma - e^{K\tilde
\beta - \gamma}},
\end{eqnarray}
where $\gamma\approx 0.577$ is the Euler gamma-constant.
One obtains:
\begin{eqnarray}
\mbox{Non-periodic}: K^2 \tilde M_2\rightarrow 2.35991, K^3\tilde M_3\rightarrow -5.105577,
K^4\tilde M_4\rightarrow 37.5258, K^5\tilde M_5\rightarrow -242.492,
\frac{\tilde M_3^2}{\tilde M_2^3}\rightarrow 1.98336;\\
\mbox{Periodic}: K^2 \tilde M^{per}_2\rightarrow 1.64493, K^3\tilde M^{per}_3\rightarrow -2.40411, K^4\tilde M^{per}_4\rightarrow 14.6114,K^5\tilde M^{per}_5\rightarrow-64.4321;
\\
\mbox{Gumbel}: K^2 \tilde M^{G}_2\rightarrow 1.64493, K^3\tilde M^{G}_3\rightarrow -2.40411, K^4\tilde M^{G}_4\rightarrow 14.6114,K^5\tilde M^{G}_5\rightarrow-64.4321,
\frac{(\tilde M^{G}_3)^2}{(\tilde M^{G}_2)^3}\rightarrow 1.29857.\\
\end{eqnarray}

Clearly, for periodic case the agreement with the Gumbel distribution
is excellent, but for non-periodic case, one can unambiguously
conclude that limiting function is NOT a Gumbel function. For
non-periodic case the function is more widely distributed
compared to the Gumbel function, as evident from the higher moments. Here
for completeness we provide numerical results for
$Z_{2n},Z^{per}_{2n}$ up to $m=5:$
\begin{eqnarray}
Z^{per}_{2n}=1+\frac{1.8379 n}{K}+\frac{-0.4112 n  + 2.5114
n^2}{K^2}+\frac{0.1002 n - 0.1548  n^2 + 2.1456
n^3}{K^3}+\nonumber\\ \frac{-0.03382 n + 1.0804  n^2 - 0.7399 n^3
+ 1.7369 n^4}{K^4}+\frac{0.01535 n + 1.05941  n^2 - 0.161727 n^3 +
0.172302 n^4 +
    0.93098 n^5}{K^5},\nonumber\\
Z_{2n}=1+\frac{1.5 n}{K}+\frac{-0.554956 n + 2.30496
n^2}{K^2}+\frac{0.049657  n + 0.343838  n^2 + 1.481505
n^3}{K^3}+\nonumber\\ \frac{-0.04741 n + 2.0329  n^2 - 1.8735 n^3
+ 1.8256  n^4}{K^4}+\frac{0.0106468n + 3.14868 n^2 -
    4.1829n^3 + 2.8980n^4 +
    0.0943039n^5}{K^5}.\nonumber
\end{eqnarray}

\subsection{D=2}

Here we will  briefly consider results of the
expansion in powers of $1/K$ for 2-dimensional case. Correlation
function below the BKT transition is given by Eq. (\ref{2dcorrfunc}),
and for square imaging area with unity aspect ratio, all
discussions of one dimensional case carry over, with substitutions
\begin{eqnarray}
\int_{0}^1 du_i \rightarrow \int_{0}^1\int_{0}^1 d^2\vec u_i, \int_{0}^1 dv_i \rightarrow \int_{0}^1\int_{0}^1 d^2 \vec v_i, G(\vec x, \vec y)=\log{|\vec x- \vec y|},
\end{eqnarray}
 and (\ref{Keta}).
In 2D case, there is one extra degree of freedom which can be
controlled in experiments, which is the aspect ratio of the
observation region. If the aspect ratio of the observation region is
very large, then the distribution function is essentially the
same as in the one-dimensional case. Below we concentrate on the
case with unity aspect ratio.   Similarly to one-dimensional
case,  the integral
\begin{equation}
\int_{0}^{1}\int_{0}^{1}dx_1 dy_1 \frac12 \log{\left((x_1-x_2)^2+(y_1-y_2)^2\right)}
\end{equation}
can be evaluated analytically, which somewhat simplifies the
numerical evaluation. However, the dimensions of integrals grow
fast with $m$, and  here we present results only up to $m=3:$
\begin{eqnarray}
Z_{2n}=1+\frac{0.805087 n}{K}+\frac{-0.0740n + 0.3342n^2}{K^2}+\frac{0.015n + 0.045n^2+0.241 n^3}{K^3}+ O(\frac1{K^4}).
\end{eqnarray}
The error in numerical coefficient due to integration is of the
order of the last reported digit. Compared to 1D case,
convergence is faster, which is consistent with our general
ideology that the role of fluctuations is larger in systems with
lower dimensions.

\section{Jack polynomials}

\label{Jack} The microcanonical  partition
functions $Z_{2n}^{(p)}$ can be computed using so-called Jack
polynomials. The Jack polynomials belong to a class of {\it
symmetric polynomials } \cite{McD,Sutherlandbook} and are a
one-parametric generalization of Schur symmetric functions and
generically can be defined as \begin{eqnarray}\label{J} J_{\lambda}({\bf
x};g)=\sum_{\mu\leq\lambda}v_{\lambda,\mu}(\beta)m_{\mu}(x), \end{eqnarray}
where the monomial symmetric function $m_{\lambda}(x)$ \begin{eqnarray}
m_{\lambda}(x)=\sum_{\sigma}x_{1}^{\lambda_{\sigma(1)}}\cdots
x_{n}^{\lambda_{\sigma(n)}} \end{eqnarray} includes sum over all
permutations $\sigma$. Here $v_{\lambda,\mu}$ are numerical
coefficients and $v_{\lambda,\lambda}=1$. The ordering in sum
(\ref{J}) means the ordering on the set of partitions
$\lambda=(\lambda_{1},\lambda_{2},\ldots,\lambda_{n})$,
$\lambda_{1}\geq\lambda_{2}\geq\ldots\geq\lambda_{n}\geq 0$ and
$\mu=(\mu_{1},\mu_{2},\ldots,\mu_{n})$,
$\mu_{1}\geq\mu_{2}\geq\ldots\geq\mu_{n}\geq 0$ of natural numbers
$\lambda$ and $\mu$.  The ordering $\mu\leq\lambda$ here means
that $\sum_{j=1}^{k}\mu_{j}\leq\sum_{j}^{k}\lambda_{j}$ for any
$k\leq n-1$. The partitions can be represented using Young
diagrams as follows: we put $\lambda_{1}$ boxes in the first row,
$\lambda_{2}$ boxes in the second, and so on. The definition of
Jack polynomials thus includes a sum over all Young diagrams for
which the number of rows is  smaller or equal to the number of
variables $n$. The Jack polynomials have the property of
orthogonality, which allows to bring the microcanonical
partition function into the following form (see \cite{Jack} for
more details on derivation) \begin{eqnarray}
Z_{2n}^{(p)}=c_{n}^{2}\sum_{\lambda}\prod_{i=1}^{n}\frac{\Gamma[\lambda_{i}+\frac{1}{2K}(n-i+1)]\Gamma[p+\lambda_{i}+\frac{1}{2K}(n-i+1)]}
{\Gamma[\lambda_{i}+1+\frac{1}{2K}(n-i)]\Gamma[p+\lambda_{i}+\frac{1}{2K}(n-i)]}.
\end{eqnarray} Here $c_{n}=\Gamma[n+1]/\Gamma^{n}[1/2K]$ and sum goes over
Young diagrams labeled by integers
$\lambda_{1}\geq\lambda_{2}\geq\ldots\lambda_{n}\geq 0$.

In particular, the  Jack polynomials expressions give the
following results for the lowest microcanonical partition
functions \begin{eqnarray} Z_{2}^{(p)}=
\frac{\sin(\frac{\pi}{2K})\Gamma(1-1/K)}{\sin\pi(\frac{1}{2K}+p)\Gamma(1-\frac{1}{2K}+p)\Gamma(1-\frac{1}{2K}-p)}
\end{eqnarray} and \begin{eqnarray}
Z_{4}^{(0)}&=&\frac{1}{\Gamma^{4}(\frac{1}{2K})}[\frac{2^{-1/K}\Gamma(\frac{1}{2}-\frac{1}{2K})\Gamma^{2}(\frac{1}{2K})\Gamma^{2}
(\frac{1}{K})}{\sqrt{\pi}\Gamma(1-\frac{1}{2K})\Gamma^{2}(1+\frac{1}{2K})}
{}_{p}F_{q}(\{1,\frac{1}{K},\frac{1}{K}\},\{
1+\frac{1}{2K},1+\frac{1}{2K} \},1)\nonumber\\
&-&\sum_{n=0}^{\infty}(\frac{\Gamma(\frac{1}{K}+n)}{\Gamma(2+n)})^{2}
{}_{p}F_{q}(\{1,1+\frac{1}{2K}+n,1+\frac{1}{2K}+n\},\{2+n,2+n\},1)],
\end{eqnarray}
where ${}_{p}F_{q}$ is  a generalized hypergeometric function
\cite{GrRy}. Apparently, for larger $n$, expressions for
$Z_{2n}^{(p)}$ become more and more cumbersome and are difficult to
use for real computations.

Jack polynomials appear particularly in connection to
Calogero-Sutherland model, i.e. the model which describes $n$
particles (the same $n$ as in formulas above) on a line
interacting via inverse square interaction. In the first quantized
form the Hamiltonian of this model can be defined as
\cite{Sutherlandbook} \begin{eqnarray}
H_{CS}=-\frac{1}{2}\sum_{j=1}^{n}\frac{\partial^{2}}{\partial
x_{j}^{2}}+\pi^{2}\sum_{1\leq i<j\leq
n}\frac{g(g-1)}{\sin^{2}\pi(x_{i}-x_{j})^{2}}. \end{eqnarray} Wave functions
for the excited states of this model are Jack polynomials.

\section{Thermodynamic Bethe Ansatz of the quantum impurity model}\label{TBA}

The model described by the partition function (\ref{zp}) can be
solved exactly by thermodynamic Bethe ansatz (TBA) \cite{FLSprl,
FLSprb}. This solution is non perturbative  in $g$. One should be
careful about the correspondence between the perturbative
expansion and the results from the non perturbative TBA
calculations. The correct correspondence between $Z(K,g)_{pert}$
and the non perturbative result of the TBA is given by
\begin{eqnarray}\label{Z}
Z_{0}(K,g)_{pert}=\frac{1}{\sqrt{2K}}\exp[\tilde{F}_{TBA}(K,g)],
\end{eqnarray}
where $\tilde{F}_{TBA}(K,g)$ is the free energy. It can be expressed
in terms of the energies $\epsilon(\theta)$ of elementary
excitations which depend on rapidity $\theta:$
\begin{eqnarray}\label{F}
\tilde{F}_{TBA}(K,g)=\int_{-\infty}^{\infty}\frac{d\theta}{2\pi}\frac{2K-1}{\cosh[(2K-1)(\theta
-\alpha) ]}\log(1+\exp(\epsilon_{+}(\theta))).
\end{eqnarray}
Here,
\begin{eqnarray}\label{alpha}
\alpha =\log(
[\frac{x}{\Gamma(\frac{1}{2K})}]^{\frac{2K}{2K-1}}\frac{2\sqrt{\pi}\Gamma(\frac{1}{4K-2})}{\Gamma(\frac{K}{2K-1})})
\end{eqnarray}
and the variable $x$ is the fugacity in the perturbative expansion
(\ref{zpert}).

\begin{figure}[ht]
\includegraphics[width=8cm]{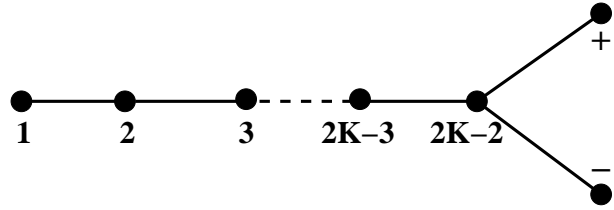}
\caption{Dynkin diagram for the algebra $D_{n}$. In the context of
sine-Gordon problem nodes $1...2K-2$ correspond to breathers,
whereas nodes $\pm$ correspond to kink/antikink. } \label{dynkin}
\end{figure}

To clarify the issue further we briefly describe the formalism of
TBA. The TBA equations are parametrized by the Dynkin diagrams
corresponding to the classical algebras $A_{n}$, $D_{n}$, $E_{n}$
\cite{Zamol}. Our case under consideration is related to the
diagrams $D_{n}$, which describes the sine-Gordon model. To each
node of the diagram we associate the particle with the mass given
by the Coxeter number of the corresponding diagram (which is equal
to $2K-2$) shown in Fig.(\ref{dynkin}). These are denoted by
$\mu_{+}=\mu_{-}$ (corresponding to the kink and antikink) and
$\mu_{j}$, $j=1...2K-2$, (note that we consider $K$ to be of the
form $n/2$, $n=3,4,5,...$) corresponding to breathers, bound
states of kink and antikink. Explicitly,
\begin{eqnarray}\label{mu}
\mu_{j}=2\mu_{+}\sin(\frac{j\pi}{4K-2}) \ \ \ \mbox{for
breathers}\ \ \ j=1...2K-2 .
\end{eqnarray}

 Due to a great
simplification of the structure of scattering matrices in
integrable theory \cite{Zamol} the free energy of the boundary
sine-Gordon problem can be expressed entirely in terms of the
energies $\epsilon_{+}$ (see Eq. (\ref{F}) above). This can
further be reduced as follows. Consider the TBA equations for the
energies:
\begin{eqnarray}
\epsilon_{r}(\theta)=\int_{-\infty}^{\infty}\frac{d\theta'}{2\pi}\frac{2K-1}{\cosh[(2K-1)(\theta
-\theta')]}\sum_{s}N_{rs}\log(1+\exp(\epsilon_{s}(\theta'))),
\end{eqnarray}
where $N_{rs}$ is the incidence matrix of the corresponding Dynkin
diagram. To clarify these notations consider few examples.

  For $K=3/2$ we have kink, antikink and one breather. Therefore
the diagram consists of the three nodes, say $+,1,-$. The vertex
$+$ (kink) is connected to the node $1$ and the vertex $1$ is
connected to the vertex $-$. Vertices $+$ and $-$ are
disconnected. The incidence matrix is therefore
\begin{eqnarray}
\left(
  \begin{array}{cc}
    1 & 0 \\
    1 & 1 \\
    0 & 1 \\
  \end{array}
\right)
\end{eqnarray}
(the number of rows is equal to the number of vertices, the number
of columns is equal to the number of bonds between these
vertices). Using this matrix we obtain the following set of
coupled TBA equations:
\begin{eqnarray}\label{K3/2}
\epsilon_{\pm}(\theta)=\int_{-\infty}^{\infty}\frac{d\theta'}{2\pi}\frac{2}{\cosh[2(\theta
-\theta')]}\log(1+\exp(\epsilon_{1}(\theta'))),\\
\epsilon_{1}(\theta)=2\int_{-\infty}^{\infty}\frac{d\theta'}{2\pi}\frac{2}{\cosh[2(\theta
-\theta')]}\log(1+\exp(\epsilon_{+}(\theta'))),
\end{eqnarray}
where we use the fact that $\epsilon_{+}=\epsilon_{-}$. Now, from
the second equation we can see that the $\epsilon_{1}(\theta)/2$
gives precisely the function $\tilde{F}$ from the Eq. (\ref{F}).
Therefore, for the case of $K=3/2$,

\begin{eqnarray}
Z_{0}(\frac{3}{2},g)_{pert}=\frac{1}{\sqrt{3}}\exp[\frac{1}{2}\epsilon_{1}(\alpha)],
\end{eqnarray} where $\epsilon_{1}$ must be determined from the set of
coupled equations (\ref{K3/2}).

 To continue, consider the case of $K=4/2$. Here we have 2
breathers, kink, and antikink. The incidence matrix is
\begin{eqnarray}
\left(
  \begin{array}{ccc}
    1 & 0 & 0 \\
    1 & 1 & 1 \\
    0 & 1 & 0 \\
    0 & 0 & 1 \\
  \end{array}
\right)
\end{eqnarray}
The set of TBA equations is therefore
\begin{eqnarray}\label{K4/2}
\epsilon_{\pm}(\theta)=\int_{-\infty}^{\infty}\frac{d\theta'}{2\pi}\frac{3}{\cosh[3(\theta
-\theta')]}\log(1+\exp(\epsilon_{2}(\theta'))),\\
\epsilon_{1}(\theta)=\int_{-\infty}^{\infty}\frac{d\theta'}{2\pi}\frac{3}{\cosh[3(\theta
-\theta')]}\log(1+\exp(\epsilon_{2}(\theta'))),\\
\epsilon_{2}(\theta)=\int_{-\infty}^{\infty}\frac{d\theta'}{2\pi}\frac{3}{\cosh[3(\theta
-\theta')]}\log(1+\exp(\epsilon_{1}(\theta'))) \nonumber \\
+2\int_{-\infty}^{\infty}\frac{d\theta'}{2\pi}\frac{3}{\cosh[3(\theta
-\theta')]}\log(1+\exp(\epsilon_{+}(\theta'))),
\end{eqnarray}
from which we conclude that the $\epsilon_{2}/3$ gives the
function $\tilde{F}$ from the Eq. (\ref{F}). Therefore, for the
case of $K=4/2$,

\begin{eqnarray}
Z_{0}(\frac{4}{2},g)_{pert}=\frac{1}{\sqrt{4}}\exp[\frac{1}{3}\epsilon_{2}(\alpha)],
\end{eqnarray} where $\epsilon_{2}$ must be determined from the set of
coupled equations (\ref{K4/2}).

For $K=5/2$ we can find that \begin{eqnarray}
Z_{0}(\frac{5}{2},g)_{pert}=\frac{1}{\sqrt{5}}\exp[\frac{1}{2}(\epsilon_{3}(\alpha)-\epsilon_{1}(\alpha))],
\end{eqnarray} where $\epsilon_{1,3}$ must be determined from the set of
corresponding coupled equations (easy to write down explicitly).

For $K=6/2$ we have \begin{eqnarray}
Z_{0}(\frac{6}{2},g)_{pert}=\frac{1}{\sqrt{6}}\exp[\frac{1}{2}(\epsilon_{4}-\epsilon_{2}+\tilde{F}(\epsilon_{1}))],
\end{eqnarray} and therefore it includes the function
$\tilde{F}(\epsilon_{1})$. This is not convenient from the point
of view of numerics since it contains an additional integration.

Continuing the above computations one can observe that for the
values of $K=2k/2$, $k=2,3,4,...$ we will always have an
additional function $\tilde{F}(\epsilon_{1})$ in the expression
for the perturbative function $Z_{0}(g,K)_{pert}$. On the other
hand for the values of $K=(2k+1)/2$, $k=1,2,3,...$ one can always
express the result for partition function in terms of energies
$\epsilon_{i}$. It seems that this is more convenient for
numerics. The result for this case is the following:
\begin{eqnarray}\label{half-odd-Z}
Z_{0}(g,
\frac{2n+1}{2})_{pert}=\frac{1}{\sqrt{2n+1}}\exp[\frac{1}{2}(\sum_{s=0}^{n-1}(-1)^{s}\epsilon_{(2n-1)-2s})],
\end{eqnarray}
where the set of energies must be determined from the solution of
coupled integral equations.

It is useful to consider the asymptotic behavior. For a large
$\theta\rightarrow\pm\infty$ values of the integration parameter,
the energies of breathers may be taken in the following form (see
Eq.(\ref{mu})):
\begin{eqnarray}
\epsilon_{s}(\alpha) =
\frac{\mu_{s}}{\mu_{+}}=2\sin[\frac{s\pi}{4K-2}]\exp[\alpha], \ \
\mbox{as} \ \ \alpha\rightarrow\infty.
\end{eqnarray}

The general strategy is clear now: one can solve numerically the
set of coupled integral equations for some values of $K=(2n+1)/2$
and compute the partition function using Eq. (\ref{half-odd-Z}).
At this point the difficulty arises: how to unambiguously define
the analytic continuation of $Z_{p}(K,g)$ into imaginary-$g$
domain. To avoid this problem it is better to use  the alternative approach
described in section \ref{thirdway}.


\begin{thebibliography}{10}
\bibitem{ballist} Readers may be concerned that
the initial part of expansion may not be purely ballistic and atomic
collisions may take place in expanding clouds.
 In the case of tight confinement such collisions are rare and have minimal effect. One can use such approximation in a variety of experimental situations, because the density of the condensate falls of rapidly during the first stages of the expansion. Effects of the
interactions on the interference signal have been recently discussed
in Refs. \cite{Xiong,Cederbaums, Masiello}. It is worth noting that
the effect of collisions can be minimized by tuning the scattering
length to zero at the moment of the release using magnetic Feshbach
resonances \cite{Feshbach}. See also discussion after Eq.
(\ref{Eq38}).
\end{thebibliography}
\end{document}